\documentclass[11pt]{article}
\usepackage[T1]{fontenc}
\usepackage[utf8]{inputenc}
\usepackage{textcomp}
\usepackage{lmodern}
\usepackage{microtype}
\usepackage[margin=1in]{geometry}
\usepackage{amsmath,amssymb,mathtools,bm}
\usepackage{graphicx}
\usepackage{xcolor}
\usepackage{tikz}
\usetikzlibrary{arrows.meta,calc,decorations.markings}
\usepackage{float}
\usepackage[section]{placeins}
\usepackage{longtable,booktabs,array,calc}
\usepackage[labelsep=period]{caption}
\usepackage{cite}
\usepackage{xurl}
\usepackage[colorlinks=true,linkcolor=blue,citecolor=blue,urlcolor=blue]{hyperref}
\usepackage{siunitx}
\usepackage{booktabs}
\usepackage{comment}
\newcommand{\coq}{\mathbin{\text{\copyright}}}
\title{Multi-stage Stern--Gerlach experiment modeled (with additional appendices)}
\author{Lihong V. Wang\\Department of Electrical Engineering\\Andrew and Peggy Cherng Department of Medical Engineering\\California Institute of Technology\\1200 E. California Blvd., MC 138-78, Pasadena, CA 91125, USA\\\texttt{LVW@caltech.edu}\\ORCID: 0000-0001-9783-4383}
\date{}
\begin{document}
\maketitle
\begin{abstract}
In the classic multi-stage Stern--Gerlach experiment conducted by Frisch
and Segrè, the Majorana (Landau--Zener) and Rabi formulae diverge afar
from the experimental observation while the physical mechanism for
electron-spin collapse remains unidentified. Here, introducing the
physical co-quantum concept provides a plausible physical mechanism and
predicts the experimental observation in absolute units without fitting
(i.e., no parameters adjusted) with a \emph{p}-value less than one per
million, which is the probability that the co-quantum theory happens to
match the experimental observation purely by chance. Further, the
co-quantum concept is corroborated by statistically reproducing exactly
the wave function, density operator, and uncertainty relation for
electron spin in Stern--Gerlach experiments.
\end{abstract}
\section{Introduction}\label{introduction}

Performed three years before the successful development of quantum
mechanics, the 1922 Stern--Gerlach experiment on silver atoms \cite{Gerlach1922}
quickly proved fundamental to quantum physics \cite{SchmidtBoecking2016,Castelvecchi2022}. The benchmark
experiment led to the quantization of all angular momenta, discovery of
electron spin, study of the measurement problem and superposition,
direct investigation of the ground-state properties of atoms without
electronic excitation, and selection of fully spin-polarized atoms
\cite{SchmidtBoecking2016}. Within a few weeks, Einstein and Ehrenfest concluded that spin
collapse cannot be interpreted by radiation, which would take 100 years
\cite{Einstein1922}. Recently, Wennerström and Westlund numerically simulated that
relaxation of 1~\(\mu\mathrm{s}\) qualitatively reproduced the double branched collapse
pattern \cite{Wennerstrom2012}, and Norsen interpreted spin collapse using the de
Broglie--Bohm pilot-wave theory \cite{Norsen2014}. The significance of the
Stern--Gerlach experiment and relevant works are detailed in a 2016
inspiring review \cite{SchmidtBoecking2016}, concluding that ``The physical mechanism
responsible for the alignment of the silver atoms remained and remains a
mystery'' and quoting Feynman, ``\ldots{} instead of trying to give you
a theoretical explanation, we will just say that you are stuck with the
result of this experiment.'' \cite{Feynman1963}.

Immediately, Heisenberg and Einstein proposed multi-stage Stern--Gerlach
experiments to explore deeper mysteries of directional quantization
\cite{SchmidtBoecking2016}. Ten years later, Phipps and Stern reported the first effort
\cite{Phipps1932}, which was unfortunately discontinued owing to Phipps'
involuntary return to the US \cite{SchmidtBoecking2016}. A year later, Frisch and Segrè
modified the same apparatus by adopting Einstein's suggestion on the use
of a single wire instead of three electromagnets to rotate spin; they
also improved magnetic shielding, slit filtering, and signal detection
\cite{SchmidtBoecking2016}. Despite the use of three layers of magnetic shielding for the
middle stage (i.e., the inner rotation chamber), the remnant or residual
fringe magnetic field was still \(0.42 \times 10^{-4}\) T (or 0.42 G).
Rather than fight the fringe magnetic field further, they took advantage
of it. The magnetic field from the wire in the middle stage cancels the
remnant field to produce a magnetic null point, around which the field
is approximated as a magnetic quadrupole; consequently, they
successfully observed nonadiabatic spin flip \cite{Frisch1933}. Note that only the
magnetic field near a null point is effective for nonadiabatic spin
flip; thus, the field far from a null point does not significantly
affect transition, and its detailed distribution is of little import.
Frisch and Segrè varied the wire current, which is the only independent
variable controlled here, over nearly two orders of magnitude
approximately uniformly on a logarithmic scale to observe the peak
fraction of spin flip and its entire range. They started and ended with
sufficiently extreme currents that yielded negligible fractions of spin
flip. Having reached a nearly zero fraction of spin flip at the highest
current might be the reason that they ceased increasing the current
further. Further, the calculation of the fraction automatically obviates
the requirement for absolute calibration. This data set suggests they
designed and executed the experiment with great care.

\begin{figure}[!htbp]
\centering
\includegraphics[width=4.53in,height=3.79667in]{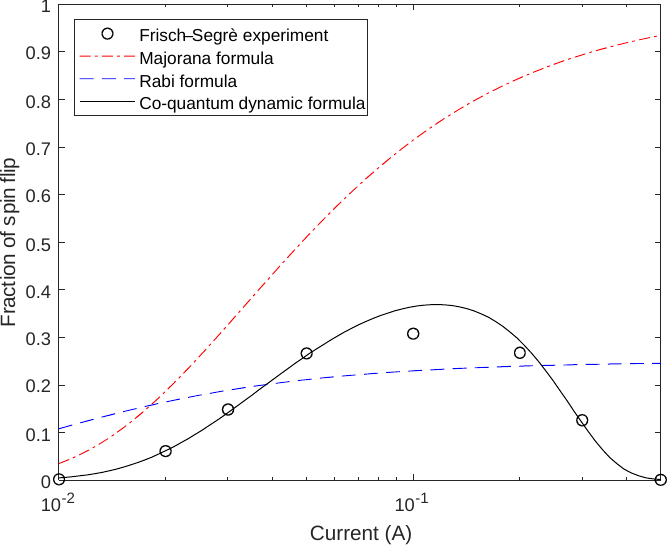}
\caption{Illustration of the divergence of the Majorana and Rabi
formulae from the Frisch--Segrè experimental observation and the
convergence of the co-quantum dynamic formula. Details are to be
discussed.}
\label{fig:1}
\end{figure}

Frisch and Segrè found that their observation \cite{Frisch1933} unexpectedly
diverges from the Majorana formula (Fig.~\ref{fig:1}) \cite{Majorana1932,Majorana2006}, which was
stimulated by the experiment of Frisch and Segrè. The Majorana formula
is a variant of the Landau--Zener formula, which is better-known despite
the concurrent publications of all four related papers in the same year
\cite{Landau1932,Zener1932,Stueckelberg1932}. For a historical comparison of the four papers, please
refer to Ref. \cite{Ivakhnenko2023}. Fermi suggested that interaction among atoms
could be responsible for the divergence, but atoms were sufficiently
sparse to be treated independently \cite{Frisch1933}. Rabi acknowledged
``Professor E. Segrè for discussions on the details of the Frisch and
Segrè experiment'', recognized the role of the nuclear magnetic moment,
and revised the Majorana formula through hyperfine coupling \cite{Rabi1936}.
Rabi's revised formula, however, did not overcome the divergence (Fig.
1).

Multi-stage Stern--Gerlach (Frisch--Segrè) experiments are much more
difficult to model than single-stage ones. Multiple stages produce far
more nuanced observation because the middle stage can vary the electron
spin orientation over a wide range after polarization by the first
stage. A correct single-stage theory must pass the more stringent test
of the multi-stage experiment. This spin-flip divergence in multi-stage
Stern--Gerlach experiments remains unresolved \cite{SchmidtBoecking2016}. One may only
speculate why the 1933 discrepancy \cite{Frisch1933} has not been resolved. The
seminal paper has not been republished in English, which might have
limited its visibility.

Here, a theory, called co-quantum dynamics (CQD) \cite{Wang2022}, is presented
to both provide a collapse mechanism and predict the Frisch--Segrè
experimental observation (Fig.~\ref{fig:1}) \cite{Frisch1933}. CQD is theoretically verified
by reproducing, for electron spin in Stern--Gerlach experiments, the
quantum mechanical wave function, density operator, and uncertainty
relation as well as, in a recent publication \cite{Wang2022}, the
Schrödinger--Pauli equation. In Methods, CQD is presented in three
subsections, including the equations of motion, branching condition, and
pre-collapse state function and prediction expression. In Results,
Stern--Gerlach experiments in both single and multiple stages are
modeled. For flow continuity, lengthy interpretations are postponed to
Discussion, and detailed mathematical derivations are presented in
Appendices (Supplement Material). Deferred to the last appendices are
the CQD derivations of the uncertainty relation, the entangled wave
function, and the observation in a two-stage Stern--Gerlach apparatus
with a varying angle between the quantization axes.

The following table (Table~\ref{tab:1}) compares briefly CQD with the
representative existing quantum mechanical theories for collapse
\cite{Carlesso2022}, e.g., the Ghirardi--Rimini--Weber model \cite{Ghirardi1986} and
continuous spontaneous localization model \cite{Pearle1989,Ghirardi1990}. CQD, based on
the classical Bloch equation (or its Landau--Lifshitz--Gilbert
derivative) and the two postulates, provides a physical instead of
phenomenological mechanism for electron spin collapse. In the presence
of an external magnetic field, the nuclear magnetic moment is
responsible for the collapse of electron spin. The absence of fitting
with any adjustable parameters and the high coefficient of determination
\(R^{2}\) (or high correlation coefficient) led to the small
\emph{p}-value (\(p < 8 \times 10^{-7}\)) \cite{Steel1960,Rahman1968}. In general,
fitting with more and more adjustable parameters, one may improve
\(R^{2}\) towards unity. While \(R^{2}\) is not penalized for the number
of adjustable parameters used relative to the number of experimental
data points available, the \emph{p}-value is. Therefore, one may achieve
an arbitrarily high \(R^{2}\) at the expense of the \emph{p}-value. The
\emph{p}-value is an objective measure of agreement between a theory and
the experiment. As a standard definition, the \emph{p}-value quantifies
the probability of observing results at least as extreme as the ones
observed given that the null hypothesis is true. For stringent
discoveries, high-energy physics, for example, requires \emph{p}
\(\leq\) \(3 \times 10^{-7}\), which corresponds to \(5\sigma\) \cite{Cousins2017}. The
LIGO observation of gravitational waves applied a similar criterion
\cite{Abbott2016}. The agreement of CQD with the experiment is at a similar level
as well. While the LIGO observed a chirp signal, which is common in
various forms in nature, the Frisch--Segrè experimental data follow an
uncommon shape, which is even more unlikely to be matched by random
chance. Therefore, the value of \(p < 8 \times 10^{-7}\) claims a
statistical significance that cannot be ignored objectively. The
probability that CQD happens to match the experimental observation so
well purely by chance is less than one in a million. It is even less
likely for an incorrect theory to match an incorrect experiment by
chance if one doubts the Frisch--Segrè experimental data. Because the
Majorana or Rabi formula, if correct, follows a monotonic trend, it
would be difficult to fathom that some experimental imperfections caused
the fraction of spin flip to increase at low currents and to decrease at
high currents. Matching a theory with the experiment so well without
using any adjustable parameters inspires conviction. Further, CQD is
corroborated by statistically reproducing exactly the wave function,
density operator, and uncertainty relation for electron spin. This
corroboration may be considered supporting evidence because an incorrect
theory would highly unlikely be able to reproduce so many fundamental
aspects of quantum mechanics.

\begin{table}[!htbp]
\centering
\caption{Comparison between representative existing quantum mechanical
theories and CQD.}
\label{tab:1}
\small
\begin{tabular}{@{}
  >{\raggedright\arraybackslash}p{(\columnwidth - 4\tabcolsep) * \real{0.3398}}
  >{\raggedright\arraybackslash}p{(\columnwidth - 4\tabcolsep) * \real{0.3361}}
  >{\raggedright\arraybackslash}p{(\columnwidth - 4\tabcolsep) * \real{0.3241}}@{}}
\toprule
\begin{minipage}[b]{\linewidth}\raggedright
\end{minipage} & \begin{minipage}[b]{\linewidth}\raggedright
\textbf{Existing theories}
\end{minipage} & \begin{minipage}[b]{\linewidth}\raggedright
\textbf{Co-quantum dynamics}
\end{minipage} \\
\midrule
\textbf{Domain} & Quantum mechanical & Semiclassical \\
\textbf{Starting equation} & Schrödinger equation & Bloch equation
(classical) \cite{Wang2022} \\
\textbf{Cause for collapse} & Phenomenological: no physical object
identified \cite{Carlesso2022} & Physical: nuclear magnetic moment identified \\
\textbf{Angular distribution of nuclear magnetic moment} & Discrete
(quantized); isotropic & Continuous; isotropic or anisotropic \\
\textbf{Collapse rate} & Preset as a constant dimensional rate (1/s) &
Scaled dynamically via a dimensionless constant (Eq.~\ref{eq:9}) \\
\textbf{Measurement uncertainty} & Inequality & Equality (Eq.~\ref{eq:186}),
yielding the inequality (Eq.~\ref{eq:187}) \\
\textbf{Quantitative prediction of multi-stage Stern--Gerlach
(Frisch--Segrè) experiment} & Not found yet in the literature except the
Majorana or Rabi formulae & Accurately (\(p < 8 \times 10^{-7}\))
without scaling or fitting, no parameters are adjusted \\
\bottomrule
\end{tabular}
\end{table}

\section{Methods}\label{methods}

\subsection{CQD equations of motion}\label{cqd-equations-of-motion}

In classical electrodynamics, the motion of a magnetic dipole moment,
\(\vec{\mu}\), is described by the Bloch equation,

\begin{equation}
\label{eq:1}
\frac{d\widehat{\mu}}{dt} = \gamma\widehat{\mu} \times \vec{B},
\end{equation}

where caret denotes a unit vector, \(t\) time, \(\gamma\) the
gyromagnetic ratio, and \(\vec{B}\) the magnetic flux
density. Majorana stated that both the classical and the
quantum-mechanical treatments on spin flip require integration of the
same differential equations \cite{Majorana1932,Majorana2006}. It is known that the
Schrödinger or von Neumann equation for a unitary two-level system can
be converted to the Bloch equation or its analog \cite{Feynman1963,Grynberg2010,Feynman1957}.

We now extend the Bloch equation to the Landau--Lifshitz--Gilbert
equation \cite{Gilbert2004},

\begin{equation}
\label{eq:2}
\frac{d\widehat{\mu}}{dt} = \gamma\widehat{\mu} \times \vec{B} - k_{i}\widehat{\mu} \times \frac{d\widehat{\mu}}{dt},
\end{equation}

where the dimensionless \({\ k}_{i}\) is called the induction factor
here. Although this equation was originally intended for condensed
matter, the underlying physical mechanism for the added term is
compatible with CQD (see Paragraph 1 in Discussion). In fact, the author
had developed CQD before realizing its connection with the
Landau--Lifshitz--Gilbert equation. If \({\ k}_{i} = 0\), the Bloch
equation is recovered.

Henceforth, subscripted \(e\) and \(n\) denote electron and nucleus,
respectively. The default atom, to match the Frisch--Segrè experiment
\cite{Frisch1933}, is potassium (\textsuperscript{39}K). The scope of the
manuscript is limited to potassium in the Stern--Gerlach or
Frisch--Segrè experiment.

The torque-averaged magnetic flux densities from
\(\vec{\mu}_{n}\) and \(\vec{\mu}_{e}\)
applied on each other are respectively (Appendix~\ref{appendix:1})

\begin{equation}
\label{eq:3}
\vec{B}_{n} = \frac{5\mu_{0}}{16\pi R^{3}}\vec{\mu}_{n}
\end{equation}

and

\begin{equation}
\label{eq:4}
\vec{B}_{e} = \frac{5\mu_{0}}{16\pi R^{3}}\vec{\mu}_{e},
\end{equation}

where \(\mu_{0}\) is the vacuum permeability (\(4\pi \times 10^{-7}\)
H/m) and \(R\) is the van der Waals atomic radius
(\(2.75 \times 10^{-10}\) m) \cite{LosAlamosPeriodicTable}. Chiefly because the nucleus is
more massive, \(\mu_{e}\) (\(9.285 \times 10^{-24}\) J/T) \(\gg\)
\(\mu_{n}\) (\(1.977 \times 10^{-27}\) J/T); thus, \(B_{e}\)
(\(558.1 \times 10^{-4}\) T) \(\gg B_{n}\) (\(0.119 \times 10^{-4}\)
T), where \(10^{-4}\) T = 1 Gauss.

CQD refers to \(\vec{\mu}_{e}\) as the principal quantum
and \(\vec{\mu}_{n}\) in the same atom as the co-quantum.
Postulate 1 states that induction between the electron and the nucleus
tends to increase \(\left| \theta_{e} - \theta_{n} \right|\), where
\(\theta\) denotes the polar angle relative to the quantization axis
(see Paragraph 1 in Discussion). We (1) apply the
Landau--Lifshitz--Gilbert equation to both \({\widehat{\mu}}_{e}\) and
\({\widehat{\mu}}_{n}\), (2) express the unit vectors in spherical
coordinates, and (3) revise the signs of the induction terms to
implement the above postulate, leading to the following CQD equations of
motion (Appendix~\ref{appendix:2}):


\begin{equation}
\label{eq:5}
{\dot{\theta}}_{e} = {- \gamma}_{e}\left\lbrack B_{y}\cos\phi_{e} + B_{n}\sin\theta_{n}\sin\left( \phi_{n} - \phi_{e} \right) \right\rbrack - \mathrm{sgn}\left( \theta_{n} - \theta_{e} \right)k_{i}\left| {\dot{\phi}}_{e} \right|\sin\theta_{e},
\end{equation}

\begin{equation}
\label{eq:6}
{\dot{\theta}}_{n} = - \gamma_{n}\left\lbrack B_{y}\cos\phi_{n} + B_{e}\sin\theta_{e}\sin\left( \phi_{e} - \phi_{n} \right) \right\rbrack - \mathrm{sgn}\left( \theta_{e} - \theta_{n} \right)k_{i}\left| {\dot{\phi}}_{n} \right|\sin\theta_{n},
\end{equation}

\begin{equation}
\label{eq:7}
{\dot{\phi}}_{e} = - \gamma_{e}\left\{ B_{z} + B_{n}\cos\theta_{n} - \cot\theta_{e}\left\lbrack B_{y}\sin\phi_{e} + B_{n}\sin\theta_{n}\cos\left( \phi_{n} - \phi_{e} \right) \right\rbrack \right\} - \frac{\mathrm{sgn}(\dot{\phi}_{e})k_{i}\,\left| {\dot{\theta}}_{e} \right|}{\sin\theta_{e}},
\end{equation}

and

\begin{equation}
\label{eq:8}
{\dot{\phi}}_{n} = - \gamma_{n}\left\{ B_{z} + B_{e}\cos\theta_{e} - \cot\theta_{n}\left\lbrack B_{y}\sin\phi_{n} + B_{e}\sin\theta_{e}\cos\left( \phi_{e} - \phi_{n} \right) \right\rbrack \right\} - \frac{\mathrm{sgn}(\dot{\phi}_{n})k_{i}\left| {\dot{\theta}}_{n} \right|}{\sin\theta_{n}}.
\end{equation}

Here, \(\phi\) denotes the azimuthal angle; \(B_{y}\) and \(B_{z}\)
represent, respectively, the \(y\) (axis of the atomic beam) and \(z\)
components of the external magnetic flux densities; \(B_{x}\) is
neglected for brevity; $\mathrm{sgn}$ denotes the sign function. When
\(\theta_{e} = 0\) or \(\pi\), Eq.~\ref{eq:7} is replaced with
\({\dot{\phi}}_{e} = 0\); when \(\theta_{n} = 0\) or \(\pi\), Eq.~\ref{eq:8} is
replaced with \({\dot{\phi}}_{n} = 0\). Primarily because the nucleus is
more massive again, \(\gamma_{e}\) (\(- 1.761 \times 10^{11}\) rad Hz/T)
in absolute value is four orders of magnitude greater than
\(\gamma_{n}\) (\(1.250 \times 10^{7}\) rad Hz/T). If \(B_{n} = 0\) and
\(k_{i} = 0\), Eqs.~\ref{eq:5} and \ref{eq:7} reduce to the equations shown by Majorana
\cite{Majorana1932,Majorana2006}.

\subsection{CQD branching condition}\label{cqd-branching-condition}

Postulate 2 states that the polar angle of the co-quantum,
\(\theta_{n}\), varies negligibly (\(\ll \pi\)) during flight in typical
Stern--Gerlach experiments, where the duration is too short for the
co-quantum to collapse (see Paragraph 2 in Discussion). The external
main field, \(B_{0}\), along the \(z\) axis is usually much stronger
than \(B_{e}\) and \(B_{n}\). While the fast motion of
\({\widehat{\mu}}_{e}\) is precession about the main field, the
secondary motion is collapse due to the induction term, which yields the
following trend from Eq.~\ref{eq:5}:

\begin{equation}
\label{eq:9}
\tan{\frac{\theta_{e}(t)}{2}\ } = \tan\frac{\theta_{e}(0)}{2}\exp\left\lbrack - \mathrm{sgn}\left( \theta_{n} - \theta_{e} \right)k_{i}\left| {\Delta\phi}_{e}(t) \right| \right\rbrack.
\end{equation}

Here, \({\Delta\phi}_{e}\) denotes the traversed azimuthal angle (i.e.,
unwrapped phase). If the Larmor frequency of the electron magnetic
moment \(\omega_{e}\) is constant, we simply have
\({\Delta\phi}_{e} = \omega_{e}t\). As time evolves, \(\theta_{e}\)
approaches either \(0\) or \(\pi\) according to the following branching
condition:

\begin{equation}
\label{eq:10}
\mathrm{sgn}\left( \theta_{n} - \theta_{e} \right) = \left\{ \begin{matrix}
1 & \mathrm{if}\, \theta_{n} > \theta_{e}, \\
0 & \mathrm{if}\, \theta_{n} = \theta_{e}, \\
-1 & \mathrm{if}\, \theta_{n} < \theta_{e}.
\end{matrix} \right.
\end{equation}

\begin{figure}[!htbp]
\centering
\includegraphics[width=2.09524in,height=1.3666in]{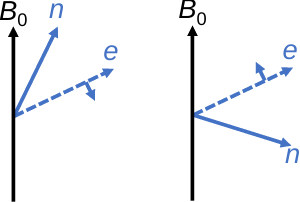}
\caption{Examples of collapse directions determined by the
branching condition in Stern--Gerlach experiments. \(B_{0}\): external
main field; \emph{e}: electron magnetic moment (principal quantum),
\({\widehat{\mu}}_{e}\); \emph{n}: nuclear magnetic moment (co-quantum),
\({\widehat{\mu}}_{n}\); Short arrows: collapse directions. While
\({\widehat{\mu}}_{e}\) precesses about \(B_{0}\) right-handedly at its
Larmor frequency (\(\omega_{e} = - \gamma_{e}B_{0}\)),
\({\widehat{\mu}}_{n}\) does left-handedly at the Larmor frequency
(\(\omega_{n} = - \gamma_{n}B_{0}\)); note that
\(\left| \frac{\omega_{e}}{\omega_{n}} \right| > 10^{4}\). For the same
given \({\widehat{\mu}}_{e}\), the collapse direction, down (left panel)
or up (right panel), depends on \({\widehat{\mu}}_{n}\) according to the
branching condition (Eq.~\ref{eq:10}). It takes on the order of \(N_{c}\)
(estimated to be on the order of \textasciitilde220 in Results) Larmor
cycles to collapse. In typical Stern--Gerlach experiments, it is assumed
that \({\widehat{\mu}}_{n}\) does not collapse, i.e., \(\theta_{n}\) is
approximately constant.}
\label{fig:2}
\end{figure}

Therefore, \({\widehat{\mu}}_{e}\) collapses to either \(+ z\) or
\(- z\) while precessing about \(B_{0}\), depending on the polar angle
of the co-quantum \(\theta_{n}\) relative to \(\theta_{e}\) (Fig.~\ref{fig:2}).

The number of precession cycles required to vary
\(\tan\left( \frac{\theta_{e}}{2} \right)\) by a factor of \(e\) is
given by

\begin{equation}
\label{eq:11}
N_{c} = \frac{1}{2\pi k_{i}}
\end{equation}

regardless of the strength of the external magnetic field. For a
constant Larmor frequency, \(\omega_{e}\), the collapse time constant is

\begin{equation}
\label{eq:12}
T_{c} = N_{c}\frac{2\pi}{\left| \omega_{e} \right|} = \frac{1}{k_{i}\left| \omega_{e} \right|}.
\end{equation}

\subsection{CQD pre-collapse state function and CQD prediction
expression}\label{cqd-pre-collapse-state-function-and-cqd-prediction-expression}

The CQD pre-collapse state function is denoted by
\(\left|{\widehat{\mu}}_{e}\coq{\widehat{\mu}}_{n} \right\rangle\),
where the co-quantum, \({\widehat{\mu}}_{n}\), is prefixed with \(\coq\)
for clarity.
\(\left|{\widehat{\mu}}_{e}\coq{\widehat{\mu}}_{n} \right\rangle\)
represents \({\widehat{\mu}}_{e}\) accompanied with
\({\widehat{\mu}}_{n}\), both governed by the CQD equations of motion.

The CQD prediction expression for Stern--Gerlach experiments is written
as

\begin{equation}
\label{eq:13}
\left|{\widehat{\mu}}_{e}\coq{\widehat{\mu}}_{n} \right\rangle = C_{+}\left( {\widehat{\mu}}_{e},{\widehat{\mu}}_{n} \right)\left| + z \right\rangle + C_{-}\left( {\widehat{\mu}}_{e},{\widehat{\mu}}_{n} \right)\exp\left( i\phi_{e} \right)\left| - z \right\rangle.
\end{equation}

The equal sign functions as a right arrow (\(\rightarrow\)) because the
right side predicts the measurement outcome. A given
\({\widehat{\mu}}_{e}\) collapses to either \(+ \widehat{z}\) or
\(- \widehat{z}\) according to the branching condition (Eq.~\ref{eq:10}). The two
real and positive \(C\) coefficients take on mutually exclusive binary
values while \(\exp\left( i\phi_{e} \right)\) captures the phase
information. If \(\theta_{n} > \theta_{e}\), then \(C_{+} = 1\) and
\(C_{-} = 0\); if \(\theta_{n} < \theta_{e}\), \(C_{+} = 0\) and
\(C_{-} = 1\). In either case, \(C_{+} \cdot C_{-} = 0\) and
\(C_{+} + C_{-} = 1\).

\section{Results}\label{results}

\subsection{Single-stage Stern--Gerlach
experiment}\label{single-stage-sterngerlach-experiment}

To describe the angular distribution of \({\widehat{\mu}}_{e}\) or
\({\widehat{\mu}}_{n}\) in an ensemble of atoms, we define the angular
probability density function, \(p(\theta,\phi)\), as the probability of
\(\widehat{\mu}\) pointing to the vicinity of \((\theta,\phi)\) per unit
infinitesimal solid angle, with the following normalization:

\begin{equation}
\label{eq:14}
\int_{0}^{\pi}{\int_{0}^{2\pi}{p(\theta,\phi)\sin\theta d\phi}d\theta} = 1.
\end{equation}

If the azimuthal distribution is isotropic, the integral reduces to
\(\int_{0}^{\pi}{p(\theta,\phi)2\pi\sin\theta d\theta} = 1\).

\begin{figure}[!htbp]
\centering
\includegraphics[width=6.5in,height=4.78403in]{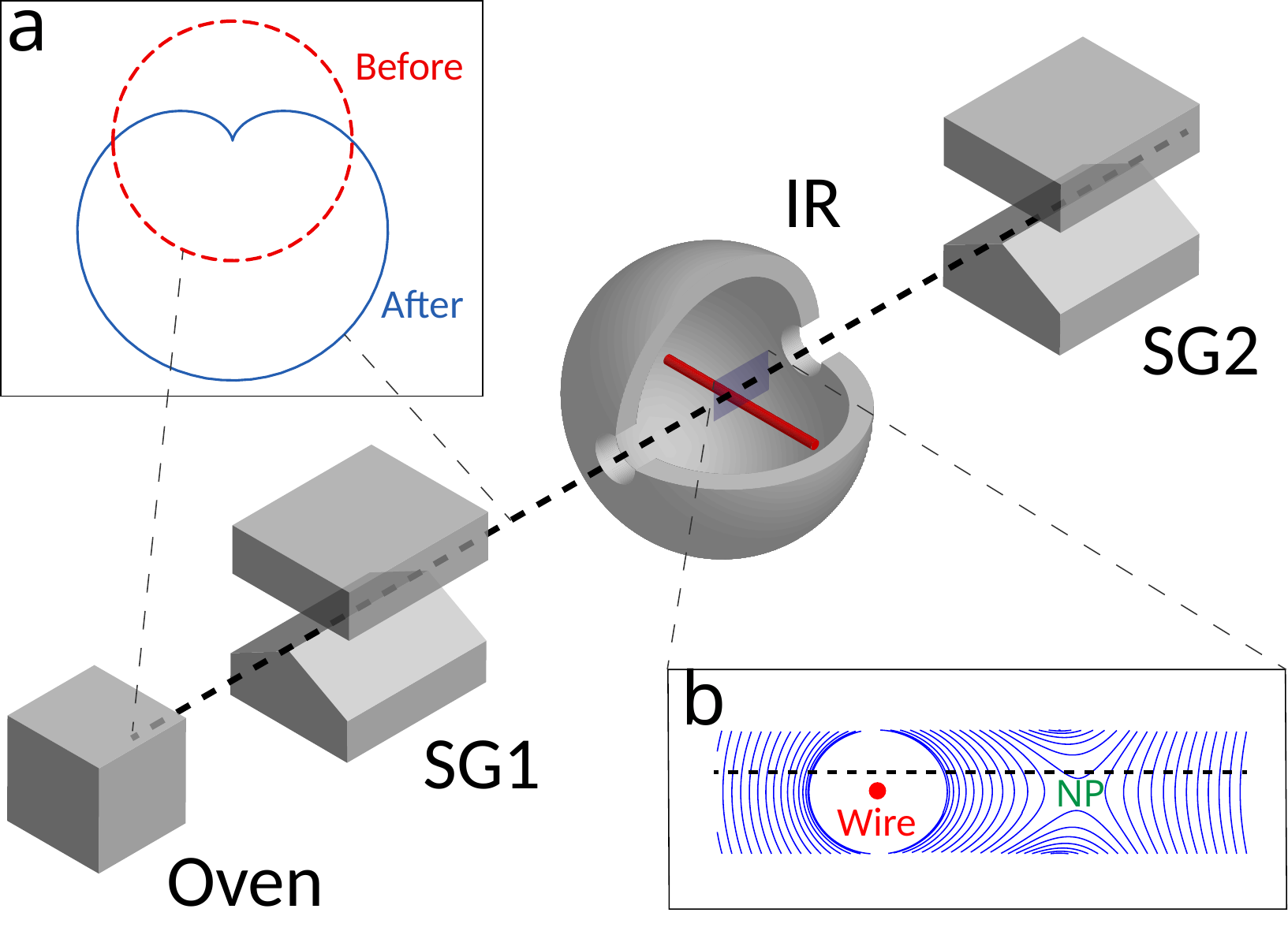}
\caption{Multi-stage Stern--Gerlach (SG) experiment conducted by
Frisch and Segrè \cite{Frisch1933}. The atomic beam from the oven is sent through
(1) Stage SG1 to collapse \({\widehat{\mu}}_{e}\) (principal quantum),
(2) the magnetically shielded inner rotation (IR) chamber to rotate
\({\widehat{\mu}}_{e}\), (3) a slit (not shown) to select a branch, and
(4) Stage SG2 to measure the fraction of spin flip. The red solid line
and filled circle represent the current-carrying wire, and the gray
sphere in cutaway view represents magnetic shielding. \textbf{Inset (a)}
Angular distributions of \({\widehat{\mu}}_{n}\) (co-quanta) before and
after Stage SG1. \textbf{Inset (b)} Magnetic field lines within the IR
chamber; NP: null point, formed by the cancelation of the magnetic field
from the wire by the vertical remnant (residual) fringe magnetic field.
Here, the vertical distance of the atomic beam from the center of the
wire, \(z_{a} = 1.05 \times 10^{-4}\) m; the most likely speed of
atoms,\(\ v = \ 800\ \)m s\textsuperscript{$-1$}; the uniformly
distributed remnant (residual) fringe magnetic flux density,
\(B_{r} = 0.42 \times 10^{-4}\) T, which is parallel with the \(+ z\)
axis (up in the figure); and the current carried by the wire, \(I\),
points along the \(- x\) axis (into the screen).}
\label{fig:3}
\end{figure}

The angular distribution of \({\widehat{\mu}}_{n}\) for atoms
immediately out of the oven is presumed to be isotropic as given by
(Fig.~\ref{fig:3}, Inset a, dashed circle)

\begin{equation}
\label{eq:15}
p_{n0}\left( \theta_{n},\phi_{n} \right) = \frac{1}{4\pi}.
\end{equation}

In a single-stage Stern--Gerlach experiment (Fig.~\ref{fig:3}, SG1), the
probabilities of collapse for a given \(\theta_{e}\) are related to the
binary coefficients through ensemble averaging of the pre-averaging
density operator defined in Appendix~\ref{appendix:3} (Eq.~\ref{eq:70}) over \(p_{n0}\). The
outcome is summarized as

\begin{equation}
\label{eq:16}
\left\langle C_{+} \right\rangle_{n}^{2} = \int_{\theta_{e}}^{\pi}{p_{n0}2\pi\sin\theta_{n}d\theta_{n}} = \cos^{2}\frac{\theta_{e}}{2}
\end{equation}

and

\begin{equation}
\label{eq:17}
\left\langle C_{-} \right\rangle_{n}^{2} = \int_{0}^{\theta_{e}}{p_{n0}2\pi\sin\theta_{n}d\theta_{n}} = \sin^{2}\frac{\theta_{e}}{2}.
\end{equation}

The angle brackets, with the subscripts denoting nuclear, represent
ensemble averaging with the integration limits determined by the
branching condition (Eq.~\ref{eq:10}). The two probabilities are proportional to
the solid angles formed by the down and up sides of the cone shaped by
the initial Bloch vector \textsuperscript{15} precessing over one cycle.
Each solid angle determines the probability of having the co-quantum on
the corresponding side of the cone.

From Eqs.~\ref{eq:16} and \ref{eq:17}, the pre-collapse state function (Eq.~\ref{eq:13}) averages to
the following familiar quantum mechanical wave function for a pure state
(Appendix~\ref{appendix:3}):

\begin{equation}
\label{eq:18}
\left|{\widehat{\mu}}_{e} \right\rangle = \cos\frac{\theta_{e}}{2}\left| + z \right\rangle + \sin\frac{\theta_{e}}{2}\exp\left( i\phi_{e} \right)\left| - z \right\rangle.
\end{equation}

If \({\widehat{\mu}}_{e}\) is also isotropically distributed as

\begin{equation}
\label{eq:19}
p_{e0}\left( \theta_{e},\phi_{e} \right) = \frac{1}{4\pi},
\end{equation}

the probabilities of collapse are predicted by averaging Eqs.~\ref{eq:16} and \ref{eq:17}
over \(p_{e0}\) (Appendix~\ref{appendix:3}):

\begin{equation}
\label{eq:20}
\left\langle C_{+} \right\rangle_{n,e}^{2} = \int_{0}^{\pi}{\cos^{2}\frac{\theta_{e}}{2}p_{e0}2\pi\sin\theta_{e}d\theta_{e}} = \frac{1}{2}
\end{equation}

and

\begin{equation}
\label{eq:21}
\left\langle C_{-} \right\rangle_{n,e}^{2} = \int_{0}^{\pi}{\sin^{2}\frac{\theta_{e}}{2}p_{e0}2\pi\sin\theta_{e}d\theta_{e}} = \frac{1}{2}.
\end{equation}

The \(e\) subscripts denote electron. The outcomes agree with the
familiar quantum mechanical prediction for a maximally mixed state of
atoms immediately out of the oven, represented by a density operator
(Eq.~\ref{eq:87}, Appendix~\ref{appendix:3}).

\subsection{Multi-stage Stern--Gerlach
experiment}\label{multi-stage-sterngerlach-experiment}

In the multi-stage Stern--Gerlach experiment conducted by Frisch and
Segrè (Fig.~\ref{fig:3}) \cite{Frisch1933}, Stage SG1 collapses \({\widehat{\mu}}_{e}\) into
two branches. The inner rotation (IR) chamber rotates
\({\widehat{\mu}}_{e}\) by an angle of \(\alpha_{r}\) using the magnetic
field shown in Inset b. A slit (not shown) selects one branch: the
\(+ z\) branch is chosen here. Stage SG2 collapses
\({\widehat{\mu}}_{e}\) and measures the fraction of spin flip.
Therefore, Stage SG1 serves as a polarizer, the IR chamber a rotator,
and Stage SG2 an analyzer.

The probability of spin flip has been predicted \cite{Majorana1932,Majorana2006} by quantum
mechanics as (see Eq.~\ref{eq:17}, set \(\theta_{e} = \alpha_{r}\))

\begin{equation}
\label{eq:22}
W_{\mathrm{qm}} = \left\langle - z \middle| \alpha_{r} \right\rangle^{2} = \sin^{2}\frac{\alpha_{r}}{2},
\end{equation}

which leads to the following Majorana formula (Appendix~\ref{appendix:4}, Eq.~\ref{eq:117})
\cite{Majorana1932,Majorana2006}:

\begin{equation}
\label{eq:23}
W_{m} = \exp\left( - \frac{\pi z_{a}}{2v}\left| \gamma_{e} \right|B_{y} \right).
\end{equation}

\begin{figure}[!htbp]
\centering
\includegraphics[width=5.02in,height=4.22in]{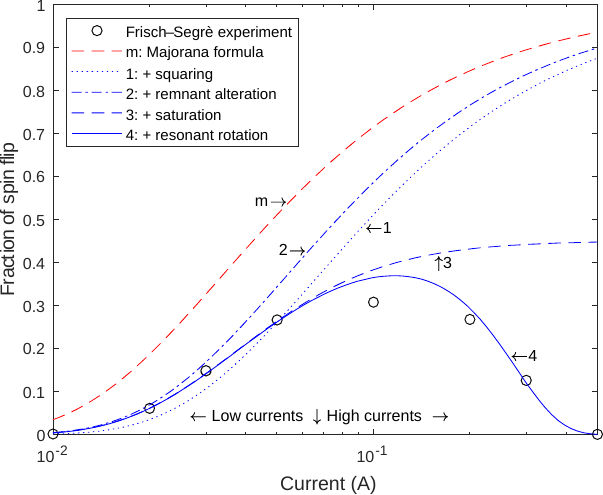}
\caption{Fraction of spin flip versus wire current. The down
arrow points to the current where
\(B_{y}^{'} = B_{n}\sin\left\langle \theta_{n} \right\rangle\) or
\(k_{0} = k_{1}\) to separate the low- and high-current regions. Curves
m and 1--4 represent \(W_{m}\) and \(W_{1} - W_{4}\), respectively.
While \(W_{m}\) diverges from the experiment with a negative \(R^{2}\),
\(W_{3}\) matches the low-current experimental observation in absolute
units without fitting with \(R^{2} = 0.9495\); further, \(W_{4}\)
matches the entire observation with improved \(R^{2} = 0.9787\) and
\(p < 8 \times 10^{-7}\). No adjustable or free parameters are used.}
\label{fig:4}
\end{figure}

Here, \(z_{a}\) is the vertical distance of the atomic beam from the
center of the wire, and \(v\) is the most likely speed of the atoms. The
spin flip is because \(B_{z}\) vanishes and reverses its sign near the
null point (Fig.~\ref{fig:3}, Inset b). Because \(B_{y}\) is inversely
proportional to the current carried by the wire, \(I\) (Eqs.~\ref{eq:92} and \ref{eq:94} in
Appendix~\ref{appendix:4}), the Majorana formula predicts a probability of spin flip
approaching 100\% with increasing currents (Fig.~\ref{fig:4}, Curve m), i.e., as
\(B_{y} \rightarrow 0\), \(W_{m} \rightarrow 1\); yet, the experimental
outcome decreases to nearly zero after peaking at 31\% (Fig.~\ref{fig:4}, circles)
\cite{Frisch1933}. Consequently, \(W_{m}\) yields a negative coefficient of
determination (\(R^{2})\). Using the dimensionless adiabaticity
parameter \(k_{m}\) (Eq.~\ref{eq:103} in Appendix~\ref{appendix:4}), one can express the above
equation concisely as
\(W_{m} = \exp\left( - {\pi k_{m}}/{2} \right)\) (Eq.~\ref{eq:116}). Rabi
revised the Majorana formula to
\({W_{m}^{{1}/{4}}}/{4}\ \)\cite{Rabi1936}, which, however,
overestimates the starting points, underestimates the peak, and
continues to diverge thereafter; as a result, the \(R^{2}\) remains
negative (Fig.~\ref{fig:1}).

In CQD, Stage SG1 varies \(\theta_{n}\) negligibly according to
Postulate 2. However, polarization selection by the slit reshapes the
co-quantum angular distribution from the original isotropic \(p_{n0}\)
(Eq.~\ref{eq:15}) to

\begin{equation}
\label{eq:24}
p_{n1}\left( \theta_{n},\phi_{n} \right) = p_{n0}\left( \theta_{n},\phi_{n} \right) \cdot 2\int_{0}^{\theta_{n}}{p_{e0}2\pi\sin\theta_{e}d\theta_{e}} = \frac{1 - \cos\left( \theta_{n} \right)}{4\pi}.
\end{equation}

Here, the pre-factor \(2\) compensates for the overall slit rejection of
the opposite polarization (Eq.~\ref{eq:20}), \(p_{e0}\) is given by Eq.~\ref{eq:19}, and
the integration limits are based on the branching condition (Eq.~\ref{eq:10}).
Because atoms with smaller \(\theta_{n}\) are deflected to the blocked
\(- z\) branch with greater probabilities, \(p_{n1}\) forms a heart
shape (Fig.~\ref{fig:3}, Inset a, solid line; Paragraph 3 in Discussion).

The heart shape is assumed to be approximately maintained throughout the
inner rotation chamber owing to the extension of Postulate 2 (see
Paragraph 2 in Discussion). The co-quanta engender the following four
effects on the principal quanta.

First, the probability of spin flip is derived by ensemble averaging
over \(p_{n1}\) instead of \(p_{n0}\) (Eq.~\ref{eq:89} with
\(\theta_{e} = \alpha_{r}\) in Appendix~\ref{appendix:3}):

\begin{equation}
\label{eq:25}
W_{\mathrm{cqd}} = \left\langle - z \middle| \alpha_{r} \right\rangle^{2} = \int_{0}^{\alpha_{r}}{p_{n1}2\pi\sin\theta_{n}d\theta_{n}} = \sin^{4}\left( \frac{\alpha_{r}}{2} \right),
\end{equation}

which equals \(W_{\mathrm{qm}}^{2}\) (Eq.~\ref{eq:22}). As shown by Curve 1 in Fig.~\ref{fig:4},
simply squaring \(W_{m}\) (Eq.~\ref{eq:23}) already brings the solution much
closer to the observation at low currents, but with an overcorrection
near \(I = 0.03\) A. This squaring effect evolves the probability of
spin flip from \(W_{m}\) to

\begin{equation}
\label{eq:26}
W_{1} = W_{m}^{2} = \exp\left( - \frac{\pi z_{a}}{v}\left| \gamma_{e} \right|B_{y} \right),
\end{equation}

where \(B_{y}\) is computed from the remnant (residual) fringe magnetic
flux density, \(B_{r}\), using Eqs.~\ref{eq:92} and \ref{eq:94} in Appendix~\ref{appendix:4}. Using the
dimensionless adiabaticity parameter \(k_{m}\) (Eq.~\ref{eq:103}), one can
express the above equation concisely as
\(W_{1} = \exp\left( - \pi k_{m} \right)\) (see Eq.~\ref{eq:154} in Appendix~\ref{appendix:5}).

Second, the \(z\) component of \(\vec{B}_{n}\) (Eq.~\ref{eq:3}),
represented by \(B_{n}\cos\left\langle \theta_{n} \right\rangle\),
offsets the upward \(B_{r}\). We substitute
\(B_{r} + B_{n}\cos\left\langle \theta_{n} \right\rangle\) (Eq.~\ref{eq:119} in
Appendix~\ref{appendix:5}) for \(B_{r}\) to update \(B_{y}\) to \(B_{y}^{'}\) (Eq.~\ref{eq:122}). The heart shape (Eq.~\ref{eq:24}) yields
\(\left\langle \theta_{n} \right\rangle = \frac{5\pi}{8}\) (Eq.~\ref{eq:118}).
The magnitude of
\(B_{n}\cos\left\langle \theta_{n} \right\rangle = - 0.045 \times 10^{-4}\)
T exceeds 10\% of \(B_{r}\) (\(0.42 \times 10^{-4}\) T), producing an
appreciable remnant-alteration effect. As shown by Curve 2 in Fig.~\ref{fig:4},
the corrected curve passes through the first two data circles and grazes
the third one. If the co-quantum distribution were isotropic,
\(\left\langle \theta_{n} \right\rangle\) would be \(\frac{\pi}{2}\);
\(B_{n}\cos\left\langle \theta_{n} \right\rangle\) would vanish, so
would the remnant-alteration effect. Effect 2 evolves \(W_{1}\) to (see
Eq.~\ref{eq:136})

\begin{equation}
\label{eq:27}
W_{2} = \exp\left( - \frac{\pi z_{a}}{v}\left| \gamma_{e} \right|B_{y}^{'} \right),
\end{equation}

where \(B_{y}^{'}\), however, is computed using
\(B_{r} + B_{n}\cos\left\langle \theta_{n} \right\rangle\) instead of
\(B_{r}\). Using the dimensionless adiabaticity parameters \(k_{0}\)
(Eq.~\ref{eq:125}), one can express the above equation concisely as
\(W_{2} = \exp\left( - \pi k_{0} \right)\) (see Eq.~\ref{eq:153}).

Third, the co-quanta saturate the rotation. As shown by Eq.~\ref{eq:27},
\(W_{2}\) increases with decreasing \(B_{y}^{'}\). However, the weakness
of \(B_{y}^{'}\) is spoiled by the transverse (\(xy\)) component of
\(\vec{B}_{n}\), denoted by
\(B_{n}\sin\left\langle \theta_{n} \right\rangle\). Substitution of
\(\sqrt{B_{y}^{'2} + \left( B_{n}\sin\left\langle \theta_{n} \right\rangle \right)^{2}}\)
for \(B_{y}^{'}\) (see Eq.~\ref{eq:141}) evolves \(W_{2}\) to

\begin{equation}
\label{eq:28}
W_{3} = \exp\left( - \frac{\pi z_{a}}{v}\left| \gamma_{e} \right|\sqrt{B_{y}^{'2} + \left( B_{n}\sin\left\langle \theta_{n} \right\rangle \right)^{2}} \right).
\end{equation}

Using the dimensionless adiabaticity parameters \(k_{0}\) (Eq.~\ref{eq:125}) and
\(k_{1}\) (Eq.~\ref{eq:126}), one can express the above equation concisely as
\(W_{3} = \exp\left( - \pi\sqrt{k_{0}^{2} + k_{0}k_{1}} \right)\) (see
Eq.~\ref{eq:152}).

As shown by Curve 3 in Fig.~\ref{fig:4}, this rotation-saturation effect clamps
the overshoot in Curve 2. The clamped curve passes through the first
four data circles. The current is divided into two regions at \(0.067\)
A, where
\(B_{y}^{'} = B_{n}\sin\left\langle \theta_{n} \right\rangle = 0.11 \times 10^{-4}\)
T. At low currents before the fourth data point (\(I = 0.05\) A and
\(B_{y}^{'} = 0.15 \times 10^{-4}\) T), \(B_{y}^{'}\) is greater than
\(B_{n}\sin\left\langle \theta_{n} \right\rangle\); at high currents,
conversely, \(B_{n}\sin\left\langle \theta_{n} \right\rangle\) becomes
dominant and saturates the curve. If the co-quantum distribution were
isotropic, then
\(\left\langle \theta_{n} \right\rangle = \frac{\pi}{2}\); consequently,
both the squaring and remnant-alteration effects (effects 1 and 2) would
vanish. In this case, the rotation-saturation effect (effect 3) alone
could not bring the Majorana solution down sufficiently in the
low-current region because as the current decreases \(B_{y}^{'}\)
increasingly overpowers \(B_{n}\); thus, the effect of the co-quanta
would become negligible.

Combining the three effects, CQD accurately predicts the low-current
observation in absolute units without fitting (i.e., no parameters are
adjusted). The coefficient of determination \(R^{2}\) for the
low-current regime reaches \(0.9495\) as computed using the natural
logarithm of the fractions of flip to suppress the exponential
dependence (Eq.~\ref{eq:28}). Therefore, effecting the three modifications to the
Majorana formula has already shown evidence for the existence of both
the co-quantum and the derived heart-shaped distribution.

Fourth, in the high-current regime, the precession of
\(\vec{B}_{n}\) generates substantial nuclear-resonant
rotation, due to precession resonance between
\(\vec{\mu}_{e}\) and \(\vec{\mu}_{n}\) when
their Larmor frequencies are matched (see Appendix~\ref{appendix:5} for details). This
effect evolves \(W_{3}\) to (Eq.~\ref{eq:167} in Appendix~\ref{appendix:5})

\begin{equation}
\label{eq:29}
W_{4} = W_{3}\exp\left( - c_{r1}I^{3} \right),
\end{equation}

where the resonant-rotation coefficient, \(c_{r1}\), is given by Eq.~\ref{eq:163}. The fraction of spin flip peaks near \(I = 0.1\) A, where
\(B_{y}^{'} = 0.074 \times 10^{-4}\) T, comparable to but less than
\(B_{n}\sin\left\langle \theta_{n} \right\rangle = 0.11 \times 10^{-4}\)
T. As shown by Curve 4 in Fig.~\ref{fig:4}, this effect increases with the current
and bends down Curve 3. At the maximum current (\(I = 0.5\) A),
\(B_{y}^{'} = 0.015 \times 10^{-4}\) T, far less than
\(B_{n}\sin\left\langle \theta_{n} \right\rangle\); the fraction of spin
flip decreases to nearly zero. Expression of the above equation based on
dimensionless parameters is discussed below Eq.~\ref{eq:150}. Using the
dimensionless adiabaticity parameters \(k_{0}\) (Eq.~\ref{eq:125}) and \(k_{1}\)
(Eq.~\ref{eq:126}) as well as \(f_{r1}\) (Eq.~\ref{eq:146}), one can express the above
equation concisely as
\(W_{4} = \exp\left( - \pi\sqrt{k_{0}^{2} + k_{0}k_{1}} - \frac{1}{2}\left\lbrack \pi k_{1}\  \right\rbrack^{2}f_{r1} \right)\)
(see Eq.~\ref{eq:151}), where \(f_{r1}\) denotes the fraction of the Larmor
period of the nuclear magnetic moment precessed during the effective
flight path-length for nuclear-resonant rotation.

Combining all four effects, CQD accurately predicts the experimental
observation in absolute units without fitting (Fig.~\ref{fig:4}, Curve 4) over the
entire domain; \(R^{2}\) is computed to be 0.9787 using the natural
logarithm of the fractions of flip to suppress the exponential
dependence (Eq.~\ref{eq:29}). Under the null hypothesis that the theoretical
prediction is uncorrelated with the observation, we estimate the
\(p\)-value to be \(< 8 \times 10^{-7}\) (Function \emph{regress} or
\emph{corr}, MATLAB, MathWorks) \cite{Steel1960,Rahman1968}. Such a small \(p\)-value
further objectively confirms the existence of both the co-quantum and
the derived heart-shaped distribution. In comparison, without taking the
logarithm of the fractions of flip, \(R^{2}\) is computed to be 0.9621.

Thus far, we have set the induction factor \(k_{i} = 0\) for the flight
in the inner rotation chamber, owing to the low field (Appendix~\ref{appendix:5}).
Including \(k_{i}\) yields the following combined probability of spin
flip (Appendix~\ref{appendix:5}, Eq.~\ref{eq:160}):

\begin{equation}
\label{eq:30}
W_{\mathrm{cqd}} = \exp\left\lbrack - \sqrt{\left( \frac{c_{r0}}{I} \right)^{2} + c_{rs}^{2}} - c_{r1}I^{3} - c_{ri}I \right\rbrack,
\end{equation}

where \(c_{r0}\), \(c_{rs}\), \(c_{r1}\), and \(c_{ri}\) represent
null-point rotation, rotation saturation, nuclear-resonant rotation, and
induction rotation, respectively. The current, \(I\), controls the
external magnetic field in the inner rotation chamber. Taken from Frisch
and Segrè \cite{Frisch1933}, the only device-specific parameters for the
predictions include \(B_{r}\) (\(0.42 \times 10^{-4}\) T), \(z_{a}\)
(\(1.05 \times 10^{-4}\) m), and \(v\) (800 m s\textsuperscript{$-1$}).
The theoretical predictions from Eqs.~\ref{eq:161}--\ref{eq:163} without adjusting any
parameters are \(c_{r0}\) = 0.054 A, \(c_{rs}\) = 0.80, and \(c_{r1}\) =
48 A\textsuperscript{$-$3}. Substitution of these coefficients into Eq.~\ref{eq:30} with \(c_{ri} = 0\) produces Curve 4 in Fig.~\ref{fig:4}, where no free
parameters are used.

Despite its small contribution in the inner rotation chamber, the
induction factor is estimated for its order of magnitude. While holding
all three other parameters constant at the predicted values, fitting
\(W_{\mathrm{cqd}}\) (Eq.~\ref{eq:30}) for the experimental data in Fig.~\ref{fig:4} yields
\(c_{ri}\sim 0.57\) A. Substitution into Eq.~\ref{eq:164} produces
\(k_{i}\sim{7.4 \times 10}^{- 4}\). Further substitution into Eq.~\ref{eq:11}
concludes that electron-spin collapse takes on the order of
\(N_{c}\sim 220\) precession cycles.

\section{Discussion}\label{discussion}

CQD postulates that the electron and nuclear magnetic moments in an
external field \(B_{0}\) along \(z\) repel in the polar direction, which
results in a revision to the sign of the induction term in the
Landau--Lifshitz--Gilbert equation. Whereas precession is governed by
the terms from the Bloch equation, collapse is modeled by the revised
induction term. If \(k_{i} = 0\), the equation of motion reduces to the
Bloch or equivalent Schrödinger equation \cite{Feynman1963,Majorana1932,Majorana2006,Grynberg2010,Feynman1957}, which
does not model collapse \cite{Norsen2017}. While precession is the dominant
motion, collapse is secondary but concurrent. Although the exact
mechanism for the repulsion is to be investigated, a conjecture is
diamagnetism extended from orbital to spin motions. Diamagnetic
magnetization, a weak but universal induction effect on all atoms,
causes repulsion \cite{Griffiths2017,Jackson1999}. The relativistic momentum density in the
Dirac wave field shows that the magnetic moment of an electron can be
attributed to a circulating flow of electric charge (Eq.~\ref{eq:34} in Appendix~\ref{appendix:1}), similar to that in orbital motions \cite{Ohanian1986}. Therefore, it is
conceivable that diamagnetism applies to spin as well. In the laboratory
reference frame, as \(\vec{\mu}_{e}\) and
\(\vec{\mu}_{n}\) precess in opposite directions, each
azimuthal encounter may be viewed as a ``collision'', causing repulsion.
Because induction is related to relative motion, the induced field on
the electron may be written as
\({\vec{B}}_{i} \propto \frac{d\left( {\widehat{\mu}}_{e} - {\widehat{\mu}}_{n} \right)}{dt}\),
and the corresponding induced torque is
\({\vec{\tau}}_{i} \propto {\widehat{\mu}}_{e} \times {\vec{B}}_{i}\).
If \({\widehat{\mu}}_{e} \times \frac{d{\widehat{\mu}}_{n}}{dt}\)
averages out, the average induced torque becomes
\({\widehat{\mu}}_{e} \times \frac{d{\widehat{\mu}}_{e}}{dt}\), matching
the induction term in the Landau--Lifshitz--Gilbert equation. As
\({\widehat{\mu}}_{e}\) nears either up or down, the average induced
torque approaches zero, providing stability. In the rotating reference
frame that rotates at \(\omega_{e}\), the external \(B_{0}\) vanishes,
\(\vec{\mu}_{e}\) becomes azimuthally stationary \cite{Rabi1954};
the rapidly precessing \(\vec{\mu}_{n}\) forms in the
time-average sense a cone-shaped magnet, which repels
\(\vec{\mu}_{e}\) towards \(\pm z\). The sign function in
the induction terms in the equations of motion is the key difference
from the standard Landau--Lifshitz--Gilbert equation and is central to
CQD. While standard damping always leads to a lower-energy state,
collapse due to the co-quantum can reach a state of either higher or
lower energy in the presence of an external magnetic field, according to
the branching condition, which agrees with the Stern--Gerlach
experimental observation. Numerical solutions to the CQD equations of
motion, to be reported separately, have illustrated collapse with the
induction term and none without. This postulate is consistent with the
Pauli exclusion principle for two identical fermions, where the two
magnetic moments repel towards anti-alignment. Therefore, one may regard
this postulate as an extension to the Pauli exclusion principle. Note
that while diamagnetism explains the collapse term, paramagnetism is
expected to perturb the precession term slightly, which is neglected
here.

CQD also postulates that the polar angle of
\(\vec{\mu}_{n}\) in flight varies negligibly. Because the
nuclear Larmor frequency is four orders of magnitude smaller (i.e.,
\(\left| \omega_{n} \right| \ll \left| \omega_{e} \right|\)), nuclear
spin collapses much more slowly than electron spin. Because no data on
the collapse rates have been found in the literature, we reference the
\(T_{1}\) relaxation times. Typical \(T_{1}\) relaxation times in
electron paramagnetic resonance are on the \(\mu\mathrm{s}\) scale \cite{Forbes2013}, consistent
with the previous estimation of the collapse time scale of
\(\vec{\mu}_{e}\) \cite{Wennerstrom2012}. In contrast, typical \(T_{1}\)
relaxation times in gas-phase nuclear magnetic resonance are on the ms
scale \cite{Marchione2016}, indicating the order-of-magnitude collapse time of
\(\vec{\mu}_{n}\). In a typical Stern--Gerlach experiment
\cite{Wennerstrom2012,Schroder1983}, the main external field \(B_{0}\) along \(z\) is at least
\(0.3\) T (\(B_{0} > B_{e} \gg B_{n}\), the Paschen--Back regime
\cite{Rabi1936}), the length of the main field is \textasciitilde{}\(35\) mm,
and the most likely atomic speed \(v\) is \textasciitilde{}\(800\) m
s\textsuperscript{$-1$}. Consequently, the flight time through the main
field is only \textasciitilde44 \(\mu\mathrm{s}\), which is long enough for
\(\vec{\mu}_{e}\) to collapse but too short for
\(\vec{\mu}_{n}\) to collapse. In fact, the fringe field on
the source side of the main field collapses
\(\vec{\mu}_{e}\) \cite{SchmidtBoecking2016}. Besides the two distinct
collapse branches due to the quantization of
\(\vec{\mu}_{e}\), no additional branches due to the
quantization of \(\vec{\mu}_{n}\) have been observed by
Frisch and Segrè \cite{Frisch1933} despite the prediction of up to eight branches
total \cite{Breit1931}. For \(N_{c}\sim\)220 (Eq.~\ref{eq:11}) estimated from the
Frisch--Segrè experimental data shown in Fig.~\ref{fig:4}, the collapse time
constants (\(T_{c}\), Eq.~\ref{eq:12} and its nuclear counterpart) at the main
field strength are computed to be \textasciitilde{}\(3 \times 10^{-8}\)
and \textasciitilde{}\(4 \times 10^{-4}\) s for
\(\vec{\mu}_{e}\) and \(\vec{\mu}_{n}\),
respectively, which are consistent with the above-mentioned
corresponding \(T_{1}\) relaxation times in orders of magnitude \cite{Forbes2013}
\cite{Marchione2016}. This postulate, extended to the weaker-field inner rotation
chamber, is consistent with the selection rule for observing an
electron-spin--resonance transition, stating that the magnetic quantum
number of the nuclear spin remains constant (i.e., \(\Delta m_{I} = 0\))
\cite{Barra2005}. The selection rule was also a major basis for Rabi's revision
to the Majorana formula \cite{Rabi1936}.

\begin{figure}[!htbp]
\centering
\includegraphics[width=1.47333in,height=1.60667in]{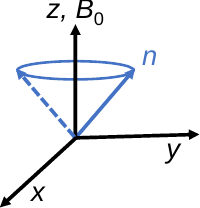}
\caption{Illustration of the cone of \({\widehat{\mu}}_{n}\)
formed by precession around the external main field, \(B_{0}\).
\emph{n}: nuclear magnetic moment (co-quantum), \({\widehat{\mu}}_{n}\).
Any electron magnetic moment (principal quantum),
\({\widehat{\mu}}_{e}\), precessing around \(B_{0}\) within the cone
collapses up, whereas \({\widehat{\mu}}_{e}\) precessing outside the
cone collapses down. For a given \(\theta_{n}\), the probability for the
atom from the oven to reach the up branch in the single-stage
Stern--Gerlach experiment is proportional to the solid angle of the
cone.}
\label{fig:5}
\end{figure}

The heart-shaped \(p_{n1}\) in Eq.~\ref{eq:24} (Fig.~\ref{fig:3}, Inset a) can be
understood in two ways. First, the integral can be perceived as the
expected transmittance through Stage SG1 for a given \(\theta_{n}\). All
principal quanta with \(\theta_{e} < \theta_{n}\) collapse to \(+ z\),
and the atoms propagate through the slit further; otherwise, the atoms
are blocked by the slit. The greater the \(\theta_{n}\) is, the greater
the transmittance, proportional to the solid angle formed by the cone
having a half angle of \(\theta_{n}\) (Fig.~\ref{fig:5}). Second, one may examine
how much principal quanta at the source around each \(\theta_{e}\)
within \(d\theta_{e}\) contribute to \(p_{n1}\). For \(\theta_{e} = 0\),
the contribution forms a perfect spherical distribution of co-quanta
because co-quanta in any direction can reach the second stage. For
\(0 < \theta_{e} < \pi\ \), the contribution forms a truncated sphere
with the cone of \(\theta_{n} < \theta_{e}\) removed because co-quanta
in this range have collapsed the principal quanta to the blocked branch.
For \(\theta_{e} = \pi\), the contribution vanishes because the
principal quanta are always in the blocked branch. Integrating these
(truncated) spheres form the final heart shape. Conversely, the
co-quantum angular distribution for the opposite branch is an inverted
heart shape. Average the two complementary shapes recovers the original
isotropic \(p_{n0}\).

A key reason for the agreement between CQD and the Frisch--Segrè
experimental observation is that the angular distribution of the
co-quantum (i.e., the nuclear magnetic moment) is changed from an
isotropic shape (Eq.~\ref{eq:15}) to a continuous heart shape (Eq.~\ref{eq:24}) due to the
polarization. The subsequent effects are illustrated using the evolution
of the curves in Fig.~\ref{fig:4}. As more effects are included, the model becomes
more and more accurate while all parameters were given (i.e., no
parameters were tuned to fit the experimental data). If the heart shape
were incorrect, the agreement would be completely off. In comparison,
the Majorana or Landau--Zener formula neglected the nuclear magnetic
moment altogether, and Rabi used an isotropic angular distribution
instead of the heart shape \cite{Rabi1936}. Note that as the wire current
approaches infinity, Rabi's formula predicts a maximum of
\(\frac{1}{2I + 1} = \frac{1}{4}\), which is well below the experimental
peak of 31\% (Fig.~\ref{fig:1}); here, \(I = \frac{3}{2}\) denotes the nuclear
spin number for potassium-39. Further, Rabi's standard hyperfine
coupling does not contain the induction terms in CQD and hence does not
model collapse. Also, the torque-averaged fields provide greater
agreement than the self-averaged fields (see Appendix~\ref{appendix:1}).

Quantum mechanics, celebrated for its countless triumphs, still pose
mysteries as discussed insightfully in recent literature \cite{Norsen2017,Adler2009,Bricmont2016,Laloe2019,Auletta2019}. The Copenhagen interpretation construes that an electron spin
is simultaneously in both eigenstates and collapses statistically upon
measurement to either \cite{Feynman1963}. The collapse is not modeled by the
original Schrödinger equation but stated separately as a measurement
postulate \cite{Norsen2017}. Debatable inconsistency has been found in thought
experiments, such as ``Schrödinger's cat'' \cite{Einstein1935,Frauchiger2018,Schrodinger1935}.

CQD potentially offers new insight. If co-quanta are isotropically
distributed, CQD has been verified with quantum mechanics by exactly
reproducing the wave function and the density operator (Appendix~\ref{appendix:3}) as
well as the uncertainty relation (Appendix~\ref{appendix:6}) and entangled wave
function (Appendix~\ref{appendix:7}). The probabilities of reaching the two eigenstates
split according to
\(\cos^{2}\frac{\theta_{e}}{2}:\sin^{2}\frac{\theta_{e}}{2}\); the wave
function is reproduced in Eq.~\ref{eq:18}. However, if the co-quanta have, for
example, a heart-shaped distribution (Eq.~\ref{eq:24}), the split becomes
\(1 - \sin^{4}\frac{\theta_{e}}{2}:\sin^{4}\frac{\theta_{e}}{2}\) (Eqs.~\ref{eq:88} and \ref{eq:89} in Appendix~\ref{appendix:3}); the wave function is revised accordingly (Eq.~\ref{eq:90}). The density operator is found to originate from a pre-averaging
counterpart with independent realizations (Appendix~\ref{appendix:3}). The measurement
uncertainty product, explained by co-quanta, depends on the initial
phase of the principal quanta and the measurement sequence, as shown by
the uncertainty equality (Eq.~\ref{eq:186}), which leads to the familiar quantum
mechanical inequality (Eq.~\ref{eq:187}). CQD has also enabled the derivation
from the classical Bloch equation to the quantum Schrödinger--Pauli
equation \cite{Wang2022}, while the latter has thus far been treated as a
postulate.

CQD can be further tested with atoms having nuclear spins of 0
(\(\mu_{n} = 0\)), which may collapse differently in Stern--Gerlach
experiments. A natural question is whether higher-order nuclear
multipoles could serve as co-quanta. Examples include
\textsuperscript{38m1}K, \textsuperscript{50}K, \textsuperscript{94}Ag,
and \textsuperscript{130}Ag, which are isotopes of the stable
\textsuperscript{39}K, \textsuperscript{107}Ag, and
\textsuperscript{109}Ag. Unfortunately, these isotopes have short
lifetimes ranging from 100s to 10s of ms. Note that free electrons have
not been used owing to the Lorentz force and orbital magnetic moment.

Since the submission of this manuscript, our team has produced several
new manuscripts to support this work. Titimbo et al. numerically modeled
the Frisch--Segrè experiment using CQD via the Bloch equation \cite{Titimbo2022},
whereas He et al. numerically modeled the experiment using CQD via the
Schrödinger equation \cite{He2022}. Both works have numerically confirmed the
analytical solution presented here and the equivalence between the the
Bloch equation and the Schrödinger equation stated by Majorana.
Interestingly, Majorana wrote the ``Bloch'' equation \cite{Majorana1932,Majorana2006}
fourteen years before Bloch published his eponymous equation \cite{Bloch1946}.
The author recently derived from the Bloch equation, which Bloch
intended for macroscopic magnetization instead of individual nuclear
magnetic moments \cite{Bloch1946}, to the Schrödinger or Schrödinger--Pauli
equation \cite{Wang2022}. Kahraman et al. demonstrated that the standard
existing treatment of hyperfine interaction, consistent with the
Breit--Rabi formula \cite{Breit1931}, cannot model the Frisch--Segrè experiment
accurately but can be improved by incorporating CQD features \cite{Kahraman2024}.
The treatment also does not agree with the Rabi formula \cite{Rabi1936}. 
Three new appendices (\ref{appendix:7}--\ref{appendix:9}) have been added about entanglement, a two–stage Stern--Gerlach apparatus with a varying angle between the quantization axes, and the vector and spherical-coordinate forms of CQD equations of motion, respectively.

While no alternative theory, to the best of our knowledge, matches the
Frisch--Segrè experiment, a recent multi-stage Stern--Gerlach experiment
on superatomic icosahedral cage-clusters Mn@Sn\textsubscript{12} also
reveals discrepancy of the Landau--Zener formula from experimental
observation \cite{Fuchs2018}.

\section{Conclusions}\label{conclusions}

CQD, based on the sign-modified Landau--Lifshitz--Gilbert equation,
provides a plausible collapse mechanism for electron spin in
Stern--Gerlach experiments. CQD models both spin evolution and collapse
by the same equations of motion. With an anisotropic angular
distribution of co-quanta, CQD revises the wave function and accurately
predicts the Frisch--Segrè experimental observation in absolute units
without fitting with adjustable parameter, achieving
\(p < 8 \times 10^{-7}\)---an objective statistical indication that
reflects both correlation and degrees of freedom. Therefore, it is
extremely unlikely that CQD happens to match the experimental
observation so well by sheer chance. Further, with an isotropic angular
distribution of co-quanta, CQD is theoretically corroborated by quantum
mechanics. Both the strong experimental evidence and the exact
quantitative agreement with quantum mechanics in diverse forms
collectively support CQD. Like statistical mechanics \cite{Einstein1949}, which
uses molecular properties to predict macroscopic properties by ensemble
averaging, CQD reproduces quantum mechanical properties by ensemble
averaging over co-quanta (Appendix~\ref{appendix:3}). If orthodox quantum mechanics is
incomplete \cite{Einstein1935}, CQD may stimulate development for a complete
theory.

\section*{Acknowledgments}

The author thanks Dr. Yixuan Tan for discussing the work, translating
the German references using Google Translate, and drawing Fig.~\ref{fig:3}; Drs.
Zhe He and Kelvin Titimbo Chaparro for verifying the mathematical
derivations and discussing the manuscript; Yixuan Tan, David Garrett,
Kelvin Titimbo Chaparro, Siddik Suleyman Kahraman, and Zhe He for
verifying CQD numerically; Prof. JT Shen, Prof. Gil Refael, and Mr. Siddik
Suleyman Kahraman for discussing the manuscript; Prof. Naisyin Wang and
Mr. Sean Wang for discussing the statistics; Dr. Victor Wang for
discussing the math; and Prof. James Ballard for editing an early
version of the manuscript.

\section*{Data availability statement}

The data that support the findings of this study are available upon
request from the authors.

\clearpage

\nocite{*}
\bibliographystyle{unsrt}
\bibliography{references}

@article{Gerlach1922,
  author = {Gerlach, W. and Stern, O.},
  title = {{Der experimentelle nachweis der richtungsquantelung im magnetfeld}},
  journal = {Zeitschrift f{\"u}r Physik},
  year = {1922},
  volume = {9},
  number = {1},
  pages = {349--352},
  doi = {10.1007/BF01326983},
  note = {doi: 10.1007/BF01326983}
}

@article{SchmidtBoecking2016,
  author = {Schmidt-B{\"o}cking, H. and Schmidt, L. and L{\"u}dde, H. J. and Trageser, W. and Templeton, A. and Sauer, T.},
  title = {{The Stern--Gerlach experiment revisited}},
  journal = {The European Physical Journal H},
  year = {2016},
  volume = {41},
  number = {4},
  pages = {327--364},
  doi = {10.1140/epjh/e2016-70053-2},
  note = {doi: 10.1140/epjh/e2016-70053-2}
}

@article{Castelvecchi2022,
  author = {Castelvecchi, D.},
  title = {{The Stern--Gerlach experiment at 100}},
  journal = {Nature Reviews Physics},
  year = {2022},
  doi = {10.1038/s42254-022-00436-4},
  note = {doi: 10.1038/s42254-022-00436-4}
}

@article{Einstein1922,
  author = {Einstein, A. and Ehrenfest, P.},
  title = {{Quantentheoretische bemerkungen zum experiment von Stern und Gerlach}},
  journal = {Zeitschrift f{\"u}r Physik},
  year = {1922},
  volume = {11},
  number = {1},
  pages = {31--34},
  doi = {10.1007/BF01328398},
  note = {doi: 10.1007/BF01328398}
}

@article{Wennerstrom2012,
  author = {Wennerstr{\"o}m, H. and Westlund, P.-O.},
  title = {{The Stern--Gerlach experiment and the effects of spin relaxation}},
  journal = {Physical Chemistry Chemical Physics},
  year = {2012},
  volume = {14},
  number = {5},
  pages = {1677--1684},
  doi = {10.1039/C2CP22173J},
  note = {doi: 10.1039/C2CP22173J}
}

@article{Norsen2014,
  author = {Norsen, T.},
  title = {{The pilot-wave perspective on spin}},
  journal = {American Journal of Physics},
  year = {2014},
  volume = {82},
  number = {4},
  pages = {337--348}
}

@book{Feynman1963,
  author = {Feynman, R. P. and Leighton, R. B. and Sands, M. L.},
  title = {{The Feynman Lectures on Physics}},
  address = {Reading, Mass.},
  publisher = {Addison-Wesley Pub. Co.},
  year = {1963}
}

@article{Phipps1932,
  author = {Phipps, T. E. and Stern, O.},
  title = {{{\"U}ber die einstellung der richtungsquantelung}},
  journal = {Zeitschrift f{\"u}r Physik},
  year = {1932},
  volume = {73},
  number = {3},
  pages = {185--191},
  doi = {10.1007/BF01351212},
  note = {doi: 10.1007/BF01351212}
}

@article{Frisch1933,
  author = {Frisch, R. and Segr{\`e}, E.},
  title = {{{\"U}ber die einstellung der richtungsquantelung. II.}},
  journal = {Zeitschrift f{\"u}r Physik},
  year = {1933},
  volume = {80},
  number = {9},
  pages = {610--616},
  doi = {10.1007/BF01335699},
  note = {doi: 10.1007/BF01335699}
}

@article{Majorana1932,
  author = {Majorana, E.},
  title = {{Atomi orientati in campo magnetico variabile}},
  journal = {Il Nuovo Cimento (1924--1942)},
  year = {1932},
  volume = {9},
  number = {2},
  pages = {43--50},
  doi = {10.1007/BF02960953},
  note = {doi: 10.1007/BF02960953}
}

@incollection{Majorana2006,
  author = {Majorana, E.},
  title = {{Oriented atoms in a variable magnetic field}},
  editor = {Bassani, G.},
  booktitle = {{Ettore Majorana: Scientific Papers}},
  address = {Bologna, Berlin},
  publisher = {Societ{\`a} Italiana di Fisica and Springer},
  year = {2006},
  pages = {125--132}
}

@article{Landau1932,
  author = {Landau, L.},
  title = {{Zur theorie der energieubertragung. II.}},
  journal = {Physikalische Zeitschrift der Sowjetunion},
  year = {1932},
  volume = {2},
  pages = {46--51}
}

@article{Zener1932,
  author = {Zener, C.},
  title = {{Non-adiabatic crossing of energy levels}},
  journal = {Proceedings of the Royal Society of London Series A},
  year = {1932},
  volume = {137},
  pages = {696}
}

@article{Stueckelberg1932,
  author = {Stueckelberg, E. C. G.},
  title = {{Theorie der unelastischen St{\"o}sse zwischen atomen}},
  journal = {Helvetica Physica Acta},
  year = {1932},
  volume = {5},
  pages = {369}
}

@article{Ivakhnenko2023,
  author = {Ivakhnenko, O. V. and Shevchenko, S. N. and Nori, F.},
  title = {{Nonadiabatic Landau--Zener--St{\"u}ckelberg--Majorana transitions, dynamics, and interference}},
  journal = {Physics Reports},
  year = {2023},
  volume = {995},
  pages = {1--89},
  doi = {10.1016/j.physrep.2022.10.002},
  note = {doi: 10.1016/j.physrep.2022.10.002}
}

@article{Rabi1936,
  author = {Rabi, I. I.},
  title = {{On the process of space quantization}},
  journal = {Physical Review},
  year = {1936},
  volume = {49},
  number = {4},
  pages = {324--328},
  doi = {10.1103/PhysRev.49.324},
  note = {doi: 10.1103/PhysRev.49.324}
}

@article{Wang2022,
  author = {Wang, L. V.},
  title = {Derivation from {Bloch} Equation to {von Neumann} Equation to {Schr{\"o}dinger--Pauli} Equation},
  journal = {Foundations of Physics},
  year = {2022},
  volume = {52},
  number = {3},
  pages = {61}
}

@article{Carlesso2022,
  author = {Carlesso, M. and Donadi, S. and Ferialdi, L. and Paternostro, M. and Ulbricht, H. and Bassi, A.},
  title = {{Present status and future challenges of non-interferometric tests of collapse models}},
  journal = {Nature Physics},
  year = {2022},
  doi = {10.1038/s41567-021-01489-5},
  note = {doi: 10.1038/s41567-021-01489-5}
}

@article{Ghirardi1986,
  author = {Ghirardi, G. C. and Rimini, A. and Weber, T.},
  title = {{Unified dynamics for microscopic and macroscopic systems}},
  journal = {Physical Review D},
  year = {1986},
  volume = {34},
  number = {2},
  pages = {470--491},
  doi = {10.1103/physrevd.34.470},
  note = {doi: 10.1103/physrevd.34.470}
}

@article{Pearle1989,
  author = {Pearle, P.},
  title = {{Combining stochastic dynamical state-vector reduction with spontaneous localization}},
  journal = {Physical Review A},
  year = {1989},
  volume = {39},
  number = {5},
  pages = {2277}
}

@article{Ghirardi1990,
  author = {Ghirardi, G. C. and Pearle, P. and Rimini, A.},
  title = {{Markov processes in Hilbert space and continuous spontaneous localization of systems of identical particles}},
  journal = {Physical Review A},
  year = {1990},
  volume = {42},
  number = {1},
  pages = {78--89},
  doi = {10.1103/physreva.42.78},
  note = {doi: 10.1103/physreva.42.78}
}

@book{Steel1960,
  author = {Steel, R. G. D. and Torrie, J. H.},
  title = {{Principles and Procedures of Statistics}},
  address = {New York},
  publisher = {McGraw Hill},
  year = {1960}
}

@book{Rahman1968,
  author = {Rahman, N.},
  title = {{A Course in Theoretical Statistics}},
  address = {New York},
  publisher = {Charles Griffin and Company},
  year = {1968}
}

@article{Cousins2017,
  author = {Cousins, R. D.},
  title = {{The Jeffreys--Lindley paradox and discovery criteria in high energy physics}},
  journal = {Synthese},
  year = {2017},
  volume = {194},
  number = {2},
  pages = {395--432}
}

@article{Abbott2016,
  author = {Abbott, B. P. and Abbott, R. and Abbott, T. D. and Abernathy, M. R. and Acernese, F. and Ackley, K. and others},
  title = {Observation of Gravitational Waves from a Binary Black Hole Merger},
  journal = {Physical Review Letters},
  year = {2016},
  volume = {116},
  number = {6},
  doi = {10.1103/physrevlett.116.061102},
  note = {doi: 10.1103/physrevlett.116.061102}
}

@book{Grynberg2010,
  author = {Grynberg, G. and Aspect, A. and Fabre, C.},
  title = {{Introduction to Quantum Optics: From the Semi-Classical Approach to Quantized Light}},
  publisher = {Cambridge University Press},
  year = {2010}
}

@article{Feynman1957,
  author = {Feynman, R. P. and Vernon, Jr., F. L. and Hellwarth, R. W.},
  title = {{Geometrical representation of the Schr{\"o}dinger equation for solving maser problems}},
  journal = {Journal of Applied Physics},
  year = {1957},
  volume = {28},
  number = {1},
  pages = {49--52}
}

@article{Gilbert2004,
  author = {Gilbert, T. L.},
  title = {{A phenomenological theory of damping in ferromagnetic materials}},
  journal = {IEEE Transactions on Magnetics},
  year = {2004},
  volume = {40},
  number = {6},
  pages = {3443--3449},
  doi = {10.1109/TMAG.2004.836740},
  note = {doi: 10.1109/TMAG.2004.836740}
}

@misc{LosAlamosPeriodicTable,
  author = {{Los Alamos National Laboratory}},
  title = {{Periodic Table of Elements}},
  howpublished = {\url{https://periodic.lanl.gov/19.shtml}},
  note = {Accessed}
}

@book{Norsen2017,
  author = {Norsen, T.},
  title = {{Foundations of Quantum Mechanics}},
  publisher = {Springer},
  year = {2017}
}

@book{Griffiths2017,
  author = {Griffiths, D. J.},
  title = {{Introduction to Electrodynamics}},
  edition = {4th},
  publisher = {Cambridge University Press},
  year = {2017}
}

@book{Jackson1999,
  author = {Jackson, J. D.},
  title = {{Classical Electrodynamics}},
  address = {USA},
  publisher = {John Wiley \& Sons},
  year = {1999}
}

@article{Ohanian1986,
  author = {Ohanian, H. C.},
  title = {{What is spin?}},
  journal = {American Journal of Physics},
  year = {1986},
  volume = {54},
  number = {6},
  pages = {500--505},
  doi = {10.1119/1.14580},
  note = {doi: 10.1119/1.14580}
}

@article{Rabi1954,
  author = {Rabi, I. I. and Ramsey, N. F. and Schwinger, J.},
  title = {{Use of rotating coordinates in magnetic resonance problems}},
  journal = {Reviews of Modern Physics},
  year = {1954},
  volume = {26},
  number = {2},
  pages = {167}
}

@incollection{Forbes2013,
  author = {Forbes, M. D. and Jarocha, L. E. and Sim, S. and Tarasov, V. F.},
  title = {{Time-resolved electron paramagnetic resonance spectroscopy: History, technique, and application to supramolecular and macromolecular chemistry}},
  editor = {Williams, I. H. and Williams, N. H.},
  booktitle = {{Advances in Physical Organic Chemistry}},
  publisher = {Elsevier},
  year = {2013},
  pages = {1--83}
}

@incollection{Marchione2016,
  author = {Marchione, A. A. and Conklin, B.},
  title = {{Gas Phase NMR for the Study of Chemical Reactions: Kinetics and Product Identification}},
  editor = {Jackowski, K. and Jaszu{\'n}ski, M.},
  booktitle = {{Gas Phase NMR}},
  address = {Cambridge},
  publisher = {Royal Society of Chemistry},
  year = {2016},
  pages = {126--151}
}

@article{Schroder1983,
  author = {Schroder, W. and Baum, G.},
  title = {{A spin flipper for reversal of polarisation in a thermal atomic beam}},
  journal = {Journal of Physics E: Scientific Instruments},
  year = {1983},
  volume = {16},
  number = {1},
  pages = {52}
}

@article{Breit1931,
  author = {Breit, G. and Rabi, I.},
  title = {{Measurement of nuclear spin}},
  journal = {Physical Review},
  year = {1931},
  volume = {38},
  number = {11},
  pages = {2082}
}

@incollection{Barra2005,
  author = {Barra, A. L. and Hassan, A. K.},
  title = {{Electron Spin Resonance}},
  editor = {Bassani, F. and Liedl, G. L. and Wyder, P.},
  booktitle = {{Encyclopedia of Condensed Matter Physics}},
  address = {Oxford},
  publisher = {Elsevier},
  year = {2005},
  pages = {58--67}
}

@article{Adler2009,
  author = {Adler, S. L. and Bassi, A.},
  title = {Is Quantum Theory Exact?},
  journal = {Science},
  year = {2009},
  volume = {325},
  number = {5938},
  pages = {275--276},
  doi = {10.1126/science.1176858},
  note = {doi: 10.1126/science.1176858}
}

@book{Bricmont2016,
  author = {Bricmont, J.},
  title = {{Making Sense of Quantum Mechanics}},
  publisher = {Springer},
  year = {2016}
}

@book{Laloe2019,
  author = {Lalo{\"e}, F.},
  title = {{Do We Really Understand Quantum Mechanics?}},
  publisher = {Cambridge University Press},
  year = {2019}
}

@book{Auletta2019,
  author = {Auletta, G.},
  title = {{The Quantum Mechanics Conundrum: Interpretation and Foundations}},
  publisher = {Springer},
  year = {2019}
}

@article{Einstein1935,
  author = {Einstein, A. and Podolsky, B. and Rosen, N.},
  title = {{Can quantum-mechanical description of physical reality be considered complete}},
  journal = {Physical Review},
  year = {1935},
  volume = {47},
  number = {10},
  pages = {777--780},
  doi = {10.1103/PhysRev.47.777},
  note = {doi: 10.1103/PhysRev.47.777}
}

@article{Frauchiger2018,
  author = {Frauchiger, D. and Renner, R.},
  title = {{Quantum theory cannot consistently describe the use of itself}},
  journal = {Nature Communications},
  year = {2018},
  volume = {9},
  number = {1},
  pages = {3711},
  doi = {10.1038/s41467-018-05739-8},
  note = {doi: 10.1038/s41467-018-05739-8}
}

@article{Schrodinger1935,
  author = {Schr{\"o}dinger, E.},
  title = {{Die gegenw{\"a}rtige situation in der quantenmechanik}},
  journal = {Naturwissenschaften},
  year = {1935},
  volume = {23},
  pages = {807}
}

@article{Titimbo2022,
  author = {Titimbo, K. and Garrett, D. C. and Kahraman, S. S. and He, Z. and Wang, L. V.},
  title = {{Numerical modeling of the multi-stage Stern--Gerlach experiment by Frisch and Segr{\`e} using co-quantum dynamics via the Bloch equation}},
  journal = {arXiv preprint arXiv:2208.13444},
  year = {2022}
}

@article{He2022,
  author = {He, Z. and Titimbo, K. and Garrett, D. C. and Kahraman, S. S. and Wang, L. V.},
  title = {{Numerical modeling of the multi-stage Stern--Gerlach experiment by Frisch and Segr{\`e} using co-quantum dynamics via the Schr{\"o}dinger equation}},
  journal = {arXiv preprint arXiv:2208.14588},
  year = {2022}
}

@article{Bloch1946,
  author = {Bloch, F.},
  title = {Nuclear Induction},
  journal = {Physical Review},
  year = {1946},
  volume = {70},
  number = {7--8},
  pages = {460--474},
  doi = {10.1103/physrev.70.460},
  note = {doi: 10.1103/physrev.70.460}
}

@article{Kahraman2024,
  author = {Kahraman, S. S. and Titimbo, K. and He, Z. and Shen, J.-T. and Wang, L. V.},
  title = {Quantum mechanical modeling of the multi-stage {Stern--Gerlach} experiment conducted by {Frisch and Segr{\`e}}},
  journal = {New Journal of Physics},
  year = {2024},
  volume = {26},
  pages = {073005}
}

@article{Fuchs2018,
  author = {Fuchs, T. M. and Sch{\"a}fer, R.},
  title = {{Double Stern--Gerlach experiments on Mn@Sn12: Refocusing of a paramagnetic superatom}},
  journal = {Physical Review A},
  year = {2018},
  volume = {98},
  number = {6},
  pages = {063411}
}

@incollection{Einstein1949,
  author = {Einstein, A.},
  title = {{Remarks concerning the essays brought together in this co-operative volume}},
  editor = {Schilpp, P. A.},
  booktitle = {{Albert Einstein: Philosopher-Scientist}},
  series = {Library of Living Philosophers},
  address = {La Salle},
  publisher = {Open Court},
  year = {1949},
  pages = {665--688}
}

@book{Bjorken1964,
  author = {Bjorken, J. D. and Drell, S. D.},
  title = {{Relativistic Quantum Mechanics}},
  address = {New York},
  publisher = {McGraw-Hill College},
  year = {1964}
}

@article{Gibbons1935,
  author = {Gibbons, J. J. and Bartlett, J. H.},
  title = {{The magnetic moment of the K$^{39}$ nucleus}},
  journal = {Physical Review},
  year = {1935},
  volume = {47},
  number = {9},
  pages = {692--694},
  doi = {10.1103/physrev.47.692},
  note = {doi: 10.1103/physrev.47.692}
}

@article{Hartree1934,
  author = {Hartree, D. R.},
  title = {{Results of calculations of atomic wave functions. II. Results for K$^+$ and Cs$^+$}},
  journal = {Proceedings of the Royal Society of London Series A},
  year = {1934},
  volume = {143},
  number = {850},
  pages = {506--517}
}

@article{Wittig2005,
  author = {Wittig, C.},
  title = {{The Landau--Zener formula}},
  journal = {Journal of Physical Chemistry B},
  year = {2005},
  volume = {109},
  number = {17},
  pages = {8428--8430},
  doi = {10.1021/jp040627u},
  note = {doi: 10.1021/jp040627u}
}

@incollection{Wilczek2014,
  author = {Wilczek, F.},
  title = {{Majorana and condensed matter physics}},
  editor = {Esposito, S.},
  booktitle = {{The Physics of Ettore Majorana: Theoretical, Mathematical, and Phenomenological}},
  publisher = {Cambridge University Press},
  year = {2014},
  pages = {279--302}
}

@article{Kofman2022,
  author = {Kofman, P. O. and Ivakhnenko, O. V. and Shevchenko, S. N. and Nori, F.},
  title = {{Majorana's approach to nonadiabatic transitions validates the adiabatic-impulse approximation}},
  journal = {arXiv preprint arXiv:2208.00481},
  year = {2022}
}

@book{Lambropoulos2007,
  author = {Lambropoulos, P. and Petrosyan, D.},
  title = {{Fundamentals of Quantum Optics and Quantum Information}},
  publisher = {Springer},
  year = {2007}
}

@article{Bell1966,
  author = {Bell, J. S.},
  title = {{On the problem of hidden variables in quantum mechanics}},
  journal = {Reviews of Modern Physics},
  year = {1966},
  volume = {38},
  number = {3},
  pages = {447--452},
  doi = {10.1103/RevModPhys.38.447},
  note = {doi: 10.1103/RevModPhys.38.447}
}

@article{Shin2019,
  author = {Shin, D. K. and Henson, B. M. and Hodgman, S. S. and Wasak, T. and Chwede{\'n}czuk, J. and Truscott, A. G.},
  title = {{Bell correlations between spatially separated pairs of atoms}},
  journal = {Nature Communications},
  year = {2019},
  volume = {10},
  number = {1},
  doi = {10.1038/s41467-019-12192-8},
  note = {doi: 10.1038/s41467-019-12192-8}
}

@article{Thomas2022,
  author = {Thomas, K. F. and Henson, B. M. and Wang, Y. and Lewis-Swan, R. J. and Kheruntsyan, K. V. and Hodgman, S. S. and others},
  title = {{A matter wave Rarity-Tapster interferometer to demonstrate non-locality}},
  journal = {arXiv preprint arXiv:2206.08560},
  year = {2022}
}

@article{Wang2023,
  author = {Wang, L. V.},
  title = {{Multi-stage Stern--Gerlach experiment modeled}},
  journal = {Journal of Physics B: Atomic, Molecular and Optical Physics},
  year = {2023},
  volume = {56},
  number = {10},
  pages = {105001},
  doi = {10.1088/1361-6455/acc149},
  note = {doi: 10.1088/1361-6455/acc149}
}
\clearpage
\setcounter{figure}{0}
\setcounter{table}{0}
\setcounter{section}{0}
\renewcommand{\thefigure}{S\arabic{figure}}
\renewcommand{\thetable}{S\arabic{table}}
\renewcommand{\thesection}{\arabic{section}}
\renewcommand{\theHsection}{appendix.\arabic{section}}
\makeatletter
\newcommand{\appendixseccntsection}{Appendix~\thesection.\quad}
\newcommand{\appendixseccntsubsection}{\thesubsection.\quad}
\newcommand{\appendixseccntsubsubsection}{\thesubsubsection.\quad}
\renewcommand{\@seccntformat}[1]{%
  \@ifundefined{appendixseccnt#1}{\csname the#1\endcsname.\quad}{\csname appendixseccnt#1\endcsname}%
}
\makeatother
\phantomsection
\addcontentsline{toc}{section}{Supplementary Material}
\FloatBarrier
\clearpage
\section[Appendix~\thesection. Derivation of torque-averaged fields]{Derivation of torque-averaged fields}\label{appendix:1}
\suppressfloats[t]

Given the focus of the Landau--Lifshitz--Gilbert equation on torque, we
derive the torque-averaged magnetic flux densities applied on the
electron and the nucleus by each other.

In relativistic quantum mechanics, the momentum density in the Dirac
wave field is given by \cite{Ohanian1986}

\begin{equation}
\label{eq:31}
\vec{G} = \frac{\hslash}{2i}\left\lbrack \psi^{\dagger}\nabla\psi - \left( \nabla\psi^{\dagger} \right)\psi \right\rbrack + \frac{\hslash}{4}\nabla \times \left( \psi^{\dagger}{\vec{\sigma}}_{4 \times 4}\psi \right).
\end{equation}

Here, \(\psi\) denotes the spatial wave function, \(\psi_{s}\),
multiplied by the spinor \({w^{1}(0) = (1,0,0,0)}^{\dagger}\);
\({\vec{\sigma}}_{4 \times 4} = \sigma_{1}\widehat{x} + \sigma_{2}\widehat{y} + \sigma_{3}\widehat{z}\),
where \(\sigma_{1} = - i\alpha_{2}\alpha_{3}\),
\(\sigma_{2} = - i\alpha_{3}\alpha_{1}\),
\(\sigma_{3} = - i\alpha_{1}\alpha_{2}\); matrices \(\alpha_{1}\),
\(\alpha_{2}\), and \(\alpha_{3}\) are from the Dirac equation \cite{Bjorken1964}:

\begin{equation}
\label{eq:32}
\alpha_{1} = \begin{pmatrix}
0 & 0 & 0 & 1 \\
0 & 0 & 1 & 0 \\
0 & 1 & 0 & 0 \\
1 & 0 & 0 & 0
\end{pmatrix},\ \alpha_{2} = \begin{pmatrix}
0 & 0 & 0 & - i \\
0 & 0 & i & 0 \\
0 & - i & 0 & 0 \\
i & 0 & 0 & 0
\end{pmatrix}, \mathrm{and}\ \alpha_{3} = \begin{pmatrix}
0 & 0 & 1 & 0 \\
0 & 0 & 0 & - 1 \\
1 & 0 & 0 & 0 \\
0 & - 1 & 0 & 0
\end{pmatrix}.
\end{equation}

While the first term on the right side of Eq.~\ref{eq:31} is attributed to the
translational motion of the electron, the second term is associated with
circulating flow of energy \cite{Ohanian1986}.

Following Ohanian \cite{Ohanian1986}, an \(s\) orbital wave function is
considered; \(\psi_{s}^{\dagger}\psi_{s}\) is set to the Gaussian
distribution,

\begin{equation}
\label{eq:33}
\rho(r) = \left( \frac{1}{{\pi a}_{0}^{2}} \right)^{\frac{3}{2}}\exp\left( - \frac{r^{2}}{a_{0}^{2}} \right),
\end{equation}

where \(r\) denotes the radial coordinate. The average radius,
\(\frac{2\, a_{0}}{\sqrt{\pi}}\), is set to the van der Waals atomic
radius, \(R\).

While the first term in Eq.~\ref{eq:31} vanishes, the second term becomes

\begin{equation}
\label{eq:34}
\vec{G} = \frac{\hslash}{2a_{0}^{2}}\rho\widehat{z} \times \vec{r}.
\end{equation}

The differential element of the magnetic dipole moment is \cite{Ohanian1986}

\begin{equation}
\label{eq:35}
d{\vec{m}}_{e}\left( \vec{r} \right) = \gamma_{e}\vec{r} \times \left( \frac{\hslash}{4}\nabla \times \left( \psi^{\dagger}\gamma^{0}{\vec{\sigma}}_{4 \times 4}\psi \right) \right)d^{3}\vec{r}.
\end{equation}

From

\begin{equation}
\label{eq:36}
\gamma^{0} = \begin{pmatrix}
1 & 0 & 0 & 0 \\
0 & 1 & 0 & 0 \\
0 & 0 & - 1 & 0 \\
0 & 0 & 0 & - 1
\end{pmatrix},
\end{equation}

we reach

\begin{equation}
\label{eq:37}
d{\vec{m}}_{e}\left( \vec{r} \right) = \gamma_{e}\vec{r} \times \vec{G}d^{3}\vec{r}.
\end{equation}

The field at location \(\vec{r}\) from
\(\vec{\mu}_{n}\) is given by \cite{Griffiths2017,Jackson1999}

\begin{equation}
\label{eq:38}
\vec{B}\left( \vec{r} \right) = \frac{\mu_{0}}{4\pi r^{3}}\left\lbrack 3\left( \vec{\mu}_{n} \cdot \widehat{r} \right)\widehat{r} - \vec{\mu}_{n} \right\rbrack + \frac{2\mu_{0}}{3}\vec{\mu}_{n}\delta\left( \vec{r} \right).
\end{equation}

The differential element of the torque from
\(\vec{\mu}_{n}\) is

\begin{equation}
\label{eq:39}
d{\vec{\tau}}_{n}\left( \vec{r} \right) = d{\vec{m}}_{e}\left( \vec{r} \right) \times \vec{B}\left( \vec{r} \right).
\end{equation}

Volumetric integration yields

\begin{equation}
\label{eq:40}
{\vec{m}}_{e}\left( \vec{r} \right) = \int_{}^{}{d{\vec{m}}_{e}} = \gamma_{e}\frac{\hslash}{2}\widehat{z},
\end{equation}

which equals \(\vec{\mu}_{e}\), and

\begin{equation}
\label{eq:41}
{\vec{\tau}}_{n} = \int_{}^{}{d{\vec{\tau}}_{n}} = \vec{\mu}_{e} \times \left( \frac{4\mu_{0}}{3\pi^{3}R^{3}}\vec{\mu}_{n} \right).
\end{equation}

Therefore, the torque-averaged \(B\) field from
\(\vec{\mu}_{n}\) applied on the electron is given by

\begin{equation}
\label{eq:42}
\vec{B}_{n} = \frac{4\mu_{0}}{3\pi^{3}R^{3}}\vec{\mu}_{n}.
\end{equation}

Now, we switch to a classical approach. On the time scale pertinent to
the precession cycle, we model the much faster motion of the \(s\)
valence electron with the probability density, \(\rho\).

The current density at position \(\vec{r}\) is given by

\begin{equation}
\label{eq:43}
\vec{j} = - e\rho\vec{\omega} \times \vec{r},
\end{equation}

where \(- e\) denotes the electron charge and
\(\vec{\omega}\) denotes the angular velocity.

The differential element of the magnetic dipole moment is

\begin{equation}
\label{eq:44}
d{\vec{m}}_{e}\left( \vec{r} \right) = \frac{1}{2}\vec{r} \times \vec{j}d^{3}\vec{r}.
\end{equation}

The differential element of the torque is (Eq.~\ref{eq:39})

\begin{equation}
\label{eq:45}
d{\vec{\tau}}_{n}\left( \vec{r} \right) = d{\vec{m}}_{e}\left( \vec{r} \right) \times \vec{B}\left( \vec{r} \right).
\end{equation}

Three distributions of \(\rho\) are considered.

First, \(\rho\) is set to the Gaussian distribution given by Eq.~\ref{eq:33}.
Volumetric integration produces

\begin{equation}
\label{eq:46}
{\vec{m}}_{e} = \int_{}^{}{d{\vec{m}}_{e}} = - \frac{1}{2}ea_{0}^{2}\vec{\omega} = - \frac{\pi}{8}eR^{2}\vec{\omega}
\end{equation}

and

\begin{equation}
\label{eq:47}
{\vec{\tau}}_{n} = \int_{}^{}{d{\vec{\tau}}_{n}} = {\vec{m}}_{e} \times \left( \frac{4\mu_{0}}{3\pi^{3}R^{3}}\vec{\mu}_{n} \right).
\end{equation}

Averaging yields

\begin{equation}
\label{eq:48}
\left\langle {\vec{\tau}}_{n} \right\rangle = \left\langle {\vec{m}}_{e} \right\rangle \times \left( \frac{4\mu_{0}}{3\pi^{3}R^{3}}\vec{\mu}_{n} \right).
\end{equation}

For an \(s\) valence electron, both the orbital angular momentum and the
orbital magnetic moment vanish; accordingly, we have
\(\left\langle {\vec{m}}_{e} \right\rangle = \vec{\mu}_{e}\),
which is due to spin only. Consequently,

\begin{equation}
\label{eq:49}
\left\langle {\vec{\tau}}_{n} \right\rangle = \vec{\mu}_{e} \times \left( \frac{4\mu_{0}}{3\pi^{3}R^{3}}\vec{\mu}_{n} \right).
\end{equation}

Therefore, we reach

\begin{equation}
\label{eq:50}
\vec{B}_{n} = \frac{4\mu_{0}}{3\pi^{3}R^{3}}\vec{\mu}_{n},
\end{equation}

which agrees with the relativistic quantum mechanical solution (Eq.~\ref{eq:42}).

Second, using the tabulated approximate 4\emph{s} wave function,
\(\psi_{s}\), for potassium (Table~\ref{tab:S1}) \cite{Gibbons1935} based on Hartree's
self-consistent field \cite{Hartree1934}, we numerically reached

\begin{equation}
\label{eq:51}
\vec{B}_{n} = \frac{0.138\mu_{0}}{\pi R^{3}}\vec{\mu}_{n},
\end{equation}

where the average radius is set to\(\ R\). Coincidentally, this solution
differs from the Gaussian solution (Eq.~\ref{eq:50}) by only 2\%.

\begin{table}[!htbp]
\centering
\caption{Normalized \(P(4s)\) for potassium.
\(\psi_{s}(0) = \sqrt{\frac{9.76}{(4\pi)}}\) and
\(\psi_{s}(r) = \frac{P}{\left( \sqrt{4\pi a_{0}}r \right)}\) for
\(r > 0\). \cite{Gibbons1935}}
\label{tab:S1}
\scriptsize
\setlength{\tabcolsep}{4pt}
\begin{tabular*}{0.9\textwidth}{@{\extracolsep{\fill}}
  S[table-format=2.3]
  S[table-format=1.4]
  S[table-format=2.3]
  S[table-format=1.4]
  S[table-format=2.3]
  S[table-format=1.4]
  S[table-format=2.3]
  S[table-format=1.4]}
\toprule
{\(\frac{r}{a_{0}}\)} & {\(P\)} & {\(\frac{r}{a_{0}}\)} & {\(P\)} & {\(\frac{r}{a_{0}}\)} & {\(P\)} & {\(\frac{r}{a_{0}}\)} & {\(P\)} \\
\midrule
0.000 & 0.0000 & 0.280 & -0.0993 & 2.800 & -0.2524 & 13.000 & -0.0634 \\
0.005 & 0.0142 & 0.300 & -0.0972 & 3.000 & -0.2926 & 14.000 & -0.0443 \\
0.010 & 0.0257 & 0.350 & -0.0830 & 3.200 & -0.3279 & 15.000 & -0.0305 \\
0.015 & 0.0349 & 0.400 & -0.0598 & 3.400 & -0.3583 & 16.000 & -0.0209 \\
0.020 & 0.0421 & 0.450 & -0.0312 & 3.600 & -0.3840 & 17.000 & -0.0138 \\
0.030 & 0.0509 & 0.500 & -0.0003 & 3.800 & -0.4052 & 18.000 & -0.0095 \\
0.040 & 0.0540 & 0.550 & 0.0307 & 4.000 & -0.4221 & 19.000 & -0.0063 \\
0.050 & 0.0527 & 0.600 & 0.0601 & 4.500 & -0.4476 & 20.000 & -0.0042 \\
0.060 & 0.0480 & 0.700 & 0.1105 & 5.000 & -0.4530 & 21.000 & -0.0028 \\
0.070 & 0.0409 & 0.800 & 0.1465 & 5.500 & -0.4430 & 22.000 & -0.0018 \\
0.080 & 0.0321 & 0.900 & 0.1679 & 6.000 & -0.4220 & 23.000 & -0.0012 \\
0.090 & 0.0220 & 1.000 & 0.1761 & 6.500 & -0.3937 & 24.000 & -0.0008 \\
0.100 & 0.0113 & 1.100 & 0.1734 & 7.000 & -0.3609 & 25.000 & -0.0005 \\
0.120 & -0.0108 & 1.200 & 0.1623 & 7.500 & -0.3264 & 26.000 & -0.0003 \\
0.140 & -0.0321 & 1.400 & 0.1226 & 8.000 & -0.2916 & 27.000 & -0.0002 \\
0.160 & -0.0511 & 1.600 & 0.0699 & 8.500 & -0.2578 & 28.000 & -0.0001 \\
0.180 & -0.0673 & 1.800 & 0.0119 & 9.000 & -0.2261 & 29.000 & -0.0001 \\
0.200 & -0.0801 & 2.000 & -0.0470 & 9.500 & -0.1967 & 30.000 & 0.0000 \\
0.220 & -0.0896 & 2.200 & -0.1040 & 10.000 & -0.1700 & 31.000 & 0.0000 \\
0.240 & -0.0958 & 2.400 & -0.1578 & 11.000 & -0.1246 & {} & {} \\
0.260 & -0.0989 & 2.600 & -0.2074 & 12.000 & -0.0896 & {} & {} \\
\bottomrule
\end{tabular*}
\end{table}

Third, \(\rho\) is set to the following top-hat distribution, which is a
zeroth-order approximation to the actual distribution:

\begin{equation}
\label{eq:52}
\rho(r) = \frac{3}{4{\pi R}^{3}}
\end{equation}

for \(r \leq R\) and \(\rho = 0\) otherwise. Repeating the derivation
starting from Eq.~\ref{eq:43} yields

\begin{equation}
\label{eq:53}
\vec{B}_{n} = \frac{5\mu_{0}}{16\pi R^{3}}\vec{\mu}_{n}.
\end{equation}

In addition to the above torque-averaged fields, for comparison, the
self-averaged fields from Eq.~\ref{eq:38} are derived:

\begin{equation}
\label{eq:54}
\vec{B}_{n} = \frac{2\mu_{0}}{3}\rho(0)\vec{\mu}_{n}.
\end{equation}

For the Gaussian, tabulated, and top-hat distributions,
\(\rho(0) = \frac{8}{\pi^{3}R^{3}}\), \(\frac{478.6}{\pi R^{3}}\), and
\(\frac{3}{4{\pi R}^{3}}\), respectively; correspondingly,
\(\vec{B}_{n} = \frac{16\mu_{0}}{3\pi^{3}R^{3}}\vec{\mu}_{n}\),
\(\frac{957.3\mu_{0}}{3\pi R^{3}}\vec{\mu}_{n}\), and
\(\frac{\mu_{0}}{2\pi R^{3}}\vec{\mu}_{n}\). These
self-averaged fields, related to the Fermi contact interaction, are
expected to be less compatible with the torque-based
Landau--Lifshitz--Gilbert and CQD equations than the above
torque-averaged fields.

The six solutions for \(\vec{B}_{n}\) differ only by a
constant factor. Eq.~\ref{eq:53}, however, predicts the experimental observation
(Fig.~\ref{fig:4}) most accurately, achieving a coefficient of determination of
\(R^{2} = 0.9787\). The alternatives produce negative \(R^{2}\),
indicating worse accuracy than modeling with a horizontal line
intercepting at the mean observation. Therefore, we choose the
torque-averaged field given by Eq.~\ref{eq:53}, rewritten below:

\begin{equation}
\label{eq:55}
\vec{B}_{n} = \frac{5\mu_{0}}{16\pi R^{3}}\vec{\mu}_{n}.
\end{equation}

Reciprocally, the torque-averaged field from
\(\vec{\mu}_{e}\) applied on the nucleus is

\begin{equation}
\label{eq:56}
\vec{B}_{e} = \frac{5\mu_{0}}{16\pi R^{3}}\vec{\mu}_{e}.
\end{equation}

\FloatBarrier
\clearpage
\section[Appendix~\thesection. Derivation of CQD equations of motion]{Derivation of CQD equations of motion}\label{appendix:2}
\suppressfloats[t]

We apply the Landau--Lifshitz--Gilbert equation to both
\({\widehat{\mu}}_{e}\) and \({\widehat{\mu}}_{n}\), yielding

\begin{equation}
\label{eq:57}
\frac{d{\widehat{\mu}}_{e}}{dt} = \gamma_{e}{\widehat{\mu}}_{e} \times \left( \vec{B} + \vec{B}_{n} \right) - k_{i}{\widehat{\mu}}_{e} \times \frac{d{\widehat{\mu}}_{e}}{dt}
\end{equation}

and

\begin{equation}
\label{eq:58}
\frac{d{\widehat{\mu}}_{n}}{dt} = \gamma_{n}{\widehat{\mu}}_{n} \times \left( \vec{B} + \vec{B}_{e} \right) - k_{i}{\widehat{\mu}}_{n} \times \frac{d{\widehat{\mu}}_{n}}{dt}.
\end{equation}

The external field is

\begin{equation}
\label{eq:59}
\vec{B} = \left( \begin{array}{l}
0 \\
B_{y} \\
B_{z}
\end{array} \right),
\end{equation}

where \(B_{x}\) is neglected for brevity. The internal torque-averaged
fields from the nucleus and the electron applied on each other (Appendix~\ref{appendix:1}) are

\begin{equation}
\label{eq:60}
\vec{B}_{n} = B_{n}{\widehat{\mu}}_{n}
\end{equation}

and

\begin{equation}
\label{eq:61}
\vec{B}_{e} = B_{e}{\widehat{\mu}}_{e}.
\end{equation}

We have

\begin{equation}
\label{eq:62}
{\widehat{\mu}}_{e} = \left( \begin{array}{l}
\sin\theta_{e}\cos\phi_{e}\, \\
\sin\theta_{e}\sin\phi_{e}\, \\
\cos\theta_{e}
\end{array} \right)
\end{equation}

and

\begin{equation}
\label{eq:63}
{\widehat{\mu}}_{n} = \left( \begin{array}{l}
\sin\theta_{n}\cos\phi_{n}\, \\
\sin\theta_{n}\sin\phi_{n}\, \\
\cos\theta_{n}
\end{array} \right),
\end{equation}

where \(\theta\) and \(\phi\) denote the polar and azimuthal angles,
respectively.

Combining the above equations yields

\begin{equation}
\label{eq:64}
{\dot{\theta}}_{e} = {- \gamma}_{e}\left\lbrack B_{y}\cos\phi_{e} + B_{n}\sin\theta_{n}\sin\left( \phi_{n} - \phi_{e} \right) \right\rbrack + k_{i}{\dot{\phi}}_{e}\sin\theta_{e},
\end{equation}

\begin{equation}
\label{eq:65}
{\dot{\theta}}_{n} = - \gamma_{n}\left\lbrack B_{y}\cos\phi_{n} + B_{e}\sin\theta_{e}\sin\left( \phi_{e} - \phi_{n} \right) \right\rbrack + k_{i}{\dot{\phi}}_{n}\sin\theta_{n},
\end{equation}

\begin{equation}
\label{eq:66}
{\dot{\phi}}_{e} = - \gamma_{e}\left\{ B_{z} + B_{n}\cos\theta_{n} - \cot\theta_{e}\left\lbrack B_{y}\sin\phi_{e} + B_{n}\sin\theta_{n}\cos\left( \phi_{n} - \phi_{e} \right) \right\rbrack \right\} - \frac{k_{i}\,{\dot{\theta}}_{e}}{\sin\theta_{e}},
\end{equation}

and

\begin{equation}
\label{eq:67}
{\dot{\phi}}_{n} = - \gamma_{n}\left\{ B_{z} + B_{e}\cos\theta_{e} - \cot\theta_{n}\left\lbrack B_{y}\sin\phi_{n} + B_{e}\sin\theta_{e}\cos\left( \phi_{e} - \phi_{n} \right) \right\rbrack \right\} - \frac{k_{i}\,{\dot{\theta}}_{n}}{\sin\theta_{n}}.
\end{equation}

To implement the second CQD postulate, we revise the sign of the last
term in each of the above four equations, producing the CQD equations of
motion (Eqs.~\ref{eq:5}--\ref{eq:8}). 
Note that azimuthal angles are not defined when the
polar angles are \(0\) or \(\pi\). Therefore, when \(\theta_{e} = 0\) or
\(\pi\), we use \({\dot{\phi}}_{e} = 0\); when \(\theta_{n} = 0\) or
\(\pi\), we use \({\dot{\phi}}_{n} = 0\).

If \(B_{x} \neq 0\), the above equations can be extended by the
following substitutions:

\[
B_{y}\cos\phi_{e} \rightarrow - B_{x}\sin\phi_{e} + B_{y}\cos\phi_{e},
\]

\[
B_{y}\cos\phi_{n} \rightarrow - B_{x}\sin\phi_{n} + B_{y}\cos\phi_{n},
\]

\[
B_{y}\sin\phi_{e} \rightarrow B_{x}\cos\phi_{e} + B_{y}\sin\phi_{e},
\]

and

\[
B_{y}\sin\phi_{n} \rightarrow B_{x}\cos\phi_{n} + B_{y}\sin\phi_{n}.
\]

A complementary vector derivation of these spherical-coordinate equations is provided in Appendix~\ref{appendix:9}.

\FloatBarrier
\clearpage
\section[Appendix~\thesection. CQD derivation of density operator and wave function]{CQD derivation of density operator and wave function}\label{appendix:3}
\suppressfloats[t]

CQD reproduces the quantum mechanical density operator and wave function
with an isotropic angular distribution of co-quanta
(\({\widehat{\mu}}_{n}\)) and extends them with an anisotropic angular
distribution of co-quanta.

For a given \({\widehat{\mu}}_{e}\), the CQD prediction expressions (see
Methods) for two independent realizations are written in dual spaces
\cite{Wang2022}:

\begin{equation}
\label{eq:68}
\left|{\widehat{\mu}}_{e}\coq{\widehat{\mu}}_{ni} \right\rangle = C_{i +}\left( {\widehat{\mu}}_{e},{\widehat{\mu}}_{ni} \right)\left| + z \right\rangle + C_{i -}\left( {\widehat{\mu}}_{e},{\widehat{\mu}}_{ni} \right)\exp(i\phi_{e})\left| - z \right\rangle
\end{equation}

and

\begin{equation}
\label{eq:69}
\left\langle {\widehat{\mu}}_{e}\coq{\widehat{\mu}}_{nj}\right|\  = C_{j +}\left( {\widehat{\mu}}_{e},{\widehat{\mu}}_{nj} \right)\left\langle + z \right|\  + C_{j -}\left( {\widehat{\mu}}_{e},{\widehat{\mu}}_{nj} \right)\exp\left( - i\phi_{e} \right)\left\langle - z \right|
\end{equation}

Integer subscripts \(i\) and \(j\) denote independent realizations
(i.e., \(i \neq j\)). Each binary coefficient represents either one or
zero according to the branching condition (Eq.~\ref{eq:10}).

The pre-averaging density operator is defined as

\begin{equation}
\label{eq:70}
\rho_{0} \coloneqq \left|{\widehat{\mu}}_{e}\coq{\widehat{\mu}}_{ni} \right\rangle\left\langle {\widehat{\mu}}_{e}\coq{\widehat{\mu}}_{nj}\right|
\end{equation}

which serves as a bridge to quantum mechanics \cite{Wang2022}. Substitution of
Eqs.~\ref{eq:68} and \ref{eq:69} results in

\begin{equation}
\label{eq:71}
\rho_{0} = \left\lbrack C_{i +}\left| + z \right\rangle + C_{i -}\exp\left( i\phi_{e} \right)\left| - z \right\rangle \right\rbrack\ \left\lbrack C_{j +}\left\langle + z \right|\  + C_{j -}\exp\left( - i\phi_{e} \right)\left\langle - z \right|\  \right\rbrack.
\end{equation}

Expansion leads to

\(\rho_{0} =\)
\(C_{i +}C_{j +}\left| + z \right\rangle\left\langle + z \right|\  + C_{i -}C_{j -}\left| - z \right\rangle\left\langle - z \right|\ \)

\begin{equation}
\label{eq:72}
+ C_{i +}C_{j -}\exp\left( - i\phi_{e} \right)\left| + z \right\rangle\left\langle - z \right|\  + C_{i -}C_{j +}\exp\left( + i\phi_{e} \right)\left| - z \right\rangle\left\langle + z \right|.
\end{equation}

If the dual vectors represented identical realizations, the cross terms
would vanish because the binary coefficients are mutually exclusive for
each realization: \(C_{i \pm} \cdot C_{i \mp} = 0\) and
\(C_{j \pm} \cdot C_{j \mp} = 0\).

If \({\widehat{\mu}}_{n}\) is random for a given
\({\widehat{\mu}}_{e}\), ensemble averaging \(\rho_{0}\) over all
realizations of \({\widehat{\mu}}_{n}\), denoted by
\(\left\langle \right\rangle_{n}\), yields

\(\rho_{1} = \left\langle \rho_{0} \right\rangle_{n} = \left\langle C_{i +}C_{j +} \right\rangle_{n}\left| + z \right\rangle\left\langle + z \right|\  + \left\langle C_{i -}C_{j -} \right\rangle_{n}\left| - z \right\rangle\left\langle - z \right|\ \)

\begin{equation}
\label{eq:73}
+ \left\langle C_{i +}C_{j -}\exp\left( - i\phi_{e} \right) \right\rangle_{n}\left| + z \right\rangle\left\langle - z \right|\  + \left\langle C_{i -}C_{j +}\exp\left( + i\phi_{e} \right) \right\rangle_{n}\left| - z \right\rangle\left\langle + z \right|.
\end{equation}

The following equations are invoked next:
\(\left\langle C_{i +} \right\rangle_{n} = \left\langle C_{j +} \right\rangle_{n}\),
denoted by \(\left\langle C_{+} \right\rangle_{n}\);
\(\left\langle C_{i -} \right\rangle_{n} = \left\langle C_{j -} \right\rangle_{n}\),
denoted by \(\left\langle C_{-} \right\rangle_{n}\). Given the
independence of the two realizations (\(i \neq j\)), we have
\(\left\langle C_{i \pm}C_{j \pm} \right\rangle_{n} = \left\langle C_{i \pm} \right\rangle_{n}\left\langle C_{j \pm} \right\rangle_{n} = \left\langle C_{\pm} \right\rangle_{n}^{2}\)
and
\(\left\langle C_{i \pm}C_{j \mp} \right\rangle_{n} = \left\langle C_{i \pm} \right\rangle_{n}\left\langle C_{j \mp} \right\rangle_{n} = \left\langle C_{+} \right\rangle_{n}\left\langle C_{-} \right\rangle_{n}\),
yielding

\(\rho_{1} = \left\langle C_{+} \right\rangle_{n}^{2}\left| + z \right\rangle\left\langle + z \right|\  + \left\langle C_{-} \right\rangle_{n}^{2}\left| - z \right\rangle\left\langle - z \right|\ \)

\begin{equation}
\label{eq:74}
+ \left\langle C_{+} \right\rangle_{n}\left\langle C_{-} \right\rangle_{n}\exp\left( - i\phi_{e} \right)\left| + z \right\rangle\left\langle - z \right|\  + \left\langle C_{-} \right\rangle_{n}\left\langle C_{+} \right\rangle_{n}\exp\left( + i\phi_{e} \right)\left| - z \right\rangle\left\langle + z \right|.
\end{equation}

Factorization yields

\begin{equation}
\label{eq:75}
\rho_{1} = \left\lbrack \left\langle C_{+} \right\rangle_{n}\left| + z \right\rangle + \left\langle C_{-} \right\rangle_{n}\exp\left( i\phi_{e} \right)\left| - z \right\rangle \right\rbrack
\left\lbrack \left\langle C_{+} \right\rangle_{n}\left\langle + z \right|\  + \left\langle C_{-} \right\rangle_{n}\exp\left( - i\phi_{e} \right)\left\langle - z \right|\  \right\rbrack.
\end{equation}

For a pure state, invoking
\(\rho_{1} = \left|{\widehat{\mu}}_{e} \right\rangle\left\langle {\widehat{\mu}}_{e}\right|\ \)
retrieves the following \emph{ket} equation:

\begin{equation}
\label{eq:76}
\left|{\widehat{\mu}}_{e} \right\rangle = \left\langle C_{+} \right\rangle_{n}\left| + z \right\rangle + \left\langle C_{-} \right\rangle_{n}\exp\left( i\phi_{e} \right)\left| - z \right\rangle.
\end{equation}

If \({\widehat{\mu}}_{n}\) follows the isotropic \(p_{n0}\) (Eq.~\ref{eq:15}),
the expected probabilities of collapse are computed from Eq.~\ref{eq:74} as
follows:

\begin{equation}
\label{eq:77}
\left\langle + z \middle| \rho_{1} \middle| + z \right\rangle = \left\langle C_{+} \right\rangle_{n}^{2} = \int_{\theta_{e}}^{\pi}{p_{n0}2\pi\sin\theta_{n}d\theta_{n}} = \cos^{2}\frac{\theta_{e}}{2}
\end{equation}

and

\begin{equation}
\label{eq:78}
\left\langle - z \middle| \rho_{1} \middle| - z \right\rangle = \left\langle C_{-} \right\rangle_{n}^{2} = \int_{0}^{\theta_{e}}{p_{n0}2\pi\sin\theta_{n}d\theta_{n}} = \sin^{2}\frac{\theta_{e}}{2}.
\end{equation}

\begin{figure}[!htbp]
\centering
\includegraphics[width=1.47333in,height=1.60667in]{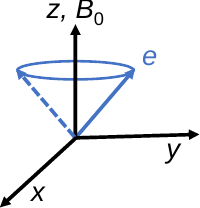}
\caption{Illustration of the cone of \({\widehat{\mu}}_{e}\)
formed by precession around the external main field, \(B_{0}\), over the
first Larmor cycle (i.e., before collapse). \(e\): electron magnetic
moment (principal quantum), \({\widehat{\mu}}_{e}\). Any nuclear
magnetic moment (co-quantum), \({\widehat{\mu}}_{n}\), precessing around
\(B_{0}\) within the cone causes \({\widehat{\mu}}_{e}\) to collapse
down, whereas \({\widehat{\mu}}_{n}\) precessing outside the cone causes
\({\widehat{\mu}}_{e}\) to collapse up. For a given polar angle
\(\theta_{e}\) of \({\widehat{\mu}}_{e}\), the probability for the atom
from the oven to reach the down branch in the first-stage Stern--Gerlach
experiment is proportional to the solid angle of the cone because
\({\widehat{\mu}}_{n}\) follows an isotropic angular distribution.
However, in subsequent-stage Stern--Gerlach experiments, the relation is
revised because \({\widehat{\mu}}_{n}\) follows an anisotropic angular
distribution, such as the heart shape (Fig.~\ref{fig:3}, Inset a, solid line; Eq.~\ref{eq:24}).}
\label{fig:S1}
\end{figure}

The integration limits are based on the branching condition (Eq.~\ref{eq:10}).
Because \(p_{n0}\) is isotropic (i.e., spherical), the two probabilities
are proportional to the solid angles formed by the down and up sides of
the cone shaped by the initial Bloch vector \textsuperscript{15}
(\({\widehat{\mu}}_{e}\)) precessing over one cycle (Fig.~\ref{fig:S1}). Each
solid angle determines the probability of having the co-quantum on the
corresponding side of the cone. In other words, the above two equations
represent the probabilities of having the co-quantum on the
corresponding side of the cone.

Consequently, we reach the familiar quantum mechanical density operator,

\(\rho_{1} = \cos^{2}\frac{\theta_{e}}{2}\left| + z \right\rangle\left\langle + z \right|\  + \sin^{2}\frac{\theta_{e}}{2}\left| - z \right\rangle\left\langle - z \right|\ \)

\begin{equation}
\label{eq:79}
+ \cos\frac{\theta_{e}}{2}\sin\frac{\theta_{e}}{2}\exp\left( - i\phi_{e} \right)\left| + z \right\rangle\left\langle - z \right|\  + \sin\frac{\theta_{e}}{2}\cos\frac{\theta_{e}}{2}\exp\left( + i\phi_{e} \right)\left| - z \right\rangle\left\langle + z \right|.
\end{equation}

Factorization of the density operator yields

\begin{equation}
\label{eq:80}
\rho_{1} = \left\lbrack \cos\frac{\theta_{e}}{2}\left| + z \right\rangle + \sin\frac{\theta_{e}}{2}\exp\left( i\phi_{e} \right)\left| - z \right\rangle \right\rbrack\left\lbrack \cos\frac{\theta_{e}}{2}\left\langle + z \right|\  + \sin\frac{\theta_{e}}{2}\exp\left( - i\phi_{e} \right)\left\langle - z \right|\  \right\rbrack.
\end{equation}

Invoking
\(\rho_{1} = \left|{\widehat{\mu}}_{e} \right\rangle\left\langle {\widehat{\mu}}_{e}\right|\ \)
for a pure state retrieves the following familiar quantum mechanical
\emph{ket} equation:

\begin{equation}
\label{eq:81}
\left|{\widehat{\mu}}_{e} \right\rangle = \cos\frac{\theta_{e}}{2}\left| + z \right\rangle + \sin\frac{\theta_{e}}{2}\exp\left( i\phi_{e} \right)\left| - z \right\rangle.
\end{equation}

Therefore, CQD statistically reproduces the quantum mechanical wave
function along with the probability amplitudes. As shown by Eq.~\ref{eq:76}, the
moduli of probability amplitudes originate from averaging the binary
coefficients in the CQD prediction expression.

If \({\widehat{\mu}}_{e}\) is also random, further ensemble averaging
\(\rho_{1}\) over all realizations, denoted by
\(\left\langle \right\rangle_{e}\), yields

\(\rho_{2} = \left\langle \rho_{1} \right\rangle_{e} = \left\langle \cos^{2}\frac{\theta_{e}}{2} \right\rangle_{e}\left| + z \right\rangle\left\langle + z \right|\  + \left\langle \sin^{2}\frac{\theta_{e}}{2} \right\rangle_{e}\left| - z \right\rangle\left\langle - z \right|\ \)

\begin{equation}
\label{eq:82}
+ \left\langle \cos\frac{\theta_{e}}{2}\sin\frac{\theta_{e}}{2}\exp\left( - i\phi_{e} \right) \right\rangle_{e}\left| + z \right\rangle\left\langle - z \right|\  + \left\langle \sin\frac{\theta_{e}}{2}\cos\frac{\theta_{e}}{2}\exp\left( + i\phi_{e} \right) \right\rangle_{e}\left| - z \right\rangle\left\langle + z \right|.
\end{equation}

If \({\widehat{\mu}}_{e}\) follows the isotropic \(p_{e0}\) (Eq.~\ref{eq:19}),
the probabilities of collapse are computed as follows:

\begin{equation}
\label{eq:83}
\left\langle + z \middle| \rho_{2} \middle| + z \right\rangle = \left\langle \cos^{2}\frac{\theta_{e}}{2} \right\rangle_{e} = \int_{0}^{\pi}{\left\lbrack \cos^{2}\frac{\theta_{e}}{2} \right\rbrack p_{e0}2\pi\sin\theta_{e}d\theta_{e}} = \frac{1}{2}
\end{equation}

and

\begin{equation}
\label{eq:84}
\left\langle - z \middle| \rho_{2} \middle| - z \right\rangle = \left\langle \sin^{2}\frac{\theta_{e}}{2} \right\rangle_{e} = \int_{0}^{\pi}{\left\lbrack \sin^{2}\frac{\theta_{e}}{2} \right\rbrack p_{e0}2\pi\sin\theta_{e}d\theta_{e}} = \frac{1}{2}.
\end{equation}

The cross terms vanish due to the azimuthal integration of
\(\exp\left( - i\phi_{e} \right)\) over a full cycle of
\({\widehat{\mu}}_{e}\):

\(\left\langle + z \middle| \rho_{2} \middle| - z \right\rangle = \left\langle \cos\frac{\theta_{e}}{2}\sin\frac{\theta_{e}}{2}\exp\left( - i\phi_{e} \right) \right\rangle_{e}\)

\begin{equation}
\label{eq:85}
= \int_{0}^{\pi}{\cos\frac{\theta_{e}}{2}\sin\frac{\theta_{e}}{2}\left\lbrack \int_{0}^{2\pi}{\exp\left( - i\phi_{e} \right)p_{e0}d\phi_{e}} \right\rbrack\sin\theta_{e}d\theta_{e}} = 0
\end{equation}

and

\(\left\langle - z \middle| \rho_{2} \middle| + z \right\rangle = \left\langle \sin\frac{\theta_{e}}{2}\cos\frac{\theta_{e}}{2}\exp\left( + i\phi_{e} \right) \right\rangle_{e}\)

\begin{equation}
\label{eq:86}
= \int_{0}^{\pi}{\sin\frac{\theta_{e}}{2}\cos\frac{\theta_{e}}{2}\left\lbrack \int_{0}^{2\pi}{\exp\left( + i\phi_{e} \right)p_{e0}d\phi_{e}} \right\rbrack\sin\theta_{e}d\theta_{e}} = 0.
\end{equation}

Therefore, we reach

\begin{equation}
\label{eq:87}
\rho_{2} = \frac{1}{2}\left| + z \right\rangle\langle + z| + \frac{1}{2}\left| - z \right\rangle\langle - z|.
\end{equation}

This familiar quantum mechanical density operator represents the mixed
state and cannot be factorized into a product of two pure-state wave
functions.

If \({\widehat{\mu}}_{n}\) follows the heart-shaped \(p_{n1}\) (Eq.~\ref{eq:24}),
the probabilities of collapse become

\begin{equation}
\label{eq:88}
\left\langle + z \middle| \rho_{1} \middle| + z \right\rangle = \left\langle C_{+} \right\rangle_{n}^{2} = \int_{\theta_{e}}^{\pi}{p_{n1}2\pi\sin\theta_{n}d\theta_{n}} = 1 - \sin^{4}\frac{\theta_{e}}{2}
\end{equation}

instead of Eq.~\ref{eq:77} and

\begin{equation}
\label{eq:89}
\left\langle - z \middle| \rho_{1} \middle| - z \right\rangle = \left\langle C_{-} \right\rangle_{n}^{2} = \int_{0}^{\theta_{e}}{p_{n1}2\pi\sin\theta_{n}d\theta_{n}} = \sin^{4}\frac{\theta_{e}}{2}
\end{equation}

instead of Eq.~\ref{eq:78}. Because \(p_{n1}\) is anisotropic (i.e.,
heart-shaped), the two probabilities are no longer simply proportional
to the solid angles. However, the above two equations still represent
the probabilities of having the co-quantum on the corresponding side of
the cone (Fig.~\ref{fig:S1}).

Accordingly, the wave function becomes

\begin{equation}
\label{eq:90}
\left|{\widehat{\mu}}_{e} \right\rangle = \sqrt{1 - \sin^{4}\frac{\theta_{e}}{2}}\left| + z \right\rangle + \sin^{2}\frac{\theta_{e}}{2}\exp\left( i\phi_{e} \right)\left| - z \right\rangle
\end{equation}

instead of Eq.~\ref{eq:81}.

\FloatBarrier
\clearpage
\section[Appendix~\thesection. Derivation of Majorana formula]{Derivation of Majorana formula}\label{appendix:4}
\suppressfloats[t]

The Majorana formula \cite{Majorana1932} or its Landau--Zener variant \cite{Landau1932,Zener1932,Stueckelberg1932}
was derived most intuitively in 2005 by Wittig \cite{Wittig2005}. Recent
rederivations of the Majorana formula can be found in Wilczek \cite{Wilczek2014}
and Kofman \cite{Kofman2022}. Majorana stated that both the classical and the
quantum-mechanical treatments require integration of the same
differential equations \cite{Majorana1932,Majorana2006}. For completeness here, we follow
Majorana's variable transformations and then abridge Wittig's solution
but with a slightly altered contour integration.

For the inner rotation chamber (Fig.~\ref{fig:3}, IR), the B field along the \(y\)
axis is approximated using a magnetic quadrupole (Fig.~\ref{fig:S2}) \cite{Majorana1932,Majorana2006}:

\begin{figure}[!htbp]
\centering
\includegraphics[width=3.385in,height=3.385in]{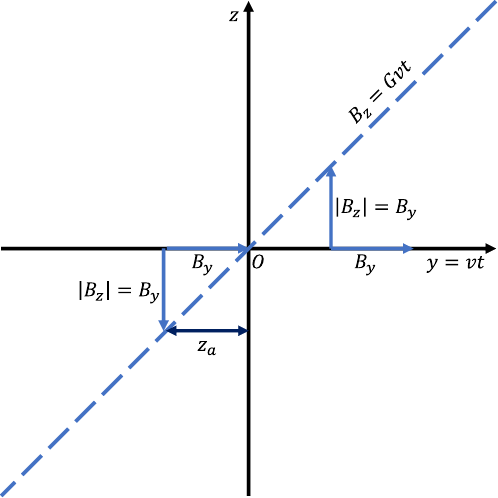}
\caption{Illustration of the magnetic field versus the \(y\)
location of the atom in the inner rotation chamber (i.e., the middle
stage). \(y\) denotes the axis of the atomic beam. Here, \(B_{y} > 0\),
and \(G = \frac{\partial B_{z}}{\partial y} > 0\). For a given current,
\(B_{y}\) is invariant with \(y\) while \(B_{z}\) is proportional to
\(y\) (i.e., negative for \(y < 0\), zero at \(y = 0\), and positive for
\(y > 0\)). At \(y = \pm z_{a}\), we have \({B_{z} = \pm B}_{y}\) or
\({\left| B_{z} \right| = B}_{y}\).}
\label{fig:S2}
\end{figure}

\begin{equation}
\label{eq:91}
B_{x} = 0,
\end{equation}

\begin{equation}
\label{eq:92}
B_{y} = Gz_{a},
\end{equation}

and

\begin{equation}
\label{eq:93}
B_{z} = Gvt.
\end{equation}

Here, \(G\) is the derivative of \(B_{z}\) with respect to \(y\) (i.e.,
the gradient magnitude of \(B_{z}\),
\(\frac{\partial B_{z}}{\partial y}\)), \(z_{a}\)
(\(1.05 \times 10^{-4}\) m) is the vertical distance of the atomic beam
from the center of the wire,\(\ v\) (800 m s\textsuperscript{$-1$}) is
the most likely speed of atoms, and \(t\) is time set to zero at the
null point of \(B_{z}\). \(G\) is given by

\begin{equation}
\label{eq:94}
G = \frac{\partial B_{z}}{\partial y} = \frac{2\pi}{\mu_{0}I}B_{r}^{2}.
\end{equation}

Here, \(I\) denotes the current carried by the wire along the \(- x\)
axis, and \(B_{r}\) (\(0.42 \times 10^{-4}\) T) denotes the uniformly
distributed remnant (residual) fringe magnetic flux density, which is
parallel with the \(+ z\) axis. The magnetic field generated by the wire
cancels the remnant field at the null point (NP) to produce an
approximate quadrupole (Fig.~\ref{fig:3}, Inset b). For a given current, \(B_{y}\)
is constant while \(B_{z}\) varies linearly with distance from the point
where \(B_{z} = 0\).

Majorana neglected the nuclear magnetic moment and induction. The Bloch
equation leads to

\begin{equation}
\label{eq:95}
{\dot{\theta}}_{e} = {- \gamma}_{e}B_{y}\cos\phi_{e}
\end{equation}

and

\begin{equation}
\label{eq:96}
{\dot{\phi}}_{e} = - \gamma_{e}\left\lbrack B_{z} - \cot\theta_{e}B_{y}\sin\phi_{e} \right\rbrack,
\end{equation}

which agree with Eqs.~\ref{eq:5} and \ref{eq:7} for \(B_{n} = 0\) and \(k_{i} = 0\).

Majorana transformed the polar and azimuthal angles into probability
amplitudes then solved the transformed equations \cite{Majorana1932,Majorana2006}. We let

\begin{equation}
\label{eq:97}
\left| \widehat{\mu} \right\rangle = \begin{pmatrix}
c_{1} \\
c_{2}
\end{pmatrix}.
\end{equation}

The Schrödinger equation becomes

\begin{equation}
\label{eq:98}
i\hslash\frac{d}{dt}\begin{pmatrix}
c_{1} \\
c_{2}
\end{pmatrix} = H\begin{pmatrix}
c_{1} \\
c_{2}
\end{pmatrix}.
\end{equation}

The Hamiltonian is

\begin{equation}
\label{eq:99}
H = - \frac{1}{2}\hslash\gamma_{e}\vec{B} \cdot \vec{\sigma}.
\end{equation}

Substituting the Pauli matrices \(\vec{\sigma}\) yields

\begin{equation}
\label{eq:100}
H = - \frac{1}{2}\hslash\gamma_{e}\left\lbrack B_{x}\begin{pmatrix}
0 & 1 \\
1 & 0
\end{pmatrix} + B_{y}\begin{pmatrix}
0 & - i \\
i & 0
\end{pmatrix} + B_{z}\begin{pmatrix}
1 & 0 \\
0 & - 1
\end{pmatrix} \right\rbrack.
\end{equation}

Merging terms gives

\begin{equation}
\label{eq:101}
H = - \frac{1}{2}\hslash\gamma_{e}\begin{pmatrix}
B_{z} & B_{x} - iB_{y} \\
B_{x} + iB_{y} & - B_{z}
\end{pmatrix}.
\end{equation}

Majorana defined the following dimensionless variables for time and
adiabaticity, respectively \cite{Majorana1932,Majorana2006}:

\begin{equation}
\label{eq:102}
\tau = \frac{1}{2}\sqrt{\left| \gamma_{e}Gv \right|} \cdot t
\end{equation}

and

\begin{equation}
\label{eq:103}
k_{m} = \frac{\left| \gamma_{e} \right|B_{y}}{\frac{Gv}{B_{y}}} = \frac{\left| \gamma_{e} \right|B_{y}^{2}}{Gv} = \frac{z_{a}}{v}\left| \gamma_{e} \right|B_{y}.
\end{equation}

The numerator, \(\left| \gamma_{e} \right|B_{y}\), in the first fraction
above represents the Larmor frequency about the \(y\) axis, whereas the
denominator, \(\frac{Gv}{B_{y}}\), represents approximately the rotation
frequency of the field.

Accordingly, the Schrödinger equation is simplified to

\begin{equation}
\label{eq:104}
\frac{d}{d\tau}\begin{pmatrix}
c_{1} \\
c_{2}
\end{pmatrix} = - i\begin{pmatrix}
2\tau c_{1} - i\sqrt{k_{m}}c_{2} \\
i\sqrt{k_{m}}c_{1} - 2\tau c_{2}
\end{pmatrix}.
\end{equation}

Majorana defined the following transformation of variables \cite{Majorana1932,Majorana2006},
which is analogous to heterodyne detection with a chirped local
oscillator to remove high-frequency signals:

\begin{equation}
\label{eq:105}
\begin{pmatrix}
c_{1} \\
c_{2}
\end{pmatrix} = \begin{pmatrix}
\exp\left( - i\tau^{2} \right)f \\
\exp\left( + i\tau^{2} \right)g
\end{pmatrix}.
\end{equation}

The Schrödinger equation becomes

\begin{equation}
\label{eq:106}
\frac{d}{d\tau}\begin{pmatrix}
f \\
g
\end{pmatrix} = \sqrt{k_{m}}\begin{pmatrix}
 - \exp\left( + 2i\tau^{2} \right)g \\
 + \exp\left( - 2i\tau^{2} \right)f
\end{pmatrix}.
\end{equation}

Eliminating variables yields

\begin{equation}
\label{eq:107}
\left( \frac{d^{2}}{d\tau^{2}} \mp 4i\tau\frac{d}{d\tau} + k_{m} \right)\begin{pmatrix}
f \\
g
\end{pmatrix} = 0.
\end{equation}

For \(\left| f( - \infty) \right| = 1\), we solve for
\(\left| f( + \infty) \right|\). Following Wittig \cite{Wittig2005}, we rewrite
the equation for \(f\) as

\begin{equation}
\label{eq:108}
4i\frac{df}{f} = \frac{d\tau}{\tau}\left( \frac{d^{2}f}{d\tau^{2}}\frac{1}{f} + k_{m} \right).
\end{equation}

Integrating over the entire flight yields

\begin{equation}
\label{eq:109}
4i\int_{- \infty}^{+ \infty}\frac{df}{f} = \int_{- \infty}^{+ \infty}{\frac{d\tau}{\tau}\left( \frac{d^{2}f}{d\tau^{2}}\frac{1}{f} + k_{m} \right)}.
\end{equation}

\begin{figure}[!htbp]
\centering
\includegraphics[width=4.335in,height=2.45333in]{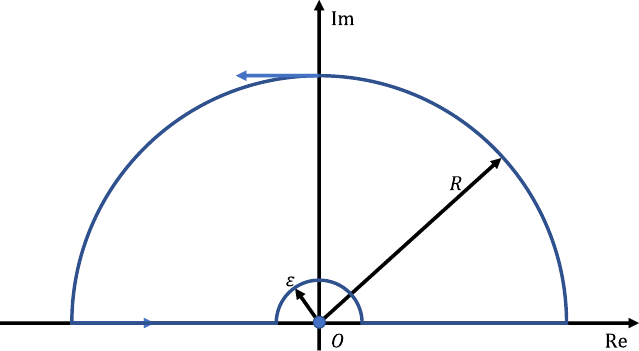}
\caption{Contour on the complex plane of \(\tau\) for the
integration. The solid circle indicates the pole at \(\tau = 0\).
Adapted from Wittig \cite{Wittig2005}.}
\label{fig:S3}
\end{figure}

We select a positively oriented and indented contour that excludes the
singularity at \(\tau = 0\) in the complex plane (Fig.~\ref{fig:S3}), whereas the
opposite orientation would yield an unphysical outcome. Consequently, we
have

\begin{equation}
\label{eq:110}
4i\ln\frac{f( + \infty)}{f( - \infty)} = \oint_{}^{}{} - \lim_{\varepsilon \rightarrow 0}\int_{\mathrm{arc}}^{} - \lim_{R \rightarrow \infty}{\int_{\mathrm{Arc}}^{}{\frac{d\tau}{\tau}\left( \frac{d^{2}f}{d\tau^{2}}\frac{1}{f} + k_{m} \right)}}.
\end{equation}

Here, \(\varepsilon\) is the radius of the small indenting semicircular
arc, \(R\) is the radius of the large semicircular arc, and \(\tau\) is
now made complex without substitution to a new complex variable (because
the typically adopted \(z\) is used already for space). Because no pole
is inside the contour, the first integral on the right side vanishes.

At \(\tau = 0\), Eq.~\ref{eq:107} gives the residue,

\begin{equation}
\label{eq:111}
\mathrm{Res}_{0} = \left. \ \left( \frac{d^{2}f}{d\tau^{2}}\frac{1}{f} + k_{m} \right) \right|_{\tau = 0}\  = 0.
\end{equation}

Thus, the second integral along the small arc on the right side of Eq.~\ref{eq:110} vanishes too.

As \(\tau \rightarrow \infty\),
\(\frac{d^{2}f}{d\tau^{2}}\frac{1}{f} \rightarrow 0\) \cite{Wittig2005};
substitution into the third integral along the large arc yields

\begin{equation}
\label{eq:112}
4i\ln\frac{f( + \infty)}{f( - \infty)} = - k_{m}\lim_{R \rightarrow \infty}{\int_{\mathrm{Arc}}^{}\frac{d\tau}{\tau}}.
\end{equation}

Letting \(\tau = R\exp(i\beta)\) leads to

\begin{equation}
\label{eq:113}
\ln\frac{f( + \infty)}{f( - \infty)} = - \frac{k_{m}}{4i}\lim_{R \rightarrow \infty}{\int_{0}^{\pi}\frac{R\exp(i\beta)id\beta}{R\exp(i\beta)}} = - \frac{\pi k_{m}}{4}.
\end{equation}

Therefore,

\begin{equation}
\label{eq:114}
\frac{\left| f( + \infty) \right|}{\left| f( - \infty) \right|} = \exp\left( - \frac{\pi k_{m}}{4} \right).
\end{equation}

Substituting \(\left| f( - \infty) \right| = 1\), we reach

\begin{equation}
\label{eq:115}
\left| f( + \infty) \right| = \exp\left( - \frac{\pi k_{m}}{4} \right).
\end{equation}

Majorana \cite{Majorana1932,Majorana2006} reasoned that because \(B_{z}\) reverses its
orientation along the flight path, the probability of spin flip,
\(W_{m}\), is given by \(\left| f( + \infty) \right|^{2}\) instead of
\(\left| g( + \infty) \right|^{2}\). The further justification that we
found is the initial adiabatic flip induced when the atom passes above
the wire \cite{Titimbo2022}. Therefore, we obtain

\begin{equation}
\label{eq:116}
W_{m} = \exp\left( - \frac{\pi k_{m}}{2} \right).
\end{equation}

Here, \(k_{m} > 0\). If \(k_{m} < 0\), one may extend the solution to
\(\exp\left( - \frac{\pi\left| k_{m} \right|}{2} \right)\). Substitution
of Eq.~\ref{eq:103} results in

\begin{equation}
\label{eq:117}
W_{m} = \exp\left( - \frac{\pi z_{a}}{2v}\left| \gamma_{e} \right|B_{y} \right).
\end{equation}

\FloatBarrier
\clearpage
\section[Appendix~\thesection. Derivation of CQD formula]{Derivation of CQD formula}\label{appendix:5}
\suppressfloats[t]

We derive the CQD formula for the probability of spin flip in the inner
rotation chamber (Fig.~\ref{fig:3}, IR) in the presence of both the quadrupole
field and the nuclear magnetic moment.

The average polar angle of \(\vec{B}_{n}\),
\(\left\langle \theta_{n} \right\rangle\), is derived from the heart
shape given by Eq.~\ref{eq:24}:

\begin{equation}
\label{eq:118}
\left\langle \theta_{n} \right\rangle = \int_{0}^{\pi}{\theta_{n}p_{n1}2\pi\sin\theta_{n}d\theta_{n}} = \frac{5\pi}{8}.
\end{equation}

To reach an approximate analytical solution, we hold \(\theta_{n}\) at
\(\left\langle \theta_{n} \right\rangle\) as a representative value
throughout the inner rotation chamber.

The presence of \(\vec{B}_{n}\) alters both the remnant
field due to the projection of \(\vec{B}_{n}\) to the \(z\)
axis, given by \(B_{n}\cos\left\langle \theta_{n} \right\rangle\), and
the transverse field due to the transverse projection of
\(\vec{B}_{n}\), represented by
\(B_{n}\sin\left\langle \theta_{n} \right\rangle\exp\left( i\phi_{n} \right)\).
To extend the Majorana solution presented in Appendix~\ref{appendix:4}, we first
substitute the remnant field as follows:

\begin{equation}
\label{eq:119}
B_{r} \rightarrow B_{r}^{'} = B_{r} + B_{n}\cos\left\langle \theta_{n} \right\rangle.
\end{equation}

Accordingly, we update the field gradient:

\begin{equation}
\label{eq:120}
G = \frac{2\pi}{\mu_{0}I}{B_{r}}^{2} \rightarrow G^{'} = \frac{2\pi}{\mu_{0}I}{B_{r}^{'}}^{2}.
\end{equation}

The quadrupole field along the \(y\) axis is still given by the same
equations but with the corrected field gradient:\cite{Majorana1932,Majorana2006}

\begin{equation}
\label{eq:121}
B_{x}^{'} = 0,
\end{equation}

\begin{equation}
\label{eq:122}
B_{y}^{'} = G^{'}z_{a},
\end{equation}

and

\begin{equation}
\label{eq:123}
B_{z}^{'} = G^{'}vt.
\end{equation}

Following Majorana \cite{Majorana1932,Majorana2006}, we define the dimensionless time as

\begin{equation}
\label{eq:124}
\tau = \frac{1}{2}\sqrt{\left| \gamma_{e}G^{'}v \right|} \cdot t
\end{equation}

and the dimensionless adiabaticity parameter due to \(B_{y}^{'}\) as

\begin{equation}
\label{eq:125}
k_{0} = \frac{\left| \gamma_{e} \right|B_{y}^{'}}{\frac{G^{'}v}{B_{y}^{'}}} = \frac{z_{a}}{v}\left| \gamma_{e} \right|B_{y}^{'}.
\end{equation}

The numerator, \(\left| \gamma_{e} \right|B_{y}^{'}\), in the first
fraction above represents the Larmor frequency about the \(y\) axis,
whereas the denominator, \(\frac{G^{'}v}{B_{y}^{'}}\), represents
approximately the rotation frequency of the field. We similarly define
the adiabaticity parameter due to the transverse field
\(B_{n}\sin\left\langle \theta_{n} \right\rangle\) as

\begin{equation}
\label{eq:126}
k_{1} = \frac{\left| \gamma_{e} \right|\left( B_{n}\sin\left\langle \theta_{n} \right\rangle \right)}{\frac{G^{'}v}{\left( B_{n}\sin\left\langle \theta_{n} \right\rangle \right)}} = \frac{z_{a}}{v}\left| \gamma_{e} \right|\frac{\left( B_{n}\sin\left\langle \theta_{n} \right\rangle \right)^{2}}{B_{y}^{'}}.
\end{equation}

We also compute the dimensionless counterpart of the Larmor frequency
\(\omega_{n}\) as

\begin{equation}
\label{eq:127}
w_{n} = \frac{d\phi_{n}}{d\tau} = \frac{d\phi_{n}}{dt}\frac{dt}{d\tau} = \frac{2}{\sqrt{\left| \gamma_{e}G^{'}v \right|}}\omega_{n},
\end{equation}

where \(\omega_{n} = \frac{d\phi_{n}}{dt}\) was invoked.

Then, we substitute the total transverse field for the Schrödinger
equation:

\begin{equation}
\label{eq:128}
B_{x} \rightarrow B_{x}^{'} + B_{n}\sin\left\langle \theta_{n} \right\rangle\cos\phi_{n}
\end{equation}

and

\begin{equation}
\label{eq:129}
B_{y} \rightarrow B_{y}^{'} + B_{n}\sin\left\langle \theta_{n} \right\rangle\sin\phi_{n}.
\end{equation}

Following the procedure presented in Appendix~\ref{appendix:4}, we revise the
Schrödinger equation to

\begin{equation}
\label{eq:130}
\frac{d}{d\tau}\begin{pmatrix}
c_{1} \\
c_{2}
\end{pmatrix} = - i\begin{pmatrix}
2\tau & \sqrt{k_{1}}\exp\left( - i\phi_{n} \right) - i\sqrt{k_{0}} \\
\sqrt{k_{1}}\exp\left( i\phi_{n} \right) + i\sqrt{k_{0}} & - 2\tau
\end{pmatrix}\begin{pmatrix}
c_{1} \\
c_{2}
\end{pmatrix}.
\end{equation}

Defining

\begin{equation}
\label{eq:131}
\begin{pmatrix}
c_{1} \\
c_{2}
\end{pmatrix} = \begin{pmatrix}
\exp\left( - i\tau^{2} \right)f(\tau) \\
\exp\left( + i\tau^{2} \right)g(\tau)
\end{pmatrix},
\end{equation}

we reach

\begin{equation}
\label{eq:132}
\frac{d}{d\tau}\begin{pmatrix}
f(\tau) \\
g(\tau)
\end{pmatrix} = \begin{pmatrix}
\left\lbrack - i\sqrt{k_{1}}\exp\left( - i\phi_{n} \right) - \sqrt{k_{0}} \right\rbrack\exp\left( + 2i\tau^{2} \right)g(\tau) \\
\left\lbrack - i\sqrt{k_{1}}\exp{i\phi_{n}} + \sqrt{k_{0}} \right\rbrack\exp\left( - 2i\tau^{2} \right)f(\tau)
\end{pmatrix}.
\end{equation}

Eliminating \(g\) yields

\begin{equation}
\label{eq:133}
4i\frac{df}{f} = \frac{d\tau}{\tau - \frac{\sqrt{k_{1}}w_{n}}{4\left\lbrack \sqrt{k_{1}} - i\sqrt{k_{0}}\exp\left( i\phi_{n} \right) \right\rbrack}}\left\{ \frac{d^{2}f}{d\tau^{2}}\frac{1}{f} + \left\lbrack k_{0} + k_{1} + 2\sqrt{k_{0}k_{1}}\sin\left( \phi_{n} \right) \right\rbrack \right\}.
\end{equation}

In the limiting case that \(k_{0} \gg k_{1}\), i.e., the extremely
low-current region, we set \(k_{1} = 0\), reducing Eq.~\ref{eq:133} to

\begin{equation}
\label{eq:134}
4i\frac{df}{f} = \frac{d\tau}{\tau}\left\{ \frac{d^{2}f}{d\tau^{2}}\frac{1}{f} + k_{0} \right\},
\end{equation}

which is the same as Eq.~\ref{eq:108}. The solution with
\(\left| f( - \infty) \right| = 1\) is given by Eq.~\ref{eq:115}:

\begin{equation}
\label{eq:135}
\left| f( + \infty) \right| = \exp\left( - \frac{\pi k_{0}}{4} \right).
\end{equation}

\begin{figure}[!htbp]
\centering
\includegraphics[width=4.83833in,height=4.05in]{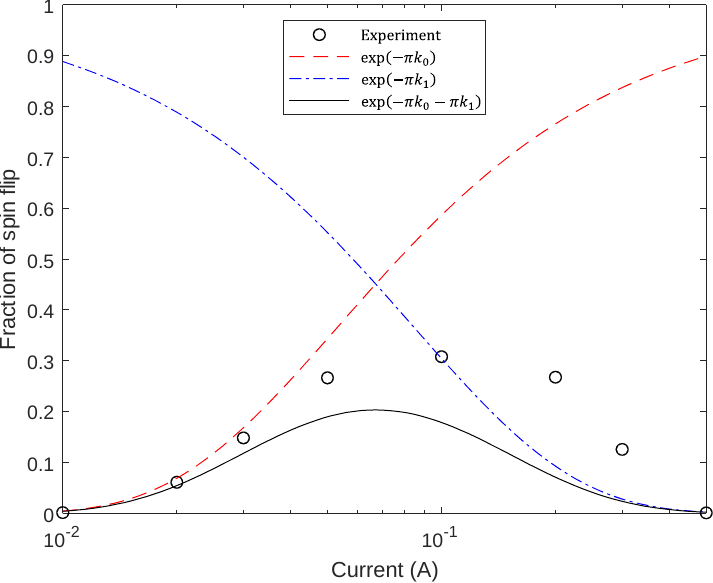}
\caption{Fraction of spin flip versus wire current. As the
current increases from 0.01 A to 0.5 A, \(k_{0}\) decreases inversely
proportionally with the current from 1.701 to 0.034, and \(k_{1}\)
increases proportionally with the current from 0.038 to 1.891. When
\(B_{y}^{'} = B_{n}\sin\left\langle \theta_{n} \right\rangle = 0.11 \times 10^{-4}\)
T, \(k_{0} = k_{1}\); the corresponding current equals \(0.067\) A,
about which the dashed and dash-dotted curves are mirror symmetric on
the semilog plot. Experiment: Frisch--Segrè experiment.}
\label{fig:S4}
\end{figure}

From Eq.~\ref{eq:25} in Results, the fraction of flip is given by
\(\left| f( + \infty) \right|^{4}\) instead of
\(\left| f( + \infty) \right|^{2}\) due to the heart-shaped \(p_{n1}\):

\begin{equation}
\label{eq:136}
W_{2} = \exp\left( - \pi k_{0} \right) = \exp\left( - \pi\frac{z_{a}}{v}\left| \gamma_{e} \right|B_{y}^{'} \right).
\end{equation}

\(W_{2}\) is shown in Fig.~\ref{fig:S4} below and Fig.~\ref{fig:4}. Because the pole in Eq.~\ref{eq:134} is at \(\tau = 0\), we call this effect null-point rotation.

Conversely, if we set \(k_{0} = 0\), Eq.~\ref{eq:133} reduces to

\begin{equation}
\label{eq:137}
4i\frac{df}{f} = \frac{d\tau}{\tau - \frac{w_{n}}{4}}\left\{ \frac{d^{2}f}{d\tau^{2}}\frac{1}{f} + k_{1} \right\},
\end{equation}

which resembles Eq.~\ref{eq:108}, however, with the pole shifted from 0 to
\(\frac{w_{n}}{4}\). The solution is likewise obtained as

\begin{equation}
\label{eq:138}
\left| f( + \infty) \right| = \exp\left( - \frac{\pi k_{1}}{4} \right).
\end{equation}

Similarly, the fraction of flip is

\begin{equation}
\label{eq:139}
W_{R} = \exp\left( - \pi k_{1} \right),
\end{equation}

which is shown in Fig.~\ref{fig:S4}. The pole \(\tau = \frac{w_{n}}{4}\) is
converted using Eq.~\ref{eq:124} and Eq.~\ref{eq:127} to dimensional quantities as
\(\omega_{ez}(t) = \omega_{n}(t)\), where \(\omega_{ez}\) denotes the
Larmor frequency of \(\vec{\mu}_{e}\) about the \(z\) axis
and \(\omega_{n}\) denotes the Larmor frequency of
\(\vec{\mu}_{n}\). Therefore, the flip is due to precession
resonance between the magnetic moments of the nucleus and the electron,
which we refer to as nuclear-resonant rotation.

There are three terms inside the square brackets in Eq.~\ref{eq:133}:
\(k_{0} + k_{1} + 2\sqrt{k_{0}k_{1}}\sin\left( \phi_{n} \right)\). A
direct combination of the first two terms yields the solid-line curve in
Fig.~\ref{fig:S4}, which predicts the fraction of flip accurately at the two ends
but only qualitatively in the intermediate region. Below, we combine the
terms more quantitatively. For an ensemble of atoms, the nuclear
magnetic moment of each atom is given a random initial phase,
\(\phi_{n0}\), at \(t = 0\).

Given Eq.~\ref{eq:25} in Results, we extend the Majorana solution (see Appendix~\ref{appendix:4}) to

\begin{equation}
\label{eq:140}
W_{\mathrm{cqd}} = \sin^{4}\left( \frac{\alpha_{r}}{2} \right) = \exp\left( - E_{r0} - E_{r1} - E_{i} \right),
\end{equation}

where \(\alpha_{r}\) represents the polar rotation by the inner rotation
chamber from the initial \(\theta_{e} = 0\) and \(E_{r0}\), \(E_{r1}\),
\(E_{i}\) represent contributions from null-point rotation,
nuclear-resonant rotation, and induction, respectively. The fourth power
is due to the heart-shaped \(p_{n1}\).

The null-point rotation exponent is determined by the quadrature-summed
(or the root-mean-squared) \(B\) field on the \(xy\) plane:

\begin{equation}
\label{eq:141}
E_{r0} = \frac{\pi z_{a}}{v}\left| \gamma_{e} \right|\sqrt{B_{y}^{'2} + \left( B_{n}\sin\left\langle \theta_{n} \right\rangle \right)^{2}}.
\end{equation}

\begin{figure}[!htbp]
\centering
\includegraphics[width=5.57333in,height=3.45in]{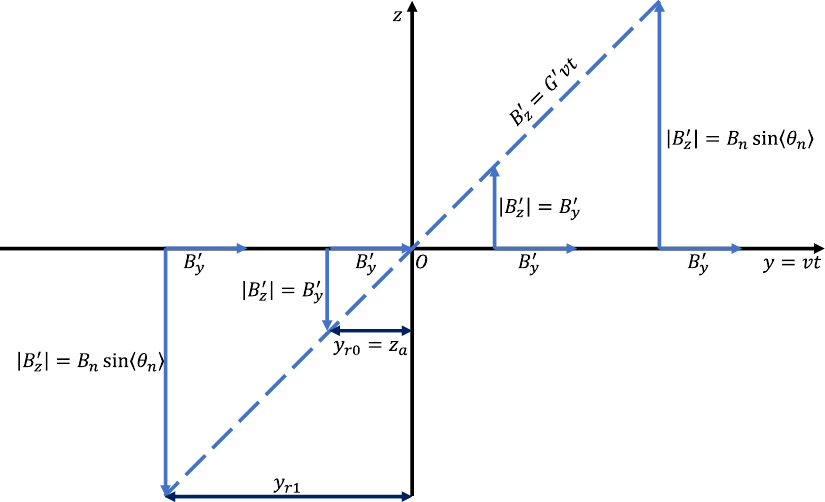}
\caption{Illustration of the magnetic field versus the \(y\)
location of the atom in the inner rotation chamber (i.e., the middle
stage). Here, \(\frac{y_{r1}}{y_{r0}} = 3\), for the current of 0.2 A.
For a given current, \(B_{y}^{'}\) is invariant with \(y\) while
\(B_{z}^{'}\) is proportional to \(y\) (i.e., negative for \(y < 0\),
zero at \(y = 0\), and positive for \(y > 0\)).
\(B_{n}\sin\left\langle \theta_{n} \right\rangle\) is the transverse
projection of \(\vec{B}_{n}\) on the \(xy\) plane.
\(G^{'} = \frac{\partial B_{z}^{'}}{\partial y}\).}
\label{fig:S5}
\end{figure}

The quadrature sum can be derived through
\(\left\langle \left| iB_{y}^{'} + B_{n}\sin\left\langle \theta_{n} \right\rangle\exp\left( i\phi_{n0} \right) \right|^{2} \right\rangle^{\frac{1}{2}}\),
where the ensemble average is over a uniform distribution of
\(\phi_{n0}\) and
\(B_{n}\sin\left\langle \theta_{n} \right\rangle\exp\left( i\phi_{n0} \right)\)
is the transverse component of \(\vec{B}_{n}\) at the null
point. One may consider \(\pi\frac{B_{y}^{'}}{G^{'}}\), yielding
\(\pi z_{a}\) (Eq.~\ref{eq:122}), as the effective flight path-length for the
null-point rotation, \(\pi y_{r0}\) (Fig.~\ref{fig:S5}). Here,
\(\pi y_{r0} = \pi z_{a} = \pi \times 0.105 \times 10^{-3} = 0.33 \times 10^{-3}\)
m, which is a constant independent of the wire current. If
\(B_{n} = 0\), the quadrature sum reduces to \(B_{y}^{'}\) as expected.
\(E_{r0}\) is responsible for Curve 3 in Fig.~\ref{fig:4}, which predicts the
experimental observation in the low-current region accurately. Further
description can be found above Eq.~\ref{eq:28}.

At high currents where \(I \geq 0.067\) A, \(B_{y}^{'}\) becomes less
than \(B_{n}\sin\left\langle \theta_{n} \right\rangle\), and
\(k_{1} \geq k_{0}\); consequently, nuclear-resonant rotation due to the
rotating transverse component of \(\vec{B}_{n}\) becomes
substantial. The resonant-rotation exponent is approximated
heuristically as

\begin{equation}
\label{eq:142}
E_{r1} = \frac{1}{2}\left\lbrack \frac{\pi y_{r1}}{v}\left| \gamma_{e} \right|B_{n}\sin\left\langle \theta_{n} \right\rangle\  \right\rbrack^{2}\left\{ \frac{\pi y_{r1}}{v}\frac{1}{T_{n}} \right\}.
\end{equation}

The term in the square bracket is analogous to the right-hand side of
Eq.~\ref{eq:141}. This estimation is inspired by the following approximation for
small fluctuations of a random variable \(X\):

\begin{equation}
\label{eq:143}
\left\langle \exp\left( X - \overline{X} \right) \right\rangle \approx \exp\left( \frac{1}{2}\left\langle \left( X - \overline{X} \right)^{2} \right\rangle \right) = \exp\left( \frac{1}{2}\mathrm{Var}\lbrack X\rbrack\  \right),
\end{equation}

where \(\mathrm{Var}\) denotes variance.

In analogy to \(\pi y_{r0} = \pi\frac{B_{y}^{'}}{G^{'}}\) defined below
Eq.~\ref{eq:141}, the effective flight path-length for nuclear-resonant rotation,
\(\pi y_{r1}\), is defined as (Fig.~\ref{fig:S5})

\begin{equation}
\label{eq:144}
\pi y_{r1} = \frac{\pi B_{n}\sin\left\langle \theta_{n} \right\rangle}{G^{'}} = \frac{\pi z_{a}B_{n}\sin\left\langle \theta_{n} \right\rangle}{B_{y}^{'}}.
\end{equation}

Because \(B_{e}\) is far greater than the field in the inner rotation
chamber, the Larmor period of the nuclear magnetic moment is
approximated to be

\begin{equation}
\label{eq:145}
T_{n} = \frac{2\pi}{\gamma_{n}B_{e}}.
\end{equation}

The curly bracket term in Eq.~\ref{eq:142} represents the fraction of the Larmor
period of the nuclear magnetic moment, denoted by \(f_{r1}\), precessed
during the flight time over \(\pi y_{r1}\):

\begin{equation}
\label{eq:146}
f_{r1} = \frac{\pi y_{r1}}{vT_{n}}.
\end{equation}

One may consider \(\frac{\pi y_{r1}}{v}\) as the effective
nuclear-resonant time. Note that \(f_{r1}\) increases proportionally
with the wire current. When \(f_{r1}\) approaches zero, the transverse
component of \(\vec{B}_{n}\) has no time to rotate or vary
within the effective flight path-length for nuclear-resonant rotation;
thus, its variance vanishes. Conversely, when \(f_{r1}\) approaches
unity, the mean transverse component of \(\vec{B}_{n}\)
nears zero; thus, the variance peaks towards
\(\left( B_{n}\sin\left\langle \theta_{n} \right\rangle \right)^{2}\).
In between, the variance is approximated by linear interpolation.

We have
\(\frac{\pi y_{r1}}{\left( \pi y_{r0} \right)} = \frac{B_{n}\sin\left\langle \theta_{n} \right\rangle}{B_{y}^{'}}\).
At the current of \(0.067\) A that divides the two regimes of current,
\(\pi y_{r1} = \pi y_{r0} = 0.33 \times 10^{-3}\) m, yielding
\(\frac{\pi y_{r1}}{\left( \pi y_{r0} \right)} = 1\). As illustrated in
Fig.~\ref{fig:S5}, at the current of 0.2 A, \(\pi y_{r1}\) equals
\(0.98 \times 10^{-3}\) m, yielding
\(\frac{\pi y_{r1}}{\left( \pi y_{r0} \right)} = 3.0.\) At the maximum
current of 0.5 A, \(\pi y_{r1}\) increases to \(2.5 \times 10^{-3}\) m,
yielding \(\frac{\pi y_{r1}}{\left( \pi y_{r0} \right)} = 7.5\);
\(f_{r1}\) reaches 0.34. In comparison, the counterpart \(f_{r0}\)
during the flight time over \(\pi y_{r0}\) is only
\(0.34 \times \frac{\pi y_{r0}}{\left( \pi y_{r1} \right)} = 0.045 \ll 1\);
thus, the transverse field is stable, i.e., its angular variation is
much less than \(2\pi\). Therefore, the null-point rotation is due to
the static and quasi-static transverse field within \(\pi y_{r0}\),
whereas the nuclear-resonant rotation is due to the rotating transverse
field within \(\pi y_{r1}\). Disrupted by the null-point rotation in
combination with the random \(\phi_{n0}\) in the ensemble of atoms (see
Eq.~\ref{eq:133}), the nuclear-resonant rotation contributes to the fraction of
flip through the variance (instead of Eq.~\ref{eq:139} in the absence of the
null-point rotation). Note that one may consider the null-point rotation
as a resonant effect but at a zero Larmor frequency.

Combining terms yields

\begin{equation}
\label{eq:147}
E_{r1} = \frac{\pi^{2}\left| \gamma_{e} \right|^{2}\gamma_{n}z_{a}^{3}}{4v^{3}}\frac{B_{e}\left( B_{n}\sin\left\langle \theta_{n} \right\rangle \right)^{5}}{B_{y}^{'3}}.
\end{equation}

Expressing \(B_{y}^{'}\) in terms of the current, \(I\), using Eqs.~\ref{eq:120} and \ref{eq:122} gives

\begin{equation}
\label{eq:148}
E_{r1} = \frac{\mu_{0}^{3}\left| \gamma_{e} \right|^{2}\gamma_{n}}{32\pi v^{3}}\frac{B_{e}\left( B_{n}\sin\left\langle \theta_{n} \right\rangle \right)^{5}}{\left( B_{r} + B_{n}\cos\left\langle \theta_{n} \right\rangle \right)^{6}}I^{3}.
\end{equation}

Using the dimensionless adiabaticity parameters \(k_{0}\) (Eq.~\ref{eq:125}) and
\(k_{1}\) (Eq.~\ref{eq:126}), we simplify Eqs.~\ref{eq:141} and \ref{eq:142} to

\begin{equation}
\label{eq:149}
E_{r0} = \pi k_{0}\sqrt{1 + \frac{k_{1}}{k_{0}}} = \pi\sqrt{k_{0}^{2} + k_{0}k_{1}}
\end{equation}

and

\begin{equation}
\label{eq:150}
E_{r1} = \frac{1}{2}\left\lbrack \pi k_{1}\  \right\rbrack^{2}\left\{ f_{r1} \right\}.
\end{equation}

Thus, \(E_{r0}\) is due to the quadrature sum of \(k_{0}\) and
\(\sqrt{k_{0}k_{1}}\), and \(E_{r1}\) is due to \(k_{1}\). Eq.~\ref{eq:146} can
be rewritten as
\(f_{r1} = \frac{\pi z_{a}}{vT_{n}}\sqrt{\frac{k_{1}}{k_{0}}}\).
Substitution of the above two equations into Eq.~\ref{eq:140} with \(E_{i} = 0\)
yields the fraction of flip

\begin{equation}
\label{eq:151}
W_{4} = \exp\left( - \pi\sqrt{k_{0}^{2} + k_{0}k_{1}} - \frac{1}{2}\left\lbrack \pi k_{1}\  \right\rbrack^{2}f_{r1} \right),
\end{equation}

which produces Curve 4 in Fig.~\ref{fig:4}. If \(f_{r1} = 0\), \(W_{4}\) reduces
to

\begin{equation}
\label{eq:152}
W_{3} = \exp\left( - \pi\sqrt{k_{0}^{2} + k_{0}k_{1}} \right),
\end{equation}

which produces Curve 3 in Fig.~\ref{fig:4}. If \(k_{1} = 0\), \(W_{3}\) reduces to

\begin{equation}
\label{eq:153}
W_{2} = \exp\left( - \pi k_{0} \right),
\end{equation}

which produces Curve 2 in Fig.~\ref{fig:4}. If \(k_{0}\) is computed without
including the correction to the remnant field (Eq.~\ref{eq:119}), \(k_{0}\)
reduces to \(k_{m}\); thus, \(W_{2}\) reduces to

\begin{equation}
\label{eq:154}
W_{1} = \exp\left( - \pi k_{m} \right),
\end{equation}

which produces Curve 1 in Fig.~\ref{fig:4}. Taking the square root of \(W_{1}\)
leads to

\begin{equation}
\label{eq:155}
W_{m} = \exp\left( - \frac{\pi k_{m}}{2} \right),
\end{equation}

which produces Curve \emph{m} in Fig.~\ref{fig:4}.

The induction exponent is estimated as

\begin{equation}
\label{eq:156}
E_{i} = 4k_{i}\int_{- \frac{T_{f}}{2}}^{+ \frac{T_{f}}{2}}{\left| {\dot{\phi}}_{e} \right|\ dt},
\end{equation}

where \(T_{f}\) denotes the entire flight time corresponding to a
path-length of 16.3 mm in the inner rotation chamber. The field
generated from the wire contributes to induction. At the nearest point
to the wire along the atomic path, the magnetic flux density is

\begin{equation}
\label{eq:157}
B_{w}(0) = \frac{\mu_{0}I}{2\pi z_{a}},
\end{equation}

which reaches \(9.5 \times 10^{-4}\) T at the maximum current of 0.5 A
and \(z_{a}\) of \(1.05 \times 10^{-4}\) m. Therefore, the strongest
\(B_{w}(0)\) is 80 times
(\(= \frac{9.5 \times 10^{-4}}{0.119} \times 10^{-4}\)) greater than
\(B_{n}\), 23 times
(\(= \frac{9.5 \times 10^{-4}}{0.42} \times 10^{-4}\)) stronger than
\(B_{r}\), but 59 times
(\(= 558 \times \frac{10^{-4}}{9.5} \times 10^{-4}\)) weaker than
\(B_{e}\). Also, \(B_{w}(0)\) is \textasciitilde300 times weaker than a
typical main field \cite{Wennerstrom2012,Schroder1983} (\(\geq 0.3\) T) in Stern--Gerlach
experiments (see Paragraph 2 in Discussion). Without including the
induction exponent, CQD predicts the Frisch--Segrè experimental
observation well already (Fig.~\ref{fig:1} or Fig.~\ref{fig:4}). Consequently, the induction
effect in the inner rotation chamber is initially neglected. However, it
is included here for completeness and for the estimation of \(k_{i}\).
Below, zero time \(t\) is set to when an atom reaches the nearest point
to the wire.

From Eq.~\ref{eq:156}, we estimate \(E_{i}\) by using the \(z\) component of the
field along the path:

\begin{equation}
\label{eq:158}
E_{i} = 4k_{i}\left| \gamma_{e} \right|\int_{- \frac{T_{f}}{2}}^{+ \frac{T_{f}}{2}}{\left| \frac{B_{w}(0)\frac{vt}{z_{a}}}{1 + \left( \frac{vt}{z_{a}} \right)^{2}} \right|\ dt} = 4k_{i}\left| \gamma_{e} \right|B_{w}(0)\frac{z_{a}}{v}\ln\left( \frac{T_{f}v}{2z_{a}} \right).
\end{equation}

Substituting Eq.~\ref{eq:157} gives

\begin{equation}
\label{eq:159}
E_{i} = k_{i}\frac{2\mu_{0}\left| \gamma_{e} \right|}{\pi v}\ln\left( \frac{T_{f}v}{2z_{a}} \right)I.
\end{equation}

Expressing in terms of the current, \(I\), yields

\begin{equation}
\label{eq:160}
W_{\mathrm{cqd}} = \exp\left\lbrack - \sqrt{\left( \frac{c_{r0}}{I} \right)^{2} + c_{rs}^{2}} - c_{r1}I^{3} - c_{ri}I \right\rbrack,
\end{equation}

where

\begin{equation}
\label{eq:161}
c_{r0} = \frac{2\pi^{2}\left| \gamma_{e} \right|\left( B_{r} + B_{n}\cos\left\langle \theta_{n} \right\rangle \right)^{2}z_{a}^{2}}{\left( \mu_{0}v \right)},
\end{equation}

\begin{equation}
\label{eq:162}
c_{rs} = \frac{\pi\left| \gamma_{e} \right|B_{n}\sin\left\langle \theta_{n} \right\rangle z_{a}}{v},
\end{equation}

\begin{equation}
\label{eq:163}
c_{r1} = \frac{\frac{\mu_{0}^{3}\left| \gamma_{e} \right|^{2}\gamma_{n}}{32\pi v^{3}}B_{e}\left( B_{n}\sin\left\langle \theta_{n} \right\rangle \right)^{5}}{\left( B_{r} + B_{n}\cos\left\langle \theta_{n} \right\rangle \right)^{6}},
\end{equation}

and

\begin{equation}
\label{eq:164}
c_{ri} = k_{i}\frac{2\mu_{0}\left| \gamma_{e} \right|}{\pi v}\ln\left( \frac{T_{f}v}{2z_{a}} \right).
\end{equation}

The four coefficients represent null-point rotation, rotation
saturation, nuclear-resonant rotation, and induction rotation of the
polar angle, respectively. While null-point rotation increases
\(W_{\mathrm{cqd}}\) with increasing current, nuclear-resonant rotation does the
opposite.

If \(k_{i} = 0,\) Eq.~\ref{eq:160} reduces to

\begin{equation}
\label{eq:165}
W_{4} = \exp\left\lbrack - \sqrt{\left( \frac{c_{r0}}{I} \right)^{2} + c_{rs}^{2}} - c_{r1}I^{3} \right\rbrack,
\end{equation}

which is equivalent to Eq.~\ref{eq:151}. Further, if \(c_{r1} = 0,\) we reach

\begin{equation}
\label{eq:166}
W_{3} = \exp\left\lbrack - \sqrt{\left( \frac{c_{r0}}{I} \right)^{2} + c_{rs}^{2}} \right\rbrack,
\end{equation}

which is another form of Eq.~\ref{eq:152}. Merging the above two equations, we
obtain

\begin{equation}
\label{eq:167}
W_{4} = W_{3}\exp\left( - c_{r1}I^{3} \right).
\end{equation}

Finally, if \(c_{rs} = 0,\) we obtain

\begin{equation}
\label{eq:168}
W_{2} = \exp\left( - \frac{c_{r0}}{I} \right),
\end{equation}

which is the same as Eq.~\ref{eq:153}.

\FloatBarrier
\clearpage
\section[Appendix~\thesection. Uncertainty relation]{Uncertainty relation}\label{appendix:6}
\suppressfloats[t]

If the initial co-quanta are isotropically distributed, CQD reproduces
the quantum mechanical uncertainty relation.

First, we predict the expectation of the spin-angular-momentum
projection, \(\left\langle s_{y} \right\rangle\), along \(y\), the
direction of the atomic beam. The CQD prediction expression for the
\(y\) axis is (Eq.~\ref{eq:13})

\begin{equation}
\label{eq:169}
\left|{\widehat{\mu}}_{e}\coq{\widehat{\mu}}_{n} \right\rangle_{y} = C_{+ y}\left( {\widehat{\mu}}_{e},{\widehat{\mu}}_{n} \right)\left| + y \right\rangle + C_{- y}\left( {\widehat{\mu}}_{e},{\widehat{\mu}}_{n} \right)\exp\left( i\phi_{ey} \right)\left| - y \right\rangle,
\end{equation}

where \(\phi_{ey}\) denotes the azimuthal angle of
\(\vec{\mu}_{e}\) about \(y\). Following Appendix~\ref{appendix:3} yields
the wave function,

\begin{equation}
\label{eq:170}
\left|{\widehat{\mu}}_{e} \right\rangle_{y} = \cos\frac{\theta_{ey}}{2}\left| + y \right\rangle + \sin\frac{\theta_{ey}}{2}\exp\left( i\phi_{ey} \right)\left| - y \right\rangle,
\end{equation}

where \(\theta_{ey}\) denotes the polar angle of
\(\vec{\mu}_{e}\) relative to \(y\). Note that
\(\left| + y \right\rangle\) here denotes ``up'' for magnetic moment
and hence ``down'' for electron spin, and
\(\left| - y \right\rangle\) denotes the opposite state.

The expectation is

\begin{equation}
\label{eq:171}
\left\langle s_{y} \right\rangle = - \frac{\hslash}{2}\cos^{2}\frac{\theta_{ey}}{2} + \frac{\hslash}{2}\sin^{2}\frac{\theta_{ey}}{2} = - \frac{\hslash}{2}\cos\theta_{ey}.
\end{equation}

Second, we measure along \(z\). The CQD prediction expression for the
\(z\) axis is (Eq.~\ref{eq:13})

\begin{equation}
\label{eq:172}
\left|{\widehat{\mu}}_{e}\coq{\widehat{\mu}}_{n} \right\rangle_{z} = C_{+ z}\left( {\widehat{\mu}}_{e},{\widehat{\mu}}_{n} \right)\left| + z \right\rangle + C_{- z}\left( {\widehat{\mu}}_{e},{\widehat{\mu}}_{n} \right)\exp\left( i\phi_{ez} \right)\left| - z \right\rangle.
\end{equation}

We similarly derive the wave function (Appendix~\ref{appendix:3}),

\begin{equation}
\label{eq:173}
\left|{\widehat{\mu}}_{e} \right\rangle_{z} = \cos\frac{\theta_{ez}}{2}\left| + z \right\rangle + \sin\frac{\theta_{ez}}{2}\exp\left( i\phi_{ez} \right)\left| - z \right\rangle,
\end{equation}

where \(\theta_{ez}\) and \(\phi_{ez}\) denote the polar and azimuthal
angles in relation to \(z\).

We compute the standard deviation, \(\Delta s_{z}\), as follows:

\begin{equation}
\label{eq:174}
\left\langle s_{z} \right\rangle = - \frac{\hslash}{2}\cos^{2}\frac{\theta_{ez}}{2} + \frac{\hslash}{2}\sin^{2}\frac{\theta_{ez}}{2} = - \frac{\hslash}{2}\cos\theta_{ez},
\end{equation}

\begin{equation}
\label{eq:175}
\left\langle s_{z}^{2} \right\rangle = \left( - \frac{\hslash}{2} \right)^{2}\cos^{2}\frac{\theta_{ez}}{2} + \left( \frac{\hslash}{2} \right)^{2}\sin^{2}\frac{\theta_{ez}}{2} = \left( \frac{\hslash}{2} \right)^{2},
\end{equation}

and

\begin{equation}
\label{eq:176}
\Delta s_{z} = \sqrt{\left\langle s_{z}^{2} \right\rangle - \left\langle s_{z} \right\rangle^{2}} = \frac{\hslash}{2}\sin\theta_{ez}.
\end{equation}

At this point, \({\widehat{\mu}}_{e}\) has collapsed to \(\pm z\); thus,
the polar angle relative to \(x\), \(\theta_{ex} = \frac{\pi}{2}\). Now,
the co-quantum distribution follows the heart shape (Eq.~\ref{eq:24}).

Third, we measure along \(x\). The CQD prediction expression for the
\(x\) axis is (Eq.~\ref{eq:13})

\begin{equation}
\label{eq:177}
\left|{\widehat{\mu}}_{e}\coq{\widehat{\mu}}_{n} \right\rangle_{x} = C_{+ x}\left( {\widehat{\mu}}_{e},{\widehat{\mu}}_{n} \right)\left| + x \right\rangle + C_{- x}\left( {\widehat{\mu}}_{e},{\widehat{\mu}}_{n} \right)\exp\left( i\phi_{ex} \right)\left| - x \right\rangle.
\end{equation}

Invoking \(\theta_{ex} = \frac{\pi}{2}\), the heart shape (Eq.~\ref{eq:24}), and
the identity \(\cos\theta_{nz} = \sin\theta_{nx}\sin\phi_{nx}\), we
follow Appendix~\ref{appendix:3} to similarly obtain for the \(\pm z\) branch

\begin{equation}
\label{eq:178}
\left\langle C_{+ x} \right\rangle_{n}^{2} = \int_{\frac{\pi}{2}}^{\pi}{\int_{0}^{2\pi}{\frac{1 \mp \cos\theta_{nz}}{4\pi}\sin\theta_{nx}d\phi_{nx}d\theta_{nx}}} = \frac{1}{2}
\end{equation}

and

\begin{equation}
\label{eq:179}
\left\langle C_{- x} \right\rangle_{n}^{2} = \int_{0}^{\frac{\pi}{2}}{\int_{0}^{2\pi}{\frac{1 \mp \cos\theta_{nz}}{4\pi}\sin\theta_{nx}d\phi_{nx}d\theta_{nx}}} = \frac{1}{2}.
\end{equation}

The even split between the two \(\pm x\) branches is because the heart
shape associated with either of the \(\pm z\) branches is rotationally
symmetric about the \(z\) axis. Thus, the wave function is

\begin{equation}
\label{eq:180}
\left|{\widehat{\mu}}_{e} \right\rangle_{x} = \frac{1}{\sqrt{2}}\left| + x \right\rangle + \frac{1}{\sqrt{2}}\exp\left( i\phi_{ex} \right)\left| - x \right\rangle.
\end{equation}

We derive the standard deviation, \(\Delta s_{x}\), as follows:

\begin{equation}
\label{eq:181}
\left\langle s_{x} \right\rangle = - \frac{\hslash}{2}\left( \frac{1}{\sqrt{2}} \right)^{2} + \frac{\hslash}{2}\left( \frac{1}{\sqrt{2}} \right)^{2} = 0,
\end{equation}

\begin{equation}
\label{eq:182}
\left\langle s_{x}^{2} \right\rangle = \left( - \frac{\hslash}{2} \right)^{2}\left( \frac{1}{\sqrt{2}} \right)^{2} + \left( \frac{\hslash}{2} \right)^{2}\left( \frac{1}{\sqrt{2}} \right)^{2} = \left( \frac{\hslash}{2} \right)^{2},
\end{equation}

and

\begin{equation}
\label{eq:183}
\Delta s_{x} = \sqrt{\left\langle s_{x}^{2} \right\rangle - \left\langle s_{x} \right\rangle^{2}} = \frac{\hslash}{2}.
\end{equation}

Fourth, combining Eqs.~\ref{eq:176} and \ref{eq:183} reaches

\begin{equation}
\label{eq:184}
\Delta s_{z}\Delta s_{x} = \left( \frac{\hslash}{2}\sin\theta_{ez} \right) \cdot \frac{\hslash}{2}.
\end{equation}

Substituting the identity,
\(\cos\theta_{ey} = \sin\theta_{ez}\sin\phi_{ez}\), into Eq.~\ref{eq:171} yields

\begin{equation}
\label{eq:185}
\frac{\hslash}{2}\left| \left\langle s_{y} \right\rangle \right| = \left\lbrack \left( \frac{\hslash}{2}\sin\theta_{ez} \right) \cdot \frac{\hslash}{2} \right\rbrack \cdot \left| \sin\phi_{ez} \right|.
\end{equation}

Combining the above two equations yields

\begin{equation}
\label{eq:186}
\Delta s_{z}\Delta s_{x} \cdot \left| \sin\phi_{ez} \right| = \frac{\hslash}{2}\left| \left\langle s_{y} \right\rangle \right|.
\end{equation}

This uncertainty equality shows that the magnitude of the uncertainty
product, \(\Delta s_{z}\Delta s_{x}\), depends on not only
\(\left\langle s_{y} \right\rangle\) but also the initial phase,
\(\phi_{ez}\), in relation to the first measurement axis. Therefore, the
order of the \(z\)-\(x\) measurements matters.

Finally, invoking \(\left| \sin\phi_{ez} \right| \leq 1\ \)reproduces
exactly the familiar quantum mechanical uncertainty inequality for
angular momenta,

\begin{equation}
\label{eq:187}
\Delta s_{z}\Delta s_{x} \geq \frac{\hslash}{2}\left| \left\langle s_{y} \right\rangle \right|,
\end{equation}

which takes on the equal sign when \(\phi_{ez} = \pm \frac{\pi}{2}\).

\FloatBarrier
\clearpage
\section[Appendix~\thesection. Entanglement]{Entanglement}\label{appendix:7}
\suppressfloats[t]

\begin{figure}[!htbp]
\centering
\includegraphics[width=6.5in,height=1.63056in]{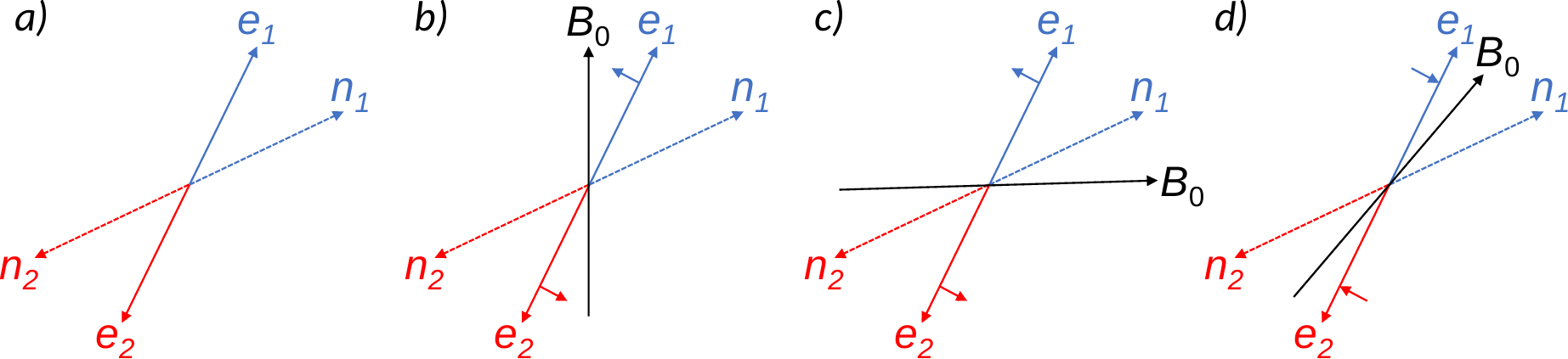}
\caption{Entanglement of the electron magnetic moments of two
atoms. \textbf{(a)} At entanglement site. \textbf{(b) -- (d)} At two
Stern--Gerlach devices with three orientations of
\({\vec{B}}_{0}\). \emph{e,} electron magnetic moment;
\emph{n}, nuclear magnetic moment; subscripts 1 and 2, atoms 1 and 2.
The short arrows indicate the directions of collapse determined by the
branching condition, where the polar angles are relative to
\({\vec{B}}_{0}\). In \emph{d}, to facilitate comparison, one
may mirror \({\widehat{\mu}}_{e}\) about \({\vec{B}}_{0}\)
because \({\widehat{\mu}}_{e}\) precesses much faster than
\({\widehat{\mu}}_{n}\).}
\label{fig:S6}
\end{figure}

CQD in its current form construes anticorrelated entanglement as a pair
of atoms having both opposing \({\widehat{\mu}}_{e}\)'s and
\({\widehat{\mu}}_{n}\)'s (Fig.~\ref{fig:S6}). The two atoms are delivered from
the entanglement site to two Stern--Gerlach devices. Once the
orientation of the external magnetic flux density
\({\vec{B}}_{0}\) is chosen, the two
\({\widehat{\mu}}_{e}\)'s are guaranteed to collapse to opposing states
according to the branching condition because
\(\theta_{e1} + \theta_{e2} = \pi\) and
\(\theta_{n1} + \theta_{n2} = \pi\). For example, if atom 1 collapses to
\(+ {\widehat{B}}_{0}\) because \(\theta_{n1} > \theta_{e1}\), atom 2
automatically collapses to \(- {\widehat{B}}_{0}\) because
\(\theta_{n2} < \theta_{e2}\) (as derived from
\(\pi - \theta_{n1} < \pi - \theta_{e1}\)). Therefore, the co-quanta
propagate with the principal quanta and determine the measurement
outcomes.

The CQD prediction expressions for the two atoms are

\begin{equation}
\label{eq:188}
\left|{\widehat{\mu}}_{e}\coq{\widehat{\mu}}_{n} \right\rangle_{1} = C_{+}\left( {\widehat{\mu}}_{e},{\widehat{\mu}}_{n} \right)\left| + {\widehat{B}}_{0} \right\rangle_{1} + C_{-}\left( {\widehat{\mu}}_{e},{\widehat{\mu}}_{n} \right)\exp\left( i\phi_{e1} \right)\left| - {\widehat{B}}_{0} \right\rangle_{1}
\end{equation}

and

\begin{equation}
\label{eq:189}
\left| - {\widehat{\mu}}_{e}\coq{- \widehat{\mu}}_{n} \right\rangle_{2} = C_{-}\left( {\widehat{\mu}}_{e},{\widehat{\mu}}_{n} \right) \left| + {\widehat{B}}_{0} \right\rangle_{2} + C_{+}\left( {\widehat{\mu}}_{e},{\widehat{\mu}}_{n} \right)\exp\left( i\phi_{e2} \right)\left| - {\widehat{B}}_{0} \right\rangle_{2}.
\end{equation}

The numeral subscripts denote the two atoms. The joint pre-collapse
state function of the atom pair is written as
\(\left|{\widehat{\mu}}_{e}\coq{\widehat{\mu}}_{n} \right\rangle_{1} \otimes\)
\(\left|{- \widehat{\mu}}_{e}\coq - {\widehat{\mu}}_{n} \right\rangle_{2}\),
where \(\otimes\) denotes tensor product \cite{Lambropoulos2007}. Because
\(C_{+} \cdot C_{-} = 0\) and \(C_{\pm} \cdot C_{\pm} = C_{\pm}\), the
joint CQD prediction expression of the atom pair is given by

\(\left|{\widehat{\mu}}_{e}\coq{\widehat{\mu}}_{n} \right\rangle_{1} \otimes \left| - {\widehat{\mu}}_{e}\coq{- \widehat{\mu}}_{n} \right\rangle_{2}\)

\begin{equation}
\label{eq:190}
= C_{+}\left| + {\widehat{B}}_{0} \right\rangle_{1} \otimes \left| - {\widehat{B}}_{0} \right\rangle_{2} + C_{-}\exp(i\Delta\phi)\left| - {\widehat{B}}_{0} \right\rangle_{1} \otimes \left| + {\widehat{B}}_{0} \right\rangle_{2},
\end{equation}

where \(\Delta\phi = \phi_{e1} - \phi_{e2}\) and the common phase is
removed.

If \(\Delta\phi\) is constant for a given experimental configuration,
ensemble averaging Eq.~\ref{eq:190}, denoted by
\(\left\langle \right\rangle_{n,e}\), yields the familiar quantum
mechanical entangled wave function,

\begin{equation}
\label{eq:191}
\left|\psi \right\rangle = \left\langle C_{+} \right\rangle_{n,e}\left| + {\widehat{B}}_{0} \right\rangle_{1} \otimes \left| - {\widehat{B}}_{0} \right\rangle_{2} + \left\langle C_{-} \right\rangle_{n,e}\exp(i\Delta\phi)\left| - {\widehat{B}}_{0} \right\rangle_{1} \otimes \left| + {\widehat{B}}_{0} \right\rangle_{2}.
\end{equation}

If both \({\widehat{\mu}}_{n}\) and \({\widehat{\mu}}_{e}\) of
individual atoms are isotropically distributed, Appendix~\ref{appendix:3} provides
\(\left\langle C_{\pm} \right\rangle_{n,e} = \frac{1}{\sqrt{2}}\),
yielding

\begin{equation}
\label{eq:192}
\left|\psi \right\rangle = \frac{1}{\sqrt{2}}\left| + {\widehat{B}}_{0} \right\rangle_{1} \otimes \left| - {\widehat{B}}_{0} \right\rangle_{2} + \frac{1}{\sqrt{2}}\exp(i\Delta\phi)\left| - {\widehat{B}}_{0} \right\rangle_{1} \otimes \left| + {\widehat{B}}_{0} \right\rangle_{2}.
\end{equation}

Here, the key to producing the above entangled wave function is the
mutual exclusivity of the binary coefficients:
\(C_{+} \cdot C_{-} = 0\). The pre-collapse ``product state'' in CQD,
\(\left|{\widehat{\mu}}_{e}\coq{\widehat{\mu}}_{n} \right\rangle_{1} \otimes\)
\(\left|{- \widehat{\mu}}_{e}\coq - {\widehat{\mu}}_{n} \right\rangle_{2}\),
averages to an entangled wave function (i.e., not a product state). One
can adapt the above derivation for correlated instead of anticorrelated
entanglement.

A future direction is to explore CQD in relation to
Bell\textquotesingle s theorem \cite{Bell1966}. An ideal experiment would
follow the above derivation, where entangled pairs of alkali-metal atoms
are delivered to two Stern--Gerlach devices with independent
quantization axes. Shin \emph{et al}. \cite{Shin2019} published in 2019 an
experiment in this direction. However, the quantization axes could not
be controlled independently, and Bose-Einstein condensate helium-4 atoms
in the long-lived metastable state \(2^{3}S_{1}\) instead of
alkali-metal atoms were used. The same group also published in 2022 an
experiment on the Bell inequality for motional degrees of freedom of
massive particles, thus far reaching a maximum CHSH-Bell parameter of
\(S\  = \ 1.1\) \cite{Thomas2022}.

\FloatBarrier
\clearpage
\section[Appendix~\thesection. Two-stage Stern--Gerlach apparatus with a varying angle between the quantization axes]{Two-stage Stern--Gerlach apparatus with a varying angle between the quantization axes}\label{appendix:8}
\suppressfloats[t]

CQD can potentially be further verified with a two-stage Stern--Gerlach
apparatus with a varying angle between the quantization axes. As usual,
the first stage polarizes the atoms to the \(+ z\) state; however, the
second stage is rotated by an arbitrary angle \(\alpha\) about the \(y\)
axis (the atomic beam axis). Below, the coordinates of the second stage
are denoted with primes.

Using the heart-shaped angular distribution of the co-quanta (Eq.~\ref{eq:24})
and invoking both \(\theta_{ez^{'}} = \alpha\) and

\begin{equation}
\label{eq:193}
\cos\theta_{nz} = \frac{1}{\sqrt{2}}\left( \cos\theta_{nz^{'}} - \sin\theta_{nz^{'}}\cos\phi_{nz^{'}} \right),
\end{equation}

CQD predicts the probability of collapsing to the \(+ z'\) state as

\begin{equation}
\label{eq:194}
\left\langle C_{+ z^{'}} \right\rangle_{n}^{2} = \int_{\alpha}^{\pi}{\int_{0}^{2\pi}{\frac{1 - \cos\theta_{nz}}{4\pi}\sin\theta_{nz^{'}}d\phi_{nz^{'}}d\theta_{nz^{'}}}} = \frac{\left( 1 + \cos\alpha \right)^{2}\,\left( 2 - \cos\alpha \right)}{4}.
\end{equation}

In comparison, the standard literature \cite{Feynman1963,Norsen2017} predicts
\(\cos^{2}\left( \frac{\alpha}{2} \right)\) or
\(\frac{\left( 1 + \cos\alpha \right)}{2}\) (Eq.~\ref{eq:16}). At \(\alpha\) of
\(0\), \(\frac{\pi}{2}\), and \(\pi\), the typical angles in the
literature, CQD and the literature predict the same probabilities of 1,
\(\frac{1}{2}\), and 0, respectively. In general, however, the ratio of
the two predictions is

\begin{equation}
\label{eq:195}
R_{+} = \frac{\left\lbrack 9 - \left( 2\cos\alpha - 1 \right)^{2} \right\rbrack}{8},
\end{equation}

which holds for \(0 \leq \alpha < \pi\) with \(\pi\) excluded to avoid
\(\frac{0}{0}\). The ratio peaks at \(\alpha\) of \(\frac{\pi}{3}\) with
\(R_{+} = \frac{9}{8}\) then decreases with increasing \(\alpha\). At
\(\alpha = \frac{11\pi}{12}\), for example, \(R_{+} = 0.05\), with CQD
predicting a much lower probability.

\FloatBarrier
\clearpage
\section[Appendix~\thesection. Vector and spherical-coordinate forms of CQD equations of motion]{Vector and spherical-coordinate forms of CQD equations of motion}\label{appendix:9}
\suppressfloats[t]

This appendix derives the spherical-coordinate equations of motion from the vector form.

\subsection{Starting equations}

We begin with the vector equations
\begin{align}
\frac{d\widehat{\mu}_e}{dt}
&=
\gamma_e\,\widehat{\mu}_e \times \left(\vec{B}+\vec{B}_n\right)
-
\mathrm{sgn}\!\left(\widehat{\mu}_n \cdot \widehat{B} - \widehat{\mu}_e \cdot \widehat{B}\right)
\mathrm{sgn}\!\left(\frac{d\widehat{\mu}_{i,e}}{dt}\cdot \widehat{B}\right)
\frac{d\widehat{\mu}_{i,e}}{dt},
\label{eq:9-start1a}
\\
\frac{d\widehat{\mu}_{i,e}}{dt}
&\equiv
k_i\,\widehat{\mu}_e \times \frac{d\widehat{\mu}_e}{dt}.
\label{eq:9-start2}
\end{align}

We take the effective field to be
\begin{equation}
\label{eq:9-beff}
\vec{B}_{\mathrm{eff}} 
= \vec{B}+\vec{B}_n
= B_y\,\widehat{y} + B_z\,\widehat{z} + B_n\,\widehat{\mu}_n.
\end{equation}
Since \(B_z\) is typically the dominant component, we take \(\widehat{z}\) as the approximate quantization axis, which yields
\begin{align}
\frac{d\widehat{\mu}_e}{dt}
&=
\gamma_e\,\widehat{\mu}_e \times \vec{B}_{\mathrm{eff}}
-
\mathrm{sgn}\!\left(\widehat{\mu}_n \cdot \widehat{z} - \widehat{\mu}_e \cdot \widehat{z}\right)
\mathrm{sgn}\!\left(\frac{d\widehat{\mu}_{i,e}}{dt}\cdot \widehat{z}\right)
\frac{d\widehat{\mu}_{i,e}}{dt}.
\label{eq:9-start1}
\end{align}

The goal is to derive the spherical-coordinate equations for \(\theta_e(t)\) and \(\phi_e(t)\).

\subsection{Spherical parameterization and local basis}

Write the electron and nuclear unit vectors as
\begin{align}
\widehat{\mu}_e
&=
\left(
\sin\theta_e \cos\phi_e,
\sin\theta_e \sin\phi_e,
\cos\theta_e
\right),
\label{eq:9-mue}
\\
\widehat{\mu}_n
&=
\left(
\sin\theta_n \cos\phi_n,
\sin\theta_n \sin\phi_n,
\cos\theta_n
\right).
\label{eq:9-mun}
\end{align}

The local spherical basis attached to \(\widehat{\mu}_e\) (Fig.~\ref{fig:9-spherical-basis}) is
\begin{align}
\widehat{e}_r
&=
\widehat{\mu}_e,
\\
\widehat{e}_\theta
&=
\left(
\cos\theta_e \cos\phi_e,
\cos\theta_e \sin\phi_e,
-\sin\theta_e
\right),
\\
\widehat{e}_\phi
&=
\left(
-\sin\phi_e,
\cos\phi_e,
0
\right).
\end{align}

Therefore,
\begin{equation}
\label{eq:9-muedot}
\frac{d\widehat{\mu}_e}{dt}
=
\dot{\theta}_e\,\widehat{e}_\theta
+
\dot{\phi}_e \sin\theta_e\,\widehat{e}_\phi.
\end{equation}

Also, using the standard right-handed spherical-basis relations,
\begin{equation}
\label{eq:9-crossbasis}
\widehat{\mu}_e \times \widehat{e}_\theta = \widehat{e}_\phi,
\qquad
\widehat{\mu}_e \times \widehat{e}_\phi = -\widehat{e}_\theta.
\end{equation}

Hence
\begin{equation}
\label{eq:9-mu-cross-mudot}
\widehat{\mu}_e \times \frac{d\widehat{\mu}_e}{dt}
=
\dot{\theta}_e\,\widehat{e}_\phi
-
\dot{\phi}_e \sin\theta_e\,\widehat{e}_\theta.
\end{equation}

Substituting Eq.~\ref{eq:9-mu-cross-mudot} into Eq.~\ref{eq:9-start2} gives
\begin{equation}
\label{eq:9-muidot}
\frac{d\widehat{\mu}_{i,e}}{dt}
=
k_i
\left(
\dot{\theta}_e\,\widehat{e}_\phi
-
\dot{\phi}_e \sin\theta_e\,\widehat{e}_\theta
\right).
\end{equation}

\subsection{Geometric sketch of the local spherical basis}

\begin{figure}[!htbp]
\centering
\begin{tikzpicture}[scale=2.0, >=Latex,
  x={(-0.5cm,-0.4cm)}, y={(1cm,0cm)}, z={(0cm,1cm)}]

  \def\R{2.0}
  \def\th{40}
  \def\ph{50}

  \coordinate (O) at (0,0,0);
  \coordinate (X) at (2.0,0,0);
  \coordinate (Y) at (0,2.0,0);
  \coordinate (Z) at (0,0,2.2);

  \pgfmathsetmacro{\Mx}{\R*sin(\th)*cos(\ph)}
  \pgfmathsetmacro{\My}{\R*sin(\th)*sin(\ph)}
  \pgfmathsetmacro{\Mz}{\R*cos(\th)}
  \coordinate (M) at (\Mx, \My, \Mz);
  \coordinate (Mxy) at (\Mx, \My, 0);

  \draw[->,thick] (O) -- (X) node[below left] {\(\widehat{x}\)};
  \draw[->,thick] (O) -- (Y) node[right] {\(\widehat{y}\)};
  \draw[->,thick] (O) -- (Z) node[above] {\(\widehat{z}\)};

  \draw[dashed, thick] (O) -- (Mxy);
  \draw[dashed, thick] (M) -- (Mxy);

  \draw[->,very thick] (O) -- (M) node[at end, anchor=south, yshift=3pt] {\(\widehat{\mu}_e=\widehat{e}_r\)};

  \draw[->, thick, domain=0:\ph, smooth, variable=\t]
    plot ({0.6*cos(\t)}, {0.6*sin(\t)}, 0);
  \node at ({0.8*cos(\ph/2)}, {0.8*sin(\ph/2)}, 0) {\(\phi_e\)};

  \draw[->, thick, domain=0:\th, smooth, variable=\t]
    plot ({0.7*sin(\t)*cos(\ph)}, {0.7*sin(\t)*sin(\ph)}, {0.7*cos(\t)});
  \node at ({0.9*sin(\th/2)*cos(\ph)}, {0.9*sin(\th/2)*sin(\ph)}, {0.9*cos(\th/2)}) {\(\theta_e\)};

  \pgfmathsetmacro{\eThX}{0.8*cos(\th)*cos(\ph)}
  \pgfmathsetmacro{\eThY}{0.8*cos(\th)*sin(\ph)}
  \pgfmathsetmacro{\eThZ}{-0.8*sin(\th)}

  \draw[->,thick] (M) -- ++(\eThX, \eThY, \eThZ)
    node[below right, align=left, inner sep=1pt]
    {\(\widehat{e}_\theta\) \\[-0.5ex] {\scriptsize \color{red} \(\dot{\theta}_e\,\widehat{e}_\theta\)}};

  \pgfmathsetmacro{\ePhX}{-0.7*sin(\ph)}
  \pgfmathsetmacro{\ePhY}{0.7*cos(\ph)}
  \pgfmathsetmacro{\ePhZ}{0}

  \draw[->,thick] (M) -- ++(\ePhX, \ePhY, \ePhZ)
    node[above right, align=left, inner sep=1pt]
    {\(\widehat{e}_\phi\) \\[-0.5ex] {\scriptsize \color{red} \(\dot{\phi}_e \sin\theta_e\,\widehat{e}_\phi\)}};

\end{tikzpicture}
\caption{Local spherical basis attached to \(\widehat{\mu}_e\). The vector \(\widehat{e}_r=\widehat{\mu}_e\) points radially, \(\widehat{e}_\theta\) is the polar-direction unit vector, and \(\widehat{e}_\phi\) is the azimuthal-direction unit vector. The time derivative \(d\widehat{\mu}_e/dt\) decomposes into \(\dot{\theta}_e\,\widehat{e}_\theta + \dot{\phi}_e \sin\theta_e\,\widehat{e}_\phi\).}
\label{fig:9-spherical-basis}
\end{figure}

\subsection{Evaluation of the sign factors}

From Eq.~\ref{eq:9-muidot},
\begin{equation}
\frac{d\widehat{\mu}_{i,e}}{dt}\cdot \widehat{z}
=
k_i
\left(
\dot{\theta}_e\,\widehat{e}_\phi \cdot \widehat{z}
-
\dot{\phi}_e \sin\theta_e\,\widehat{e}_\theta \cdot \widehat{z}
\right).
\end{equation}

Since
\begin{equation}
\widehat{e}_\phi \cdot \widehat{z} = 0,
\qquad
\widehat{e}_\theta \cdot \widehat{z} = -\sin\theta_e,
\end{equation}
we obtain
\begin{equation}
\label{eq:9-sign2arg}
\frac{d\widehat{\mu}_{i,e}}{dt}\cdot \widehat{z}
=
k_i\,\dot{\phi}_e \sin^2\theta_e.
\end{equation}

Therefore, for \(k_i>0\) and \(0<\theta_e<\pi\),
\begin{equation}
\label{eq:9-sign2}
\mathrm{sgn}\!\left(\frac{d\widehat{\mu}_{i,e}}{dt}\cdot \widehat{z}\right)
=
\mathrm{sgn}(\dot{\phi}_e).
\end{equation}
Next,
\begin{equation}
\label{eq:9-sign1arg}
\widehat{\mu}_n \cdot \widehat{z} - \widehat{\mu}_e \cdot \widehat{z}
=
\cos\theta_n - \cos\theta_e.
\end{equation}

Because \(\cos\theta\) is strictly decreasing on \(0<\theta<\pi\),
\begin{equation}
\label{eq:9-monotonic}
\mathrm{sgn}\!\left(\cos\theta_n-
\cos\theta_e\right)
=
-\mathrm{sgn}(\theta_n-
\theta_e).
\end{equation}

Thus,
\begin{equation}
\label{eq:9-sign1}
\mathrm{sgn}\!\left(\widehat{\mu}_n \cdot \widehat{z} - \widehat{\mu}_e \cdot \widehat{z}\right)
=
-\mathrm{sgn}(\theta_n-
\theta_e).
\end{equation}

Substituting Eqs.~\ref{eq:9-sign2} and \ref{eq:9-sign1} into Eq.~\ref{eq:9-start1}, we get
\begin{equation}
\label{eq:9-start-simplified}
\frac{d\widehat{\mu}_e}{dt}
=
\gamma_e\,\widehat{\mu}_e \times \vec{B}_{\mathrm{eff}}
+
\mathrm{sgn}(\theta_n-
\theta_e)\,\mathrm{sgn}(\dot{\phi}_e)
\frac{d\widehat{\mu}_{i,e}}{dt}.
\end{equation}

Using Eq.~\ref{eq:9-muidot},
\begin{equation}
\label{eq:9-expanded-main}
\frac{d\widehat{\mu}_e}{dt}
=
\gamma_e\,\widehat{\mu}_e \times \vec{B}_{\mathrm{eff}}
+
\mathrm{sgn}(\theta_n-
\theta_e)\,\mathrm{sgn}(\dot{\phi}_e)
k_i
\left(
\dot{\theta}_e\,\widehat{e}_\phi
-
\dot{\phi}_e \sin\theta_e\,\widehat{e}_\theta
\right).
\end{equation}

Note the following alternative: 
\begin{equation}
\label{eq:9-sign2b}
\mathrm{sgn}(\dot{\phi}_e)
=\mathrm{sgn}(-\gamma_e B_z).
\end{equation}

\subsection{Projection of the effective field onto the local basis}

Write
\begin{equation}
\vec{B}_{\mathrm{eff}}
=
B_r\widehat{e}_r + B_\theta\widehat{e}_\theta + B_\phi\widehat{e}_\phi.
\end{equation}

Then
\begin{equation}
\label{eq:9-mu-cross-b}
\widehat{\mu}_e \times \vec{B}_{\mathrm{eff}}
=
B_\theta\widehat{e}_\phi - B_\phi\widehat{e}_\theta.
\end{equation}

So we need only \(B_\theta\) and \(B_\phi\).

\subsubsection[Computation of B phi]{Computation of \(B_\phi\)}

First,
\begin{align}
\widehat{y}\cdot \widehat{e}_\phi
&=
\cos\phi_e,
\\
\widehat{z}\cdot \widehat{e}_\phi
&=
0,
\\
\widehat{\mu}_n \cdot \widehat{e}_\phi
&=
\sin\theta_n \sin(\phi_n-
\phi_e).
\end{align}

Therefore,
\begin{equation}
\label{eq:9-bphi}
B_\phi
=
\vec{B}_{\mathrm{eff}}\cdot \widehat{e}_\phi
=
B_y \cos\phi_e
+
B_n \sin\theta_n \sin(\phi_n-
\phi_e).
\end{equation}

\subsubsection[Computation of B theta]{Computation of \(B_\theta\)}

Next,
\begin{align}
\widehat{y}\cdot \widehat{e}_\theta
&=
\cos\theta_e \sin\phi_e,
\\
\widehat{z}\cdot \widehat{e}_\theta
&=
-\sin\theta_e.
\end{align}

Also,
\begin{align}
\widehat{\mu}_n \cdot \widehat{e}_\theta
&=
\left(
\sin\theta_n \cos\phi_n,
\sin\theta_n \sin\phi_n,
\cos\theta_n
\right)
\cdot
\left(
\cos\theta_e \cos\phi_e,
\cos\theta_e \sin\phi_e,
-\sin\theta_e
\right)
\nonumber\\
&=
\sin\theta_n \cos\theta_e
\left(
\cos\phi_n \cos\phi_e + \sin\phi_n \sin\phi_e
\right)
-
\cos\theta_n \sin\theta_e
\nonumber\\
&=
\sin\theta_n \cos\theta_e \cos(\phi_n-
\phi_e)
-
\cos\theta_n \sin\theta_e.
\end{align}

Hence
\begin{align}
B_\theta
&=
\vec{B}_{\mathrm{eff}}\cdot \widehat{e}_\theta
\nonumber\\
&=
B_y \cos\theta_e \sin\phi_e
-
B_z \sin\theta_e
+
B_n
\left[
\sin\theta_n \cos\theta_e \cos(\phi_n-
\phi_e)
-
\cos\theta_n \sin\theta_e
\right].
\label{eq:9-btheta}
\end{align}

\subsection[Derivation of the equation for theta e dot]{Derivation of the equation for \(\dot{\theta}_e\)}

From Eqs.~\ref{eq:9-muedot}, \ref{eq:9-expanded-main}, and \ref{eq:9-mu-cross-b},
\begin{align}
\dot{\theta}_e\,\widehat{e}_\theta
+
\dot{\phi}_e \sin\theta_e\,\widehat{e}_\phi
&=
\gamma_e
\left(
B_\theta \widehat{e}_\phi - B_\phi \widehat{e}_\theta
\right)
+
\mathrm{sgn}(\theta_n-
\theta_e)\,\mathrm{sgn}(\dot{\phi}_e)
k_i
\left(
\dot{\theta}_e\,\widehat{e}_\phi
-
\dot{\phi}_e \sin\theta_e\,\widehat{e}_\theta
\right).
\end{align}

Comparing the \(\widehat{e}_\theta\) components,
\begin{equation}
\label{eq:9-theta-before-abs}
\dot{\theta}_e
=
-\gamma_e B_\phi
-
\mathrm{sgn}(\theta_n-
\theta_e)\,\mathrm{sgn}(\dot{\phi}_e)
k_i \dot{\phi}_e \sin\theta_e.
\end{equation}

Since
\begin{equation}
\mathrm{sgn}(\dot{\phi}_e)\,\dot{\phi}_e = \left|\dot{\phi}_e\right|,
\end{equation}
Eq.~\ref{eq:9-theta-before-abs} becomes
\begin{equation}
\dot{\theta}_e
=
-\gamma_e B_\phi
-
\mathrm{sgn}(\theta_n-
\theta_e)
k_i \left|\dot{\phi}_e\right|\sin\theta_e.
\end{equation}

Now substitute Eq.~\ref{eq:9-bphi}:
\begin{equation}
\label{eq:9-theta-final}
\dot{\theta}_e
=
-\gamma_e
\left[
B_y \cos\phi_e
+
B_n \sin\theta_n \sin(\phi_n-
\phi_e)
\right]
-
\mathrm{sgn}(\theta_n-
\theta_e)
k_i \left|\dot{\phi}_e\right|\sin\theta_e.
\end{equation}

This is the first target equation.

\subsection[Derivation of the equation for phi e dot]{Derivation of the equation for \(\dot{\phi}_e\)}

Now compare the \(\widehat{e}_\phi\) components:
\begin{equation}
\label{eq:9-phi-before-div}
\dot{\phi}_e \sin\theta_e
=
\gamma_e B_\theta
+
\mathrm{sgn}(\theta_n-
\theta_e)\,\mathrm{sgn}(\dot{\phi}_e)
k_i \dot{\theta}_e.
\end{equation}

Divide by \(\sin\theta_e\):
\begin{equation}
\label{eq:9-phi-before-sub}
\dot{\phi}_e
=
\gamma_e \frac{B_\theta}{\sin\theta_e}
+
\mathrm{sgn}(\theta_n-
\theta_e)\,\mathrm{sgn}(\dot{\phi}_e)
k_i \frac{\dot{\theta}_e}{\sin\theta_e}.
\end{equation}

From Eq.~\ref{eq:9-btheta},
\begin{align}
\frac{B_\theta}{\sin\theta_e}
&=
B_y \frac{\cos\theta_e}{\sin\theta_e}\sin\phi_e
-
B_z
+
B_n
\left[
\sin\theta_n \frac{\cos\theta_e}{\sin\theta_e}\cos(\phi_n-
\phi_e)
-
\cos\theta_n
\right]
\nonumber\\
&=
-B_z
-
B_n \cos\theta_n
+
\cot\theta_e
\left[
B_y \sin\phi_e
+
B_n \sin\theta_n \cos(\phi_n-
\phi_e)
\right].
\label{eq:9-btheta-over-sin}
\end{align}

We substitute Eq.~\ref{eq:9-btheta-over-sin} into Eq.~\ref{eq:9-phi-before-sub}:
\begin{align}
\dot{\phi}_e
&=
-\gamma_e
\left\{
B_z
+
B_n \cos\theta_n
-
\cot\theta_e
\left[
B_y \sin\phi_e
+
B_n \sin\theta_n \cos(\phi_n-
\phi_e)
\right]
\right\}
\nonumber\\
&\quad
+
\mathrm{sgn}(\theta_n-
\theta_e)\,\mathrm{sgn}(\dot{\phi}_e)
k_i \frac{\dot{\theta}_e}{\sin\theta_e}.
\label{eq:9-phi-intermediate}
\end{align}

To rewrite the last term in the target form, we impose the branching condition, i.e., the generalized Pauli exclusion principle, that drives \(\theta_e\) away from \(\theta_n\), namely
\begin{equation}
\label{eq:9-relaxation}
\mathrm{sgn}(\dot{\theta}_e)
=
-\mathrm{sgn}(\theta_n-
\theta_e).
\end{equation}

Then
\begin{equation}
\label{eq:9-theta-abs-relation}
\mathrm{sgn}(\theta_n-
\theta_e)\,\dot{\theta}_e
=
-\left|\dot{\theta}_e\right|.
\end{equation}

Using Eq.~\ref{eq:9-theta-abs-relation}, Eq.~\ref{eq:9-phi-intermediate} becomes
\begin{equation}
\label{eq:9-phi-final}
\dot{\phi}_e
=
-\gamma_e
\left\{
B_z
+
B_n \cos\theta_n
-
\cot\theta_e
\left[
B_y \sin\phi_e
+
B_n \sin\theta_n \cos(\phi_n-
\phi_e)
\right]
\right\}
-
\frac{\mathrm{sgn}(\dot{\phi}_e)\,k_i \left|\dot{\theta}_e\right|}{\sin\theta_e}.
\end{equation}

This is the second target equation.



\subsection{Remarks on accuracy and assumptions}

The projection algebra above is exact for the modified vector equation in which the sign factors are evaluated along \(\widehat{z}\) rather than along the exact direction \(\widehat{B}\). The final absolute-value form additionally requires the branching condition.

The effective field is taken to be
\begin{equation}
\vec{B}_{\mathrm{eff}}
=
B_y\,\widehat{y}
+
B_z\,\widehat{z}
+
B_n\,\widehat{\mu}_n .
\end{equation}

The \(z\)-axis is used as the approximate quantization axis. This corresponds to replacing \(\widehat{B}\) by \(\widehat{z}\) in the sign factors of Eq.~\ref{eq:9-start1a}. Since the sign function is discontinuous, this replacement is valid only on branches where the relevant signs are unchanged. A sufficient condition for the first sign factor is
\begin{equation}
\left|
\left(
\widehat{\mu}_n-\widehat{\mu}_e
\right)
\cdot
\left(
\widehat{B}-\widehat{z}
\right)
\right|
<
\left|
\cos\theta_n-\cos\theta_e
\right|.
\end{equation}
A sufficient condition for the second sign factor is
\begin{equation}
\left|
\left(
\dot{\theta}_e\,\widehat{e}_\phi
-
\dot{\phi}_e\sin\theta_e\,\widehat{e}_\theta
\right)
\cdot
\left(
\widehat{B}-\widehat{z}
\right)
\right|
<
\left|
\dot{\phi}_e\sin^2\theta_e
\right|.
\end{equation}

The spherical-coordinate form is valid on the electron coordinate patch
\begin{equation}
0<\theta_e<\pi.
\end{equation}
This restriction is required because the derivation uses \(\phi_e\), \(\cot\theta_e\), and \(1/\sin\theta_e\), which are singular at the poles \(\theta_e=0\) and \(\theta_e=\pi\).
Because the azimuthal angle is undefined at these singular points, we set  \({\dot{\phi}}_{e} = 0\).

The polar angles are taken in their principal ranges,
\begin{equation}
\theta_e,\theta_n\in[0,\pi].
\end{equation}
With this convention, the monotonicity of \(\cos\theta\) on \([0,\pi]\) gives
\begin{equation}
\mathrm{sgn}\!\left(
\cos\theta_n-\cos\theta_e
\right)
=
-\mathrm{sgn}\!\left(
\theta_n-\theta_e
\right).
\end{equation}

The sign reduction
\begin{equation}
\mathrm{sgn}\!\left(
\frac{d\widehat{\mu}_{i,e}}{dt}\cdot\widehat{z}
\right)
=
\mathrm{sgn}(\dot{\phi}_e)
\end{equation}
requires
\begin{equation}
k_i>0,
\qquad
0<\theta_e<\pi.
\end{equation}
Indeed,
\begin{equation}
\frac{d\widehat{\mu}_{i,e}}{dt}\cdot\widehat{z}
=
k_i\dot{\phi}_e\sin^2\theta_e,
\end{equation}
so the sign agrees with \(\mathrm{sgn}(\dot{\phi}_e)\) only when \(k_i>0\) and \(\sin\theta_e\ne0\).

To convert the last term in Eq.~\ref{eq:9-phi-intermediate} into the absolute-value form in Eq.~\ref{eq:9-phi-final}, one must impose the additional branching condition
\begin{equation}
\mathrm{sgn}(\dot{\theta}_e)
=
-\mathrm{sgn}\!\left(
\theta_n-\theta_e
\right).
\end{equation}
It is not obtained by projecting Eq.~\ref{eq:9-start1a} alone. Without this additional condition, the direct consequence of the vector equation is Eq.~\ref{eq:9-phi-intermediate}, not Eq.~\ref{eq:9-phi-final}.

If \(\theta_n\) is treated as fixed, the branching condition above drives \(\theta_e\) away from \(\theta_n\) in polar angle. If the nuclear moment is also evolving, the corresponding condition for increasing polar-angle separation is instead
\begin{equation}
\mathrm{sgn}\!\left(
\theta_n-\theta_e
\right)
\left(
\dot{\theta}_n-\dot{\theta}_e
\right)
>
0.
\end{equation}

We use the convention
\begin{equation}
\mathrm{sgn}(0)=0.
\end{equation}
Therefore, the following
\begin{equation}
\theta_n=\theta_e,
\qquad
\dot{\phi}_e=0,
\qquad
\dot{\theta}_e=0
\end{equation}
must be treated separately.

\subsection{Vector-level nuclear counterpart by symmetry exchange}

By symmetry, we obtain the vector-form nuclear equations of motion by exchanging the roles of the electron and nuclear magnetic moments, namely \(e \leftrightarrow n\). Applying this transformation to Eqs.~\eqref{eq:9-start1a}–\eqref{eq:9-start2}, we obtain
\begin{align}
\frac{d\widehat{\mu}_n}{dt}
&=
\gamma_n\,\widehat{\mu}_n \times \left(\vec{B}+\vec{B}_e\right) 
-
\mathrm{sgn}\!\left(\widehat{\mu}_e \cdot \widehat{B} - \widehat{\mu}_n \cdot \widehat{B}\right)
\mathrm{sgn}\!\left(\frac{d\widehat{\mu}_{i,n}}{dt}\cdot \widehat{B}\right)
\frac{d\widehat{\mu}_{i,n}}{dt},
\label{eq:9-start1b}
\\
\frac{d\widehat{\mu}_{i,n}}{dt}
&\equiv
k_i\,\widehat{\mu}_n \times \frac{d\widehat{\mu}_n}{dt}.
\label{eq:9-start2b}
\end{align}
The structure remains identical, with the sign function reflecting the reversed ordering of projections along \(\widehat{B}\). This preserves the directional consistency of the co-quantum coupling under the exchange symmetry.

\clearpage
\section*{Peer-reviewed version of record}

This manuscript, except Appendices~\ref{appendix:7}--\ref{appendix:9} and their referents, is the version of the article submitted by the author to Journal of Physics B
before copy editing.\, IOP Publishing Ltd is not responsible for any
errors or omissions in this version of the manuscript, or any version
derived from it.\, The Version of Record \cite{Wang2023} is available online at
\url{https://doi.org/10.1088/1361-6455/acc149}.

\end{document}